# Sum-capacity of Interference Channels with a Local View: Impact of Distributed Decisions

Vaneet Aggarwal, Youjian Liu and Ashutosh Sabharwal

**Abstract**

Due to the large size of wireless networks, it is often impractical for nodes to track changes in the complete network state. As a result, nodes have to make distributed decisions about their transmission and reception parameters based on their local view of the network. In this paper, we characterize the impact of distributed decisions on the global network performance in terms of achievable sum-rates. We first formalize the concept of local view by proposing a protocol abstraction using the concept of local message passing. In the proposed protocol, nodes forward information about the network state to other neighboring nodes, thereby allowing network state information to trickle to all the nodes. The protocol proceeds in rounds, where all transmitters send a message followed by a message by all receivers. The number of rounds then provides a natural metric to quantify the extent of local information at each node.

We next study three network connectivities, Z-channel, a three-user double Z-channel and a reduced-parametrization $K$-user stacked Z-channel. In each case, we characterize achievable sum-rate with partial message passing leading to three main results. First, in many cases, nodes can make distributed decisions with only local information about the network and can still achieve the same sum-capacity as can be attained with global information irrespective of the actual channel gains. We label such schemes as universally optimal. Second, for the case of three-user double Z-channel, we show that universal optimality is not achievable if the per node information is below a threshold. In fact, distributed decisions can lead to unbounded losses compared to full information case for some channel gains. Third, using reduced parametrization $K$-user channel, we show that very few protocol rounds are needed for the case of very weak or very strong interference. However, in other regimes, $O(K)$ rounds are essential to achieve sum-capacity.

## I. Introduction

Due to mobility, the network connectivity and channel gains in a wireless network are constantly time-varying. For nodes to make optimal decisions about their transmission and reception parameters, like rate, power, codebooks and decoders, they require full knowledge of the state of the network (defined as the network connectivity and the channel gains on each link) to compute the capacity region and thus their own operational point. However, in large wireless networks, centralizing complete information about the network state implies prohibitive overhead and thus, seldom performed in any practical network. As a result, nodes have to make distributed decisions about their transmission and reception parameters based on their limited local view of the network state. In this situation, the driving question is *can and when do distributed decisions lead to globally optimal network operation?*

In this paper, we consider single-hop interference channels [3–5] where the receiver for each transmitter has a direct connection to its receiver but otherwise the network connectivity is arbitrary. The network state is not known to any node in the network. As a result, none of the nodes know the set of jointly achievable rates and the associated capacity-achieving transmission schemes. In contrast, prior work in quantifying the network capacity implicitly assumes that each node in the network knows the full network state perfectly, e.g. [1–3].

To understand network performance with partial information about the network, we need a natural metric to quantify *extent* of network information each node has about the network. Towards that end, we

V. Aggarwal is with Department of Electrical Engineering, Princeton University. Y. Liu is with Department of Electrical and Computer Engineering, University of Colorado. A. Sabharwal is with Department of Electrical and Computer Engineering, Rice University. The authors were partially supported by NSF DMS-0701226, CCF-0635331, CCF-0728955, and ECCS-0725915, by ONR under grant N00173-06-1-G006, and by AFOSR under grant FA9550-05-1-0443.

propose a protocol abstraction in the form of a local message passing protocol, where the nodes propagate messages related to network state information. The protocol abstraction is inspired by the fact that in a network, the only feasible mechanism available for nodes to learn any information is to pass messages to their neighbors. In fact, local message passing is the building block in all network protocols, like medium access, routing and gossiping. Inspired by belief propagation algorithm commonly used in LDPC decoding [6], the proposed message-passing algorithm proceeds in rounds, where one round consists of a forward and a reverse phase. In the forward phase, each transmitter sends a message and in the reverse phase, each receiver responds with a message. Each message constitutes of only the new information and thus is similar to extrinsic information in belief propagation. While there are many parallels between the proposed algorithm and belief propagation, we will not explore them further in this paper.

The message passing protocol exposes the fundamental capacity problem of interest. With each round the nodes learn more about the network but they do not have full network information till the message passing protocol has terminated. As a result, in the intermediate rounds before the termination of the protocol, not only the nodes have incomplete network information, they may have mismatched view of the network. Thus, we propose to use the number of protocol rounds as a proxy to quantify *extent* of network information each node has about the network state.

The characterization of capacity under partial network state information is non-trivial since the exact network capacity with full information is still unsolved. Thus, we consider three special cases: Z-channel, three-user double Z-channel (two Z channels stacked on top of each other), and a $K$-user stacked Z-channel. In each case, we focus on the sum-rate point on the capacity boundary. Each of the three network connectivities has the largest network diameter for a given number of users, and thus the message passing requires maximum number of rounds to ensure that all nodes have full network information. For the cases when the protocol has not terminated, the resulting network problem is often labeled as *hidden node problem* in network protocol literature [19]. While there is a rich body of literature to design protocols to counter the issues related to hidden nodes, the authors are not aware of any information theoretic capacity analyses with hidden nodes.

We seek *universally optimal strategies*, where each node decides its action based only on its own view of the network but the resulting network sum-rate is equal to the sum-capacity with global information at all the nodes for all network states. Our results are derived for both deterministic and Gaussian channels, and are summarized as follows.

1) *Z-channel*: For the Z-channel, the message-passing protocol requires three full rounds to terminate. In the case of deterministic Z-channel, we show that a unique universally optimal scheme exists with only one and a half rounds of message-passing protocol, even though one of the transmitters does not know all the channel gains. The key feature of the scheme is that transmitters are *politely* as greedy as possible but do not hurt transmission of any other flow about which they have knowledge. The deterministic case is extended to the case of Gaussian Z-channel, where we show that the sum-capacity within 2 bits can be obtained with one and a half rounds of message passing. Note that at least two full rounds are required for all nodes to learn the full network in a Z-channel.
2) *Double Z-channel*: For the case of deterministic double Z-channel, four rounds of message passing are needed. We build on the Z-channel result to propose a distributed rate allocation for double Z-channel and characterize the resulting sum-rate after one and a half rounds, and two and a half rounds, which are in general different. Our result shows the growth in achievable sum-rate with more information about the network. In this case, two and a half rounds suffice to obtain a universally optimal scheme. However, the scheme is no longer unique. For the case of one and a half rounds, we also show a converse result that there exists no distributed scheme which can be universally optimal. In fact, our proposed scheme is optimal for some network states but can have arbitrarily large losses in other cases. Thus, this is the first indication that partial information can significantly reduce network capacity and that loss in network capacity is unavoidable, i.e, every strategy will be suboptimal in certain regime of channel gains.

   In order to prove the above results, the capacity region for the deterministic three user double-Z

channel is found in this paper for a general class of deterministic channels which is later specialized to the deterministic models. The above results have further been extended to a Gaussian model. For the Gaussian double Z-channel, the sum capacity within 4 bits can be achieved with 2.5 rounds of message passing. However for 1.5 rounds, sum-rate optimality can only be guaranteed for a subset of channel gains. To derive the Gaussian result, we derived novel outer bounds for the sum capacity of the Gaussian double Z-channel. Interestingly, we find that treating interference as noise which is optimal for weak interference for Z-channel is only optimal for double-Z channel in the cases of very weak interference.

3) *K-user Z-Channel*: For a general $K$-user stacked Z-channel, we consider a reduced parametrization where all direct links have identical gain and all cross-links are of the same value. Thus, there are only three unknown parameters, size of network $K$, direct link gain and the cross link gain. For both deterministic and Gaussian case, we show that one round is sufficient to achieve sum-capacity if the ratio $\alpha$ of cross-link gain to direct link gain is less than 1/2 (very weak interference) or greater than 2 (very strong interference). In the first case ($\alpha \leq 1/2$), flows treat interference as noise and thus learning about other parts of the network is not useful. In the second case ($\alpha \geq 2$), interference can be completely cancelled out and thus each node can be greedy without requiring any information from other nodes. For $\alpha \in (1/2, 2/3]$, no more than two rounds are required to achieve optimality. And for all other values of $\alpha \in (2/3, 2)$, $O(K)$ rounds are required to achieve optimality. This suggests that network could measure itself adaptively by using more rounds only if certain channel gains are detected in the first round. For the $K$-user case, extension to Gaussian case is provided for a subset of $\alpha$ values. We find that for very weak interference and very strong interference, one round of message passing suffices to achieve the sum capacity for symmetric Gaussian stacked Z-channel.

We make two salient observations about the message passing protocol and its relation to other metrics. First, after $d$ full rounds of message passing, each transmitter knows all channels which are $2d$ hops away and each receiver knows links up to $2d-1$ hops away. After $d.5$ rounds, receiver information increases to $2d+1$ hops. Thus, number of protocol rounds directly relate to a common method to specify local information in network protocol analyses. Second, the messages can be easily related to practical network operation. For example, 1.5 rounds is a common choice in cellular systems, translating to beacons from the base-stations, feedback by the mobile units followed by last half round of a message from the base-station indicate rate and power decisions.

The rest of the paper is organized as follows. In Section II, we describe the channel models. In Section III, we give a general message passing protocol for a $K$ user interference channel. In Section IV, we characterize the sum-rate with partial information at the nodes for a deterministic and Gaussian Z-channel. In Section V, we find the capacity region for a deterministic double Z interference channel and outer bounds for Gaussian deterministic channel. The capacity region is found for a general class of deterministic channels on the lines of [12] which is then specialized to a specific deterministic channel model considered in the paper. We also derive results for 1.5 and 2.5 rounds of message passing for both deterministic and Gaussian channels. In Section VI, we consider a symmetric K-user one-sided channel with unknown $K$ and find the increase in the sum capacity for varying rounds of message passing algorithm. Section VII concludes the paper.

## II. Problem Formulation

### A. Channel Models

We will consider two models for interference channels with $K$ transmitters $\{\mathsf{T}_i\}_{i=1}^{K}$ and $K$ receivers $\{\mathsf{D}_i\}_{i=1}^{K}$: a deterministic model [3] and the additive noise Gaussian model as described follows. We assume that both transmitters and receivers can transmit messages, that is the network is bi-directional in nature.

*1) Deterministic Model:* In a deterministic interference channel, the inputs of $k^{\text{th}}$ transmitter at time $i$ are denoted by $X_{ki} = \left[X_{ki_1}\ X_{ki_2}\ \ldots X_{ki_q}\right]^T \in \{0,1\}^q$, $k = 1, 2, \cdots, K$, such that $X_{ki_1}$ and $X_{ki_q}$ are the most and the least significant bits, respectively. The received signal of user $j$, $j = 1, 2, \cdots, K$, at time $i$ is denoted by the vector $Y_{ji} = \begin{bmatrix} Y_{ji_1} \ Y_{ji_2} \ \ldots \ Y_{ji_q} \end{bmatrix}^T \in \{0,1\}^q$. Specifically, the received signal $Y_{ji}$, of an interference channel is given by

$$Y_{ji} \;=\; \bigoplus_{k=1}^{K} \mathbf{S}^{q-n_{kj}} X_{ki} \tag{1}$$

where $\oplus$ denotes the XOR operation, and $\mathbf{S}^{q-n_{kj}}$ is a $q \times q$ shift matrix with entries $S_{m,n}$ that are non-zero only for $(m,n) = (q - n_{kj} + n, n), n = 1, 2, \ldots, n_{jk}$. We will also use $X_k^n$, $Y_k^n$, etc. to denote $(X_{k1}, \cdots, X_{kn})$, $(Y_{k1}, \cdots, Y_{kn})$, etc. Associated with each transmitter $k$ and receiver $j$ is a non-negative integer $n_{kj}$ that defines the number of bit levels of $\mathbf{X}_k$ observed at receiver $j$. The maximum number of bits supported by any link is $q = \max_{k,j}(n_{kj})$. The network can be represented by a square matrix $H$ whose $(i,j)^{th}$ entry is $H_{ij} = n_{ij}$. We note that $H$ need not be symmetric.

*2) Gaussian Model:* In a Gaussian interference channel, the inputs of $k^{\text{th}}$ transmitter at time $i$ are denoted by $X_{ki} \in \mathbb{C}$, $k = 1, 2, \cdots, K$, and the outputs at $j^{\text{th}}$ receiver in time $i$ can be written as $Y_{ji} \in \mathbb{C}$, $j = 1, 2, \cdots, K$. The received signal $Y_{ji}$, $j = 1, 2, \cdots, K$ is given by

$$Y_{ji} \;=\; \sum_{k=1}^{K} h_{kj} X_{ki} + Z_{ji}, \tag{2}$$

where $h_{kj} \in \mathbb{R}^+$ is the channel gains associated with each transmitter $k$ and receiver $j$, and $Z_{ji}$ are additive white complex Gaussian random variables of unit variance. We will also use $X_k^n$, $Y_k^n$, etc. to denote $(X_{k1}, \cdots, X_{kn})$, $(Y_{k1}, \cdots, Y_{kn})$, etc. Further, the input $X_{ki}$ has an average power constraint of unity, i.e. $\mathbb{E}(|X_{ki}|^2) \leq 1$ (where $\mathbb{E}$ denotes the expectation of the random variable). Only in Section III, we will make an exception where no power constraint will be imposed on messages sent by transmitters or receivers.

Like the deterministic case, we represent the network state by a square matrix $H$ whose $(i,j)^{th}$ entry is $H_{ij} = |h_{ij}|^2$. Thus we will use the matrix $H$ for both the deterministic and the Gaussian model, where the usage will be clear from the context.

### *B. Per Node Local View*

Our objective is to understand the impact of nodes' decisions on network sum-rate, when the decisions are based on their partial information about the matrix $H$. For transmitters we will denote this partial information about the network as $N_k$ and as $N'_k$ for the receivers. If the nodes know nothing about the network matrix $H$ (i.e no information about its size or entries), then $N_k = N'_k = \Phi$ (empty set), which is equal to assuming that there is no other node in the network. On the other hand, if the nodes know everything about the network, then $N_k = N'_k = H$ and is also the most commonly assumed scenario in most information-theoretic analyses [4, 15, 18].

We now define network state and network connectivity. We assume that that there is a direct link between every transmitter $\mathsf{T}_i$ and its intended receiver $\mathsf{D}_i$. On the other hand, if a cross-link between transmitter $i$ and receiver $j$ does not exist, then $H_{ij} \equiv 0$. Given a network, its connectivity is a set of edges $\mathsf{E} = \{(\mathsf{T}_i, \mathsf{D}_j)\}$ such that a link $\mathsf{T}_i - \mathsf{D}_j$ is not identically zero. Then the set of network states, $\mathcal{G}$, is the set of all weighted graphs defined on $\mathsf{E}$. For the deterministic model, the set of network states can be written as

$$\mathcal{G}(\mathsf{E}) = \{H : H_{ij} \equiv 0 \text{ if } (\mathsf{T}_i, \mathsf{D}_j) \notin \mathsf{E} \text{ else } H_{ij} \in \{0, 1, \ldots, q\}\},$$

and in the Gaussian model as

$$\mathcal{G}(\mathsf{E}) = \{H : H_{ij} \equiv 0 \text{ if } (\mathsf{T}_i, \mathsf{D}_j) \notin \mathsf{E} \text{ else } H_{ij} \geq 0\}.$$

Note that the channel gain can be zero but not guaranteed[1] to be if the node pair $(\mathsf{T}_i, \mathsf{D}_j) \in \mathsf{E}$.

Our main focus is the case where $N_k, N'_k$ for each $k$ is only a subset of the whole matrix. Thus, each node knows the network matrix partially. In fact, it is quite possible that nodes know only a few entries of the matrix and do not know the size of the whole matrix $H$, i.e network size. As we will see later, this partial knowledge of the network matrix at each node leads to the case where each node's knowledge about the network is mismatched from other nodes in the network. That is, each node possibly knows a different part of the whole matrix $H$. We will study the achievable sum-rate as the network information at each node grows from no-information to full information. In Section III, we will define a special trajectory of sequence of growing network information which is directly connected to protocols in practical systems and is also related to commonly used metric of 'number of hops' to denote amount of side information at each node. To aid analysis, we will assume that all nodes are provided some side information, $\mathsf{SI}$, about the network state before the onset of the protocol. Thus, nodes may have non-zero information about the network before even a single message is sent. We next define sum-capacity.

### C. Sum Capacity

First consider the $K$-user deterministic interference channel. For each user $k$, let message index $m_k$ be uniformly distributed over $\{1, 2, ..., 2^{nR_k}\}$. The message is encoded as $X_k^n$ using the encoding functions $e_k(m_k|N_k, \mathsf{SI}) : \{1, 2, \ldots, 2^{nR_k}\} \mapsto \{0, 1\}^{nq}$, which depend on the local view, $N_k$, and side information about the network, $\mathsf{SI}$. The message is decoded at the receiver using the decoding function $d_k(Y_k^n|N'_k, \mathsf{SI}) : \{0, 1\}^{nq} \mapsto \{1, 2, \ldots, 2^{nR_k}\}$, where $N'_k$ is the receiver local view and $\mathsf{SI}$ is the side information. The corresponding probability of decoding error $\lambda_k(n)$ is defined as $\Pr[m_k \neq d_k(Y_k^n|N'_k, \mathsf{SI})]$. A rate tuple $(R_1, R_2, \cdots, R_K)$ is said to be achievable if there exists a sequence of codes such that the error probabilities $\lambda_1(n), \cdots \lambda_K(n)$ go to zero as $n$ goes to infinity. The closure of the set of achievable rate tuples is defined as the capacity region $\mathcal{C}$.

Now, consider the $K$-user Gaussian interference channel. For each user $k$, we again assume that the message index $m_k$ is uniformly distributed over $\{1, 2, ..., 2^{nR_k}\}$. Further, we use the same notation for encoding and decoding functions. Thus, encoding functions are $e_k(m_k|N_k, \mathsf{SI}) : \{1, 2, ..., 2^{nR_k}\} \to \mathbb{C}^n$ and decoding functions are $g_k(Y_k^n|N'_k, \mathsf{SI}) : \mathbb{C}^n \to \{1, 2, ..., 2^{nR_k}\}$. The corresponding probability of decoding error $\lambda_k(n)$ defined as $\Pr[m_k \neq g_k(Y_k^n|N'_k, \mathsf{SI})]$. A rate tuple $(R_1, R_2, \cdots, R_K)$ is said to be achievable if there exists a sequence of codes such that the error probabilities $\lambda_1(n), \cdots \lambda_K(n)$ go to zero as $n$ goes to infinity. Again, the closure of the set of achievable rate tuples is defined capacity region $\mathcal{C}$.

The sum capacity in both cases is defined as

$$C_{sum} = \sup \left\{ \sum_{i=1}^{K} R_i : (R_1, \cdots, R_K) \in \mathcal{C} \right\}. \tag{3}$$

We note that all encoding and decoding functions depend only the local and side information at the transmitters and receivers about the network. In this case, the nodes have to operate with the local knowledge $N_k, N'_k$ and the side information $\mathsf{SI}$ so that the probability of error at the receivers go to zero as $n$ goes to infinity for all $H \in \mathcal{G}(\mathsf{E})$, leading to a compound channel capacity formulation. In this paper, optimal sum capacity refers to the sum capacity with the full state information at all the nodes.

When the local information about the network is mismatched, the nodes can take actions which can work against each other and in the process reduce the sum-rate of the network. This issue of making *distributed* decisions about rate, power, codebooks and decoder is fundamental in most networks and is the main topic of study in this paper.

[1]The model is inspired by fading channels, where the existence of a link is based on its average channel gain. On the average the link gain may be above noise floor but its instantaneous value can be below noise floor.

## III. Learning Network State Information

In this section, we describe a protocol which uses local message passing to propagate network state information to the nodes in the network. The protocol is described in terms of entries of matrix $H = [H_{ij}]$ and thus applies to both deterministic and Gaussian models. Our motivation is to find the most relevant cases of local view $\{N_k, N'_k\}$ we should consider.

### *A. Why Learn the Network ?*

Before we dive into the details of learning the network state information, it is important to understand why we might need to even estimate and propagate network state information. One could adopt a "non-coherent" approach, where no resources are wasted in estimating any channel or network connectivity, and nodes code such that reliable communication is possible without any network state information, i.e, $N_k = N'_k = \Phi$. However, in compound capacity formulation, the capacity region with no information (local or side) about the network is a singleton, where the only possible rate tuple is all zero-rate tuple. This follows directly from point to point Gaussian channel where the compound capacity is zero if the link state is unknown to the transmitter since in worst case the link gain can be zero. Thus, to achieve a non-zero rate, the network information at nodes should be non-trivial.

The obvious next question is what cases of network information $\{N_k, N'_k\}_k$ should one consider. For a $K$-user network, the matrix has $K^2$ entries, which implies that the per node information can be any of $2^{K^2}$ cases. Thus, there are $2^{2K^3}$ possible combination of side information cases. This large number of cases quickly becomes intractable. However, we contend that most of these side information cases are not of practical interest.

A common metric to capture extent of network view at each node is number of hops (e.g see [20, and references therein]). That is equivalent to each node $k$ knowing a sub-matrix of $H$. The metric is motivated by message-passing algorithms which broadcast and forward information about local state. A clear advantage of this metric that it greatly reduces the number of local information sub-cases one needs to consider and there is a direct relation with actual protocols which gather this side information. We propose to adopt a related metric which is equally concise and tightly related to protocols in many operational networks.

### *B. Message Passing Protocol*

For nodes to learn and propagate the network state, they have to communicate with each other. This inter-node communication is possible only with nodes to which there is a direct link, i.e, messages have to be exchanged locally and those messages are then processed and propagated to other nodes. This obvious construct of local message passing is central to all multi-hop network protocols. In our development, the only practical reality we will be concerned with is that direct communication is possible only between neighbors and its impact on amount of network state information at each node. Hence, we will simplify some of the implementation complexities as follows.

The proposed message passing protocol proceeds in rounds, where each full round has two phases: a *forward* phase where all transmitters broadcast a message and a *reverse* phase where all receivers broadcast a message each. We assume that all messages are scheduled so that there are no "collisions" at any of the nodes in receiving mode due to simultaneous transmissions. Finally, the broadcast messages can only be heard by nodes to which the sending nodes has direct links (the links that are in the network connectivity $\mathsf{E}$), thus no extra feedback or Genie channels are available.

The message broadcasted by the transmitter $k$ in round $t$ (transmitters are data sources) is labeled $m_{k,t}$, which is received by all the receivers $j$ who have direct links to transmitter $k$. Analogously, the message broadcasted by the receiver $k$ at round $t$ is labeled $M_{k,t}$, which is received by all the transmitters $j$ who have a direct link to receiver $k$.

Recall that each node's information about the network is represented by either $N_k$ (for transmitters) or $N'_k$ (for receivers). Instead of assuming that the nodes have no information to start with (i.e $N_k = N'_k = \Phi$), we will consider special sub-cases where all nodes have some minimal side information $\mathsf{SI}$ about the network. The assumption of side information is largely for analytical simplification. This will only change the contents of messages and not the message-computation and passing rules. For the following description, we will assume that at time $t = 0$, each node knows the size of the network or $\mathsf{SI} = \{K\}$. Thus, the nodes know the size of the matrix $H$ but do not know any of its entries. An alternate case will also be considered where the size of network, $K$, will be assumed unknown at the onset of the protocol but nodes will have a priori knowledge of the form of the matrix $H$.

The message passing protocol with knowledge of side information $\mathsf{SI} = \{K\}$ is described below.

1) **Round 1 (Forward)**: Since none of the entries in the matrix $H$ are known, the first message from each transmitter is a known training signal along with the transmitter identity. Thus $m_{k,1} = \{\psi_k, \mathsf{T}_k\}$, where $\psi_k$ is the training signal from transmitter $k$. At the end of the transmitter messages, receiver $j$ knows the non-zero elements of column $j$ of matrix $H$ learnt via channel estimation (however may not know the value of $j$).
   **Round 1 (Reverse)**: The receiver $k$ broadcasts $M_{k,1} = \bigcup_{i \in \mathsf{E}_k} \{(H_{i,k}, \mathsf{T}_i, \mathsf{D}_k)\}$, where $\mathsf{E}_k$ is the set of vertices connected to receiver $k$. Transmitter $\mathsf{T}_j$ can receive $M_{k,1}$ if it has a direct link to receiver $k$. This completes the first round.
2) **Round** $t > 1$: In round $t > 1$, nodes only forward new information which is computed as follows. In the forward phase for transmitters, the broadcast message is

$$m_{k,t} = \bigcup_{j \in \mathsf{J}_k} M_{j,t-1} \setminus \bigcup_{t'=2}^{t-1} m_{k,t'} \setminus \bigcap_{j \in \mathsf{J}_k} \left\{ \bigcup_{t'=1}^{t-1} M_{j,t'} \right\}, \tag{4}$$

   where $\mathsf{J}_k$ is the set of vertices connected to transmitter $k$. The message $m_{k,t}$ is a concatenated version of its received messages from previous round minus the messages it has broadcasted in previous transmissions and those that are already known to all of its neighbors.
   In response, the receivers broadcasts following in the reverse phase

$$M_{k,t} = \bigcup_{j \in \mathsf{E}_k} m_{j,t} \setminus \bigcup_{t'=1}^{t-1} M_{k,t'} \setminus \bigcap_{j \in \mathsf{E}_k} \left\{ \bigcup_{t'=2}^{t} m_{j,t'} \right\}. \tag{5}$$

   The message $M_{k,t}$ is the concatenation of its received message minus its previously broadcasts messages and after removing what is known to all its neighboring transmitters. The messages $m_{k,t}$ and $M_{k,t}$ are similar to the extrinsic information in belief propagation with the main difference being that the messages are broadcasts.
3) **Stopping Rule**: If a transmitter or receiver has no new updates, it sends a NULL message $\phi$ in its assigned time-slot. Thus, nodes only forward information when new information is received and send "nothing" otherwise. When all the neighbors of a node send a NULL message, each node stops sending any new messages (even NULL messages).

The above protocol is similar to message passing for belief propagation on factor graphs, often used in LDPC decoding [6]. Belief propagation is closer to gossiping in networks [7], where a node can talk only one node at any time. Our proposed approach exploits the broadcast nature of wireless, and hence is closer to broadcasting in networks [8]. While connections between broadcast-based generalization of belief propagation [9] and proposed message passing are of interest, they are beyond the scope of this paper.

Before we proceed, we quickly note our use of word "message." In this paper, we use messages for transmissions which may or may *not* depend on information bits. In contrast, the usual information-theoretic parlance, message often only refers to the "raw" information bits (the information-bearing message) sent from sender to the receiver [21]. In our case, the messages do carry information but about the network state. Thus, they are similar to control messages, like training, feedback, ARQ etc, in networks.

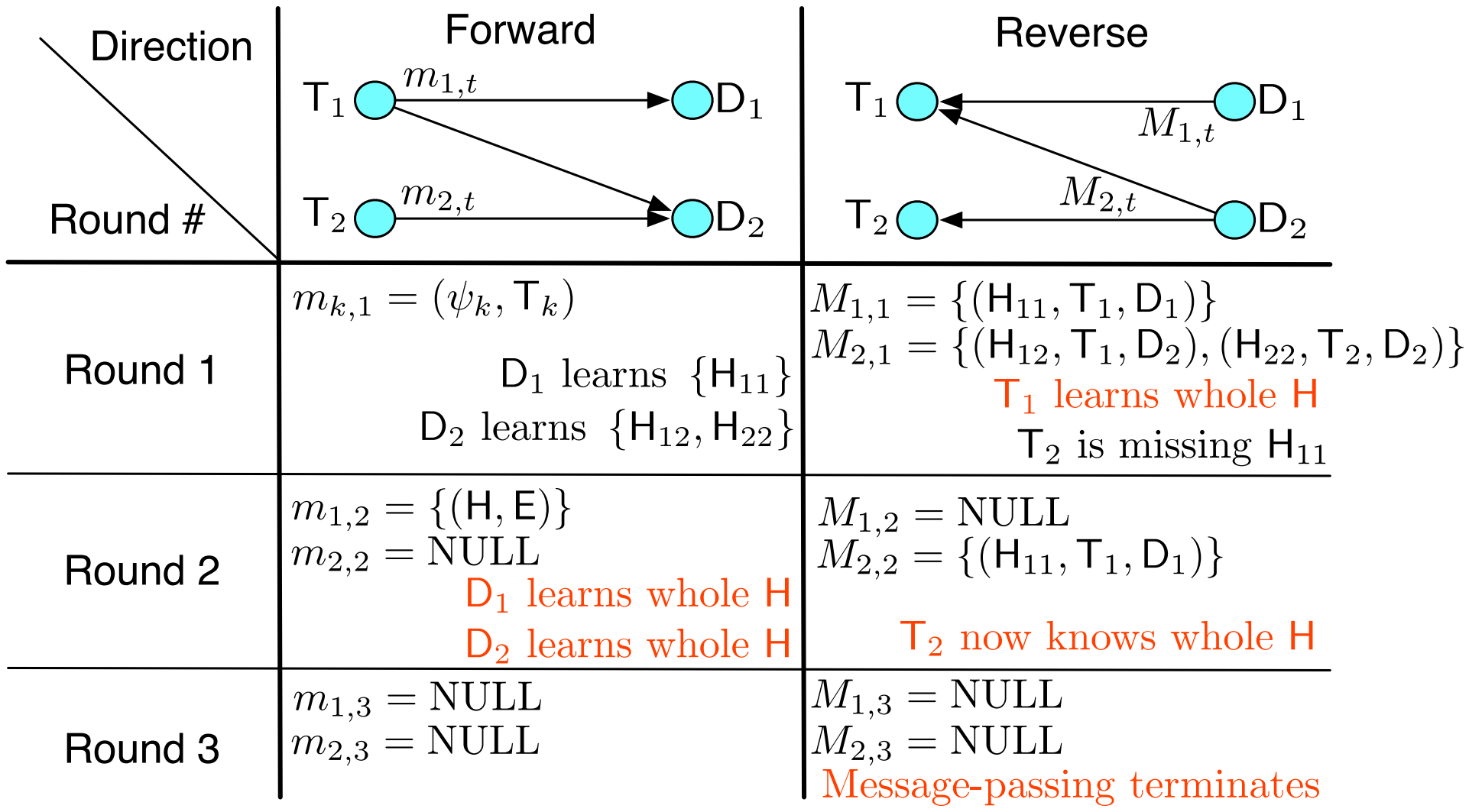

Fig. 1. *Full three rounds of message-passing protocol in Z-channel.*

### *C. Notes on Message Passing Protocol*

It is instructive to consider an example to understand the properties of the message passing protocol; Figure 1 shows the steps for the Z-channel. Some key facts are as follows:

1) *Form of Messages*: Each non-null message is an unordered set of channel coefficients along with the location in the matrix indicated by the transmitter-receiver identity. This ensures that the set intersections and union in (4) and (5) are well defined.
2) *Guaranteed Termination in $K+1$ rounds*: Since the channel estimation is assumed to be perfect and messages face no losses, the message passing protocol is guaranteed to converge in $K+1$ rounds. However, the actual number of rounds depends on the graph diameter. If the interference network is fully connected, then message passing terminates after 2 rounds. However, for a stacked Z-network (see Section VI), full $K+1$ rounds are required.
3) *Impact of Side Information*: The messages can be reduced in length if any side information about the network is known. In the preceding discussion, we assumed that the network size is known and hence the size of the matrix is assumed known. As a result, we do not need to estimate the network size.

   In Section VI, we will consider a class of $K$-user interference channel which is of the form of stacked Z-channels. For the $K$-user channel, we will assume that the network is parameterized by only three parameters $K$, $H_{11}$ and $H_{12}$ where direct channel gains are $H_{11}$ and $H_{12}$ are all cross channel gains in the network. All the users know that the network is parameterized by three parameters but do not know the value of these parameters or their placement in the network. Thus, $\mathsf{SI} = \{\text{stacked symmetric Z-channel connectivity}\}$ is the side knowledge about network connectivity (will be described more precisely in Section VI). Hence, we assume that these are the only three parameters that the nodes need to learn besides knowing their relative position in the network (how many users above/below a particular user). In the first round (forward), all the receivers except the first would know $H_{11}$ and $H_{12}$ and hence they can directly tell the transmitters $H_{11}$ and $H_{12}$. Thus, the only other information which will not be available after the first round is the relative location in the network which depends on the total number of nodes $K$. Let the set of node identities of the transmitters that the receiver $\mathsf{D}_j$ is connected to be $\mathcal{T}_j$. The message passing protocol simplifies as follows.

   **Round 1 (Forward)**: Since none of the entries in the matrix $H$ are known, the first message from each transmitter is a known training signal along with the transmitter identity. Thus $m_{k,1} = \{\psi_k, \mathsf{T}_k\}$, where $\psi_k$ is the training signal from transmitter $k$ and $\mathsf{T}_k$ its node identity.

**Round 1 (Reverse)**: The receiver 1 broadcasts $M_{1,1} = \{H_{11}\} \bigcup M'_{1,1}$ while the receiver $k \geq 1$ broadcasts $M_{k,1} = \{(H_{11}, H_{12})\} \bigcup M'_{k,1}$, where $M'_{i,1} = \{(\mathsf{D}_i, \mathcal{T}_i)\}$. Transmitter $j$ can receive $M_{k,1}$ if it has a direct link to receiver $k$.

All the further rounds proceed as before replacing $M_{k,1}$ with $M'_{k,1}$, except that $m_{1,2}$ also includes broadcasting $H_{12}$ so that $D_1$ knows the cross channel gain. This is possible since Node 1 will know that its position in the network after first round based on its side information $\mathsf{SI}$.

4) *Full and Partial Network Information*: The case of full information is equivalent to the case of message passing operating till termination. With fewer than maximum number of rounds, at least one of the nodes may not have full network information. For example, in Figure 1, less than 2 rounds imply that there is some node (e.g. $\mathsf{T}_1$) which does not have full information.
5) *Mismatched Local Views*: The protocol naturally exposes one of the key issues in networks that different nodes will have different information about the network state if the protocol is not carried to its completion. For large networks, taking the protocol to completion would imply a large number of rounds, and is thus impractical. For example, after 1.5 rounds (first full round + only forward phase in second round) in Z-channel, $\mathsf{T}_2$ does not know $\mathsf{T}_1 \to \mathsf{D}_1$ channel, while other nodes know the whole network. We will thus use number of rounds as a *proxy* for amount of local information, with each extra round providing increased information about the network.
6) *Relation to Hop Length*: By observing the time-line in message passing in Figure 1, it is straightforward to conclude that different nodes learn about the whole network at different times. Each full round allows transmitters to learn two extra hops of information and receivers one extra hop. In $d$ rounds, a transmitter will know all routes which are $2d$-hop long rooted at that transmiter, and a receiver knows routes of length $2d-1$. That is why, in Figure 1, $\mathsf{T}_2$ needs 2 rounds to learn about $\mathsf{H}_{11}$, since it is three hops away.
7) *Relation to Practical Protocols*: The messages can be translated into practical network operation. First round is training, like physical layer preamble in most networks. And rest of the rounds can be understood as channel feedback, like often studied in [10]. In practice, networks operate with very *few* rounds of message passing. For cellular networks, 1.5 rounds is a common choice, roughly translating to beacons from base-stations, feedback from mobiles and a rate allocation decision from the base-station. Thus, the case of 1.5 rounds will be of special interest to us.

### D. Optimal Strategies

We now formally define the concept of universally optimal strategy with partial information. Suppose that the transmitter $i$ knows local information $N_i$, the receiver $i$ knows local information $N'_i$ and all the nodes know side information $\mathsf{SI}$.

**Definition 1** (Universal Optimality)**.** *A universally optimal strategy with partial information at nodes ($N_i$ at transmitter $i$, $N'_i$ at receiver $i$ and $\mathsf{SI}$ at all the nodes) is defined as the strategy that each of the transmitter $i$ uses based on its local information $N_i$ and side information $\mathsf{SI}$, such that following holds. The strategy yields a sequence of codes having rates $R_i$ at the transmitter $i$ such that the error probabilities at the receiver, $\lambda_1(n), \cdots \lambda_K(n)$, go to zero as $n$ goes to infinity, satisfying*

$$\sum_i R_i = C_{sum}$$

*for all the sets of network states consistent with the side information. Here $C_{sum}$ is the sum-capacity of the whole network with the full information.*

**Definition 2** (Approximate Universal Optimality)**.** *An approximate universally optimal strategy with partial information at nodes ($N_i$ at transmitter $i$, $N'_i$ at receiver $i$ and $\mathsf{SI}$ at all the nodes) is defined as the strategy that each of the transmitter $i$ uses based on its local information $N_i$ and side information $\mathsf{SI}$, such that* *following holds. There exist a sequence of codes having rates $R_i$ at the transmitter $i$ such that the error probabilities at the receivers, $\lambda_1(n), \cdots \lambda_K(n)$, go to zero as $n$ goes to infinity, satisfying*

$$\sum_i R_i \geq C_{sum} - \tau$$

*for all the sets of network states consistent with the side information. Here $C_{sum}$ is the sum-capacity of the whole network with the full information and $\tau$ is a fixed constant independent of the channel gains.*

Thus, an (approximate) universally optimal strategy is one where for all network states consistent with the side information, decisions based on local information and the side information lead to (approximately) globally optimal solutions. We will use the notion of universal optimality for the deterministic model. Since we do not know the exact capacity region for the Gaussian interference channel, the notion of universal optimality will be replaced by approximate universal optimality.

In this paper, we will assume that the local information at the nodes is obtained by a message passing protocol when run for $d$ or $d.5$ rounds for $d \geq 0$. In the case of $d.5$ rounds, the last 0.5 round of message passing represents the message from the transmitters to the receivers but no message in the reverse direction. This is to ensure that the receivers know more than the transmitters so that reliable decoding can take place.

## IV. Two User Z-channel

The smallest possible network of interest is a two-user interference channel. There are three network connectivities in this case, a fully connected bi-partite graph (interference channel), a Z-channel (one cross-link is missing) and two decoupled flows (two point-to-point links). We assume that all the nodes know that there are a total of two nodes in the network, or $\mathsf{SI} = \{K = 2\}$. From the message passing protocol in Figure 1, it is clear that 1.5 rounds are sufficient for every node in the network to learn the whole of the state information for both fully connected bipartite graph and two decoupled flow connectivities. So the strategies decided with local view result in globally optimal strategy decisions. Hence we will focus our interest on the Z-channel, where 1.5 rounds do not result in full information at all the nodes; $\mathsf{T}_2$ does not know $H_{11}$ after 1.5 rounds.

### A. Deterministic Z-Channel

In a deterministic Z-channel, $K = 2$ and $n_{21} = 0$. Specifically, the received signal $Y_{ji}$, $j = 1, 2$, of a Z-channel is given by

$$Y_{1i} = \mathbf{S}^{q-n_{11}} X_{1i} \quad (6a)$$

$$Y_{2i} = \mathbf{S}^{q-n_{12}} X_{1i} \oplus \mathbf{S}^{q-n_{22}} X_{2i} \quad (6b)$$

The network state message passing is described in Figure 1. After two full rounds, every node in the network knows the complete network state, i.e. the matrix $H$ is known completely to all four nodes. In this case, the achievable capacity region is also known exactly [5, 11, 12].

**Theorem 1** ([5, 11, 12])**.** *The deterministic channel capacity region for a two-user Z-channel is the set of nonnegative rates $(R_1, R_2)$ satisfying*

$$R_1 \leq n_{11}$$

$$R_2 \leq n_{22}$$

$$R_1 + R_2 \leq \max(n_{22}, n_{12}, n_{11}, n_{11} + n_{22} - n_{12}) \quad (7a)$$

Since our main interest is in the case of partial information, we ask if fewer than two full rounds suffice to achieve sum capacity. The following theorem proves that 1.5 rounds are sufficient to achieve sum capacity for all two-user $H$, i.e. a universally optimal rate allocation exists for all $H$. Note that we assume that the network size of $K = 2$ is known to all the nodes. Further, in 1.5 rounds of message passing all nodes know that $n_{21} = 0$. Thus, all nodes know that the connectivity is that of a Z-channel.

**Theorem 2.** *The sum capacity for a Z-channel can be achieved without completing the full message passing algorithm. To be precise, only the first 1.5 rounds are required to achieve the full-knowledge sum-capacity with the side information that the network size is $K = 2$.*

*Proof:* With the side information and 1.5 rounds of message passing, all the nodes know if they are the top user or the bottom user of the Z-channel. In Appendix A, we show that each transmitter can use only their local view to decide their rate and codebooks such that full-knowledge sum-capacity is achievable. ∎

**Corollary 1.** *There exist a universally optimal strategy with the local information $N_i$ and $N_i'$ provided by 1.5 rounds of message passing and the side information that $K = 2$.*

*Proof:* With 1.5 rounds of message passing and side information $K = 2$, all the nodes know the network connectivity. For all the connectivity choices, a strategy can be found by the nodes based on the channel gains they know that would achieve the full-information sum-capacity. For example, if the network is fully-connected, the nodes can use the node indices to order themselves since the labels $T_i$ are unique and thus the nodes can compute an optimal strategy. If the network is a Z-channel, the strategy of Z-channel described in the proof of Theorem 2 can be used. ∎

Thus, there is no loss in the performance even if the second transmitter $\mathsf{T}_2$ does not know about the whole network state and schemes with partial information exist which are *universally optimal* for all Z-channel $\mathsf{H} \in \mathcal{G}(\mathsf{E}_Z)$, where $\mathsf{E}_Z$ is the $\mathsf{E}_Z$ is the Z-channel connectivity. It is very instructive to closely study the structure of the rate allocation scheme.

Since transmitter $\mathsf{T}_2$ does not know the direct channel $\mathsf{H}_{11}$, it in fact chooses to *ignore* the presence of the other transmitter. As a result, $\mathsf{T}_2$ acts in a greedy fashion and sends at full rate of $\mathsf{H}_{22}$ bits. On the other hand, $\mathsf{T}_1$ knows the whole network and that it interferes with $\mathsf{D}_2$. It can then choose a transmission scheme which sends information *below* the noise floor of $\mathsf{T}_2$'s transmission and if possible, *above* $\mathsf{T}_2$'s signal. Figure 2(a)-(b) depicts the allocations for the case of deterministic channels.

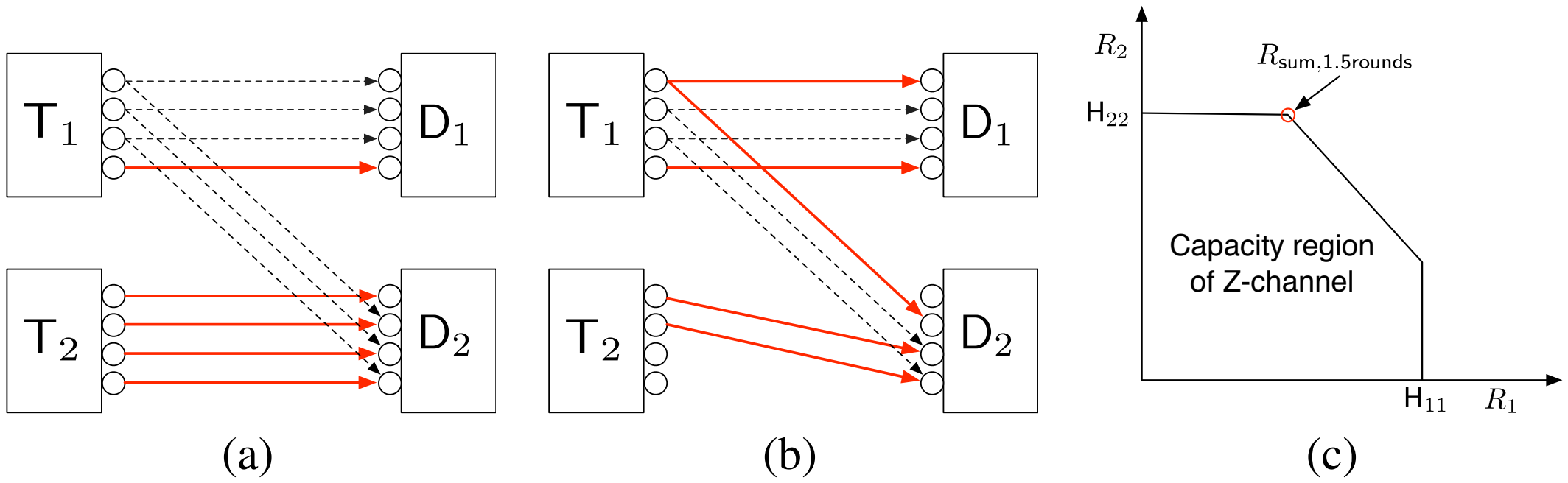

Fig. 2. *(a) and (b): Allocation for two possible deterministic Z-channels, and (c) corner point achieved by the rate allocation with partial information. [Note: solid red lines mean levels in use and dashed lines imply levels which are turned off.]*

The schemes which achieve optimal sum capacity follow two rules. First, if a transmitter does not have enough information about other links, it acts *greedily* and sends at its maximum possible link rate. This is the case for the $\mathsf{T}_2$, which does not know about $\mathsf{T}_1 \rightarrow \mathsf{D}_1$ link. Second, if the link does know who it is causing interference to and by how much, then it ensures that it only sends at rates and powers, which do *not* impede on the success of other flows. This is the case for transmitter $\mathsf{T}_1$, which knows that it is causing interference at receiver $\mathsf{D}_2$. In short, the transmitters act greedily and politely by maximizing their individual rate but constraining themselves *not* to hurt other flows about which they have sufficient information.

We next show that the rate allocation in the above strategy is the only unique possible way to get a universally optimal strategy with 1.5 rounds of message passing.

**Theorem 3.** *For the network connectivity of Z-channel, there exists a unique distributed rate allocation strategy that is universally optimal with 1.5 rounds of message passing protocol and is given by the strategy of Theorem 2.*

*Proof:* To show the uniqueness, we first consider the strategy of the second user. Since the second user does not know $n_{11}$ $\mathsf{T}_2$ has to transmit at $n_{22}$ to avoid being sub-optimal for the case of $n_{11} = 0$. This implies that irrespective of the channel gains $n_{22}$ and $n_{12}$ no strategy that involves the rate of the second user other than $n_{22}$ can be universally optimal.

Now given that the above $\mathsf{T}_2$ strategy of sending at full rate, the first user cannot assign any other rate than in the proof of Theorem 3, since only this rate would result in the optimality of the sum rate given that the second user is sending at $n_{22}$. ■

Perhaps the most interesting aspect of Theorem 2 is that *locally optimal rates achieve globally optimal sum capacity*. The reason is that the proposed distributed scheme achieves a corner point on the capacity region of Z-channel, as shown in Figure 2(c), which explains the uniqueness as shown in Theorem 3. Thus an interesting observation is that *corner points need less information about the network at some of the nodes*. We will observe this fact again for a bigger network in Section V.

### *B. Gaussian Z-Channel*

In a Gaussian Z-channel, $K = 2$, and $h_{21} = 0$. The received signal $Y_{ji}$, $j = 1, 2$, of a Z-channel is given by

$$Y_{1i} = h_{11}X_{1i} + Z_{1i} \tag{8a}$$

$$Y_{2i} = h_{12}X_{1i} + h_{22}X_{2i} + Z_{2i}. \tag{8b}$$

Also, let $\mathsf{SNR}_i = |h_{ii}|^2$, $i \in \{1, 2\}$ and $\mathsf{INR}_2 = |h_{12}|^2$. We have assumed that all $h_{ij}$'s are real and positive and hence use $h_{ii} = \sqrt{\mathsf{SNR}_i}$, $h_{12} = \sqrt{\mathsf{INR}_2}$.

After two full rounds, every node in the network knows the complete network state, i.e. the matrix $H$ is known completely to all four nodes. In this case, an upper bound on the capacity region is given in the following Theorem.

**Theorem 4** ([4, 16–18])**.** *The channel capacity region for a two-user Gaussian Z-channel is upper bounded by the region formed by the set of nonnegative rates* $(R_1, R_2)$ *satisfying*

$$R_1 \leq \log(1 + \mathsf{SNR}_1) \tag{9a}$$

$$R_2 \leq \log(1 + \mathsf{SNR}_2) \tag{9b}$$

*If* $\mathsf{INR}_2 \leq \mathsf{SNR}_1$,

$$R_1 + R_2 \leq \log(1 + \mathsf{SNR}_1) + \log\left(1 + \frac{\mathsf{SNR}_2}{1 + \mathsf{INR}_2}\right). \tag{10}$$

*If* $\mathsf{INR}_2 \geq \mathsf{SNR}_1$,

$$R_1 + R_2 \leq \log(1 + \mathsf{SNR}_2 + \mathsf{INR}_2). \tag{11}$$

We now focus on 1.5 rounds of message passing in Gaussian channel. In Section IV-A, we showed that for deterministic Z channel, the sum capacity can be achieved with 1.5 rounds of message passing. We now show that sum capacity within 2 bits can be achieved for a two-user Gaussian Z-channel with 1.5 rounds of message passing. Thus, there exist an approximate universally optimal strategy with $\tau = 2$.

**Theorem 5.** *The sum capacity for a Gaussian two user Z-channel can be achieved within two bits with the local information $N_i$ and $N_i'$ at the nodes obtained with 1.5 rounds of message passing and the side information about network size, i.e,* $\mathsf{SI} = \{K = 2\}$.

*Proof:* With 1.5 rounds of message passing, the two nodes know the network connectivity, i.e, if they are the first user (that causes interference to the other) or the second.

The second transmitter uses a codebook of rate

$$R_2(\mathsf{SNR}_2, \mathsf{INR}_2) = \log\left(1 + \mathsf{SNR}_2(1 + \mathsf{INR}_2) / (1 + 2\mathsf{INR}_2)\right), \tag{12}$$

with a power level of $P_2 = 1$ to transmit. The first user however uses a common and a private message with rates

$$R_{1,c}(\mathsf{SNR}_1, \mathsf{SNR}_2, \mathsf{INR}_2) = \begin{cases} 0 & \text{if } \mathsf{INR}_2 < \mathsf{SNR}_2 \\ \log\left(1 + \min\left(\mathsf{SNR}_1, \frac{\mathsf{INR}_2}{1+\frac{\mathsf{SNR}_2\mathsf{INR}_2}{1+\mathsf{SNR}_2+2\mathsf{INR}_2}}\right)\right) & \text{if } \mathsf{INR}_2 \geq \mathsf{SNR}_2, \mathsf{INR}_2 \geq \mathsf{SNR}_1 \\ \log\left(1 + \frac{\mathsf{INR}_2^2}{1+2\mathsf{INR}_2+\mathsf{SNR}_2(1+\mathsf{INR}_2)}\right) & \text{if } \mathsf{INR}_2 \geq \mathsf{SNR}_2, \mathsf{INR}_2 < \mathsf{SNR}_1 \end{cases} \tag{13}$$

$$R_{1,p}(\mathsf{SNR}_1, \mathsf{SNR}_2, \mathsf{INR}_2) = \begin{cases} 0 & \text{if } \mathsf{INR}_2 \geq \mathsf{SNR}_2, \mathsf{INR}_2 \geq \mathsf{SNR}_1 \\ \log\left(1 + \mathsf{SNR}_1 / (1 + \mathsf{INR}_2)\right) & \text{otherwise} \end{cases} \tag{14}$$

Further, the power levels of

$$P_{1,c}(\mathsf{SNR}_1, \mathsf{SNR}_2, \mathsf{INR}_2) = \begin{cases} 0 & \text{if } \mathsf{INR}_2 < \mathsf{SNR}_2 \\ 1 & \text{if } \mathsf{INR}_2 \geq \mathsf{SNR}_2, \mathsf{INR}_2 \geq \mathsf{SNR}_1 \\ \mathsf{INR}_2/(1+\mathsf{INR}_2) & \text{if } \mathsf{INR}_2 \geq \mathsf{SNR}_2, \mathsf{INR}_2 < \mathsf{SNR}_1 \end{cases} \tag{15}$$

$$P_{1,p}(\mathsf{SNR}_1, \mathsf{SNR}_2, \mathsf{INR}_2) = \begin{cases} 0 & \text{if } \mathsf{INR}_2 \geq \mathsf{SNR}_2, \mathsf{INR}_2 \geq \mathsf{SNR}_1 \\ 1/(1+\mathsf{INR}_2) & \text{otherwise} \end{cases} \tag{16}$$

are used for the common and private parts, respectively. The common and the private parts are added together and transmitted.

In Appendix B, we prove that the above rates can be decoded and yield the sum capacity within 2 bits. ∎

**Corollary 2.** *There exist an approximately universally optimal strategy with $\tau = 2$ with the local information $N_i$ and $N_i'$ provided by 1.5 rounds of message passing and the side information that $K = 2$.*

*Proof:* With 1.5 rounds of message passing and side information $K = 2$, all nodes know the global connectivity. For all the connectivity choices, a strategy can be found by the nodes based on the channel gains they know that would achieve the global information sum capacity within 2 bits. For example, if the network is fully-connected, then the nodes can use the node identities to order themselves and can use an appropriate (approximately) optimal strategy. If the network is a Z-channel, the strategy of Z-channel described in Theorem 5 can be used. ∎

For a general Gaussian Z-channel, the capacity region is not known exactly. There exist achievable schemes that can achieve the region as shown in Figure 3. The achievable point corresponding to maximum $R_2$ when $R_1 = \log(1 + h_{11}^2)$ is known exactly [18] but the maximal rate point corresponding to $R_2 = \log(1 + h_{22}^2)$ is only known approximately. Here, we achieve a point approximate to the maximal rate point corresponding to $R_2 = \log(1 + h_{22}^2)$.

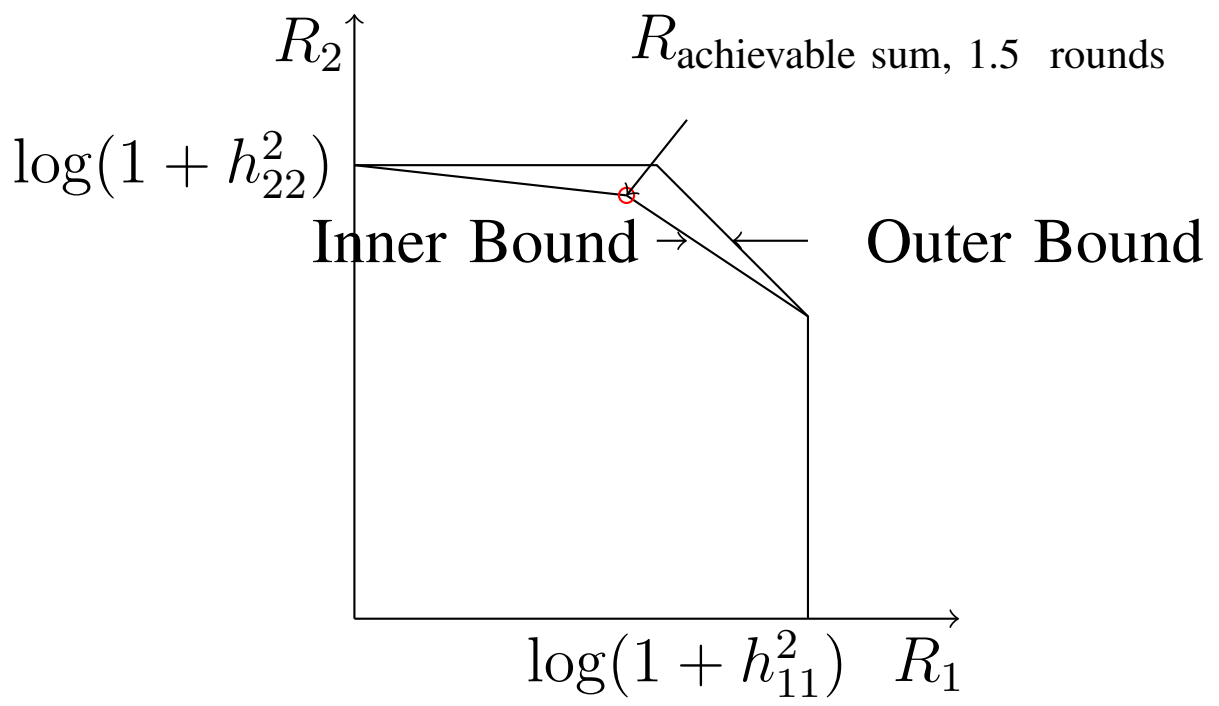

Fig. 3. Outer and Inner Bounds in Gaussian Z-channel.

While there is no loss in rate due to lack of knowledge in the deterministic Z-channel, it is no longer true for the case of Gaussian Z-channel. For example, exact capacity region is known for some regimes, like the strong interference case. However, the optimal sum rate point is not achieved by the distributed scheme with 1.5 rounds of knowledge. In this case, lack of full knowledge at $\mathsf{T}_2$ implies it is not aware that it is part of a strong interference channel. As a result, the $\mathsf{T}_2$ backs off a little on the rate thinking that it may not be able to cancel all interference.

## V. Three User double Z-channel

In this section, we will describe our results for three user double Z-channel. We will first find the capacity region for a general class of deterministic channels in which the interference at the receivers is a deterministic function of the inputs and the received signal is a deterministic function of the direct signal and the interference, and then specialize it to the deterministic model of Section II. We then derive new genie-aided outer bounds for the Gaussian double Z-channel. We will further provide an achievable strategy with 1.5 rounds of message passing algorithm which achieves the sum capacity in some cases and has an unbounded gap in certain other cases. We also show that the loss is unavoidable in the sense that there does not exist any strategy with local information that can be universally optimal even with additional side information of network connectivity, $\mathsf{SI} = \mathsf{E}_{3Z}$ ($\mathsf{E}_{3Z}$ is the network connectivity representing three-user double Z-channel), which is more than the side information $\mathsf{SI} = \{K = 3\}$. The achievability is extended to the Gaussian model. Further, we show that 1.5 round strategy can be tweaked for the first user to derive a 2.5 strategy which is optimal for the deterministic case and is approximately optimal in terms of the sum capacity for the Gaussian case with side information $\mathsf{SI} = \{K = 3\}$. This also proves that there exist a universally optimal/approximately universally optimal strategy with the local information obtained by 2.5 rounds of message passing and the side information of $\{K = 3\}$. That is because with 2.5 rounds of messages, in all network connectivities except double-Z channel, all the nodes would know the whole state in 2.5 rounds and can hence take the same decision as the centralized optimal solution.

### A. Channel Models and Messaging Passing

In a deterministic double Z-channel, $K = 3$ and $n_{13} = n_{21} = n_{31} = n_{32} = 0$. Specifically, the received signal $Y_{ji}$, $j = 1, 2, 3$, of a double Z-channel is given by

$$Y_{1i} = \mathbf{S}^{q-n_{11}} X_{1i} \tag{17a}$$
$$Y_{2i} = \mathbf{S}^{q-n_{12}} X_{1i} \oplus \mathbf{S}^{q-n_{22}} X_{2i} \tag{17b}$$
$$Y_{3i} = \mathbf{S}^{q-n_{23}} X_{2i} \oplus \mathbf{S}^{q-n_{33}} X_{3i} \tag{17c}$$

In a Gaussian double Z-channel, and $h_{13} = h_{21} = h_{31} = h_{32} = 0$. The received signal $Y_{ji}$, $j = 1, 2, 3$,

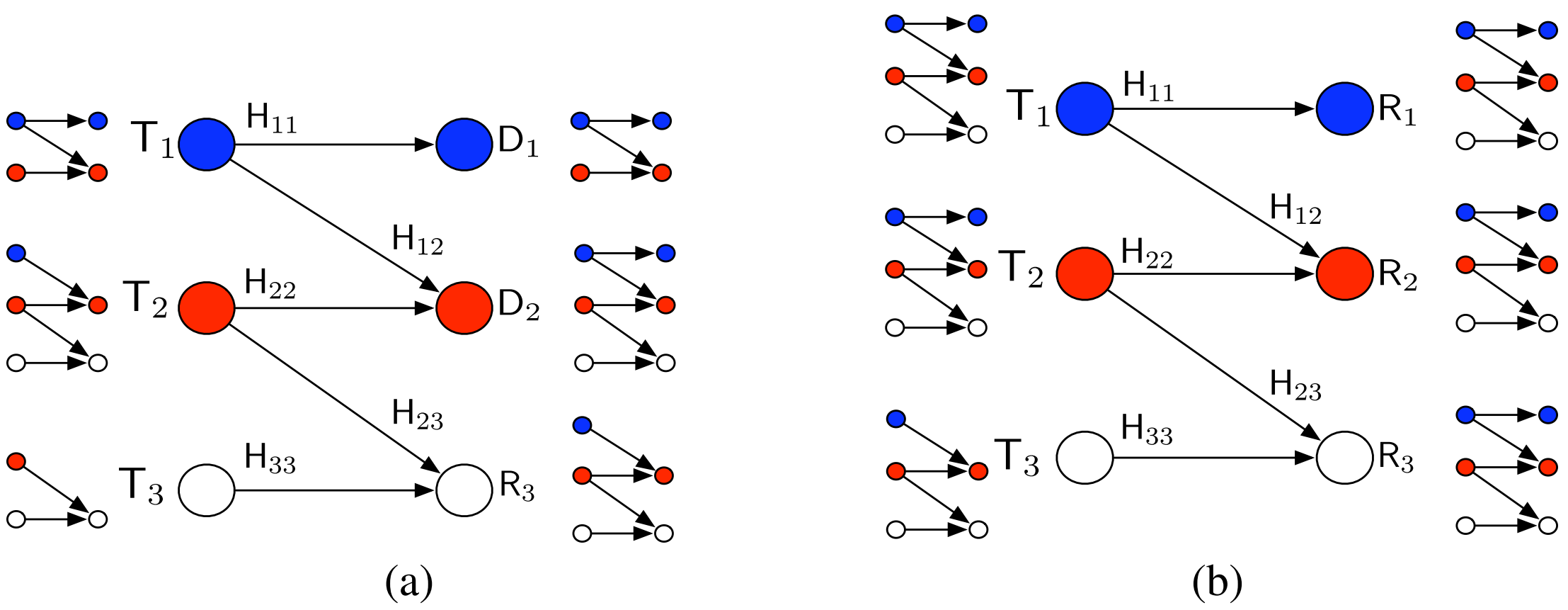

Fig. 4. Node's view at different nodes in three user double Z interference channel, after (a) 1.5 and (b) 2.5 rounds.

of a double Z-channel is given by

$$Y_{1i} = h_{11}X_{1i} + Z_{1i} \quad (18a)$$

$$Y_{2i} = h_{12}X_{1i} + h_{22}X_{2i} + Z_{2i} \quad (18b)$$

$$Y_{3i} = h_{23}X_{2i} + h_{33}X_{3i} + Z_{3i} \quad (18c)$$

Also, let $\mathsf{SNR}_i = |h_{ii}|^2$, $i \in \{1,2,3\}$ and $\mathsf{INR}_{i+1} = |h_{i(i+1)}|^2$, $i \in \{1,2\}$. Further, we assume that all $h_{ij}$'s are real and positive.

If the side information is $\mathsf{SI} = \{K = 3\}$, message passing converges in 4 rounds. The details of the protocol are as follows (To simplify the notation, the node identities appended to each channel gain are not shown and implied by the channel subscripts.).

1) **Round 1**: $m_{k,1} = \psi_k$. $M_{1,1} = \{H_{11}\}$, $M_{2,1} = \{H_{12}, H_{22}\}$ and $M_{3,1} = \{H_{23}, H_{33}\}$.
2) **Round 2**: $m_{1,2} = \{H_{11}, H_{12}, H_{22}\}$, $m_{2,2} = \{H_{12}, H_{22}, H_{23}, H_{33}\}$ and $m_{3,2} = \phi$. $M_{1,2} = \phi$, $M_{2,2} = \{H_{11}, H_{23}, H_{33}\}$ and $M_{3,2} = \{H_{12}, H_{22}\}$.
3) **Round 3**: $m_{1,3} = \{H_{23}, H_{33}\}$, $m_{2,3} = \{H_{11}\}$ and $m_{3,3} = \phi$. $M_{1,3} = M_{2,3} = \phi$ and $M_{3,3} = \{H_{11}\}$.
4) **Round 4**: No new information is to be sent by any transmitter and hence the algorithm halts by transmitters sending a silent message $\phi$.

Note that $H_{ij}$ is replaced with $n_{ij}$ in the deterministic or $h_{ij}^2$ for the Gaussian model.

The local view of each node after 1.5 and 2.5 rounds of message passing is shown in Figure 4. It is clear that with fewer rounds of message passing, the local view of each node is less than full. The challenge for the node is that they have to make decisions on their transmission parameters based only on their local view. Thus, we will ask how close we can get to sum-capacity with 1.5 and 2.5 rounds of message passing. Before we derive sum-capacity with partial information, we will need full capacity region for the case of full information. In the next section, we derive the new capacity results.

### *B. Capacity Regions With Full Information*

The capacity region for a general three-user interference channel is open. It has been solved in certain special cases of deterministic model in [11]. In this section, we provide the capacity region for a three user double-Z channel, which has not been considered before.

**Theorem 6** (Double-Z Deterministic Channel Capacity Region)**.** *The deterministic channel capacity region*

*for a three user double Z interference channel is the set of nonnegative rates* $(R_1, R_2, R_3)$ *satisfying*

$$R_i \leq n_{ii},\ i = 1, 2, 3 \tag{19a}$$
$$R_1 + R_2 \leq \max(n_{11}, n_{12}, n_{22}, n_{11} + n_{22} - n_{12}) \tag{19b}$$
$$R_2 + R_3 \leq \max(n_{22}, n_{23}, n_{33}, n_{22} + n_{33} - n_{23}) \tag{19c}$$
$$R_1 + R_2 + R_3 \leq \max(n_{33}, n_{23}) + (n_{11} - n_{12})^+ + \max(n_{12}, n_{22} - n_{23}). \tag{19d}$$

*Proof:* The proof is provided in Appendix C for a class of deterministic channels which is much broader than the deterministic model, along the lines of the class of deterministic channels for two user interference channels proposed in [12]. This region can be specialized for the deterministic model in (17a)-(17c) to get the above result. ■

Note that the bounds on $R_i$ are the single user bounds, and the $R_i + R_j$ bounds are due to the two two-user Z-channels, one consisting of $\{\mathsf{T}_1, \mathsf{T}_2, \mathsf{D}_1, \mathsf{D}_2\}$ and the other consisting of $\{\mathsf{T}_2, \mathsf{T}_3, \mathsf{D}_2, \mathsf{D}_3\}$. Finally, the sum bound $R_1 + R_2 + R_3$ is due to common transmitter-receiver pair $\{\mathsf{T}_2, \mathsf{D}_2\}$ in the two Z-channels.

For the Gaussian channel, we provide an outer bound to the capacity region for the three user Z-channel. We divide the region of the channels to four cases depicting the strong/weak interference from the first two transmitters.

1) $\mathsf{INR}_2 \geq \mathsf{SNR}_1$ and $\mathsf{INR}_3 \geq \mathsf{SNR}_2$: In this case, an outer bound on the rate region is given as follows.

$$R_1 \leq \log(1 + \mathsf{SNR}_1) \tag{20a}$$
$$R_2 \leq \log(1 + \mathsf{SNR}_2) \tag{20b}$$
$$R_3 \leq \log(1 + \mathsf{SNR}_3) \tag{20c}$$
$$R_1 + R_2 \leq \log(1 + \mathsf{SNR}_2 + \mathsf{INR}_2) \tag{20d}$$
$$R_2 + R_3 \leq \log(1 + \mathsf{SNR}_3 + \mathsf{INR}_3) \tag{20e}$$

2) $\mathsf{INR}_2 \geq \mathsf{SNR}_1$ and $\mathsf{INR}_3 \leq \mathsf{SNR}_2$: In this case, an outer bound on the rate region is given as follows.

$$R_1 \leq \log(1 + \mathsf{SNR}_1) \tag{21a}$$
$$R_2 \leq \log(1 + \mathsf{SNR}_2) \tag{21b}$$
$$R_3 \leq \log(1 + \mathsf{SNR}_3) \tag{21c}$$
$$R_1 + R_2 \leq \log(1 + \mathsf{SNR}_2 + \mathsf{INR}_2) \tag{21d}$$
$$R_2 + R_3 \leq \log(1 + \mathsf{SNR}_2) + \log(1 + \frac{\mathsf{SNR}_3}{1 + \mathsf{INR}_3}) \tag{21e}$$

Further, if $(\mathsf{INR}_2 + 1)\mathsf{INR}_3 \leq \mathsf{SNR}_2$

$$R_1 + R_2 + R_3 \leq \log(1 + \frac{\mathsf{SNR}_3}{1 + \mathsf{INR}_3}) + \log(1 + \mathsf{INR}_2 + \mathsf{SNR}_2) \tag{22}$$

else if $(\mathsf{INR}_2 + 1)\mathsf{INR}_3 \geq \mathsf{SNR}_2$

$$R_1 + R_2 + R_3 \leq \log(1 + \mathsf{INR}_3 + \mathsf{SNR}_3) + \log(1 + \mathsf{INR}_2). \tag{23}$$

3) $\mathsf{INR}_2 \leq \mathsf{SNR}_1$ and $\mathsf{INR}_3 \geq \mathsf{SNR}_2$: In this case, an outer bound on the rate region is given as follows.

$$R_1 \leq \log(1+\mathsf{SNR}_1) \tag{24a}$$
$$R_2 \leq \log(1+\mathsf{SNR}_2) \tag{24b}$$
$$R_3 \leq \log(1+\mathsf{SNR}_3) \tag{24c}$$
$$R_1+R_2 \leq \log(1+\mathsf{SNR}_1)+\log\left(1+\frac{\mathsf{SNR}_2}{1+\mathsf{INR}_2}\right) \tag{24d}$$
$$R_2+R_3 \leq \log(1+\mathsf{SNR}_3+\mathsf{INR}_3) \tag{24e}$$

4) $\mathsf{INR}_2 \leq \mathsf{SNR}_1$ and $\mathsf{INR}_3 \leq \mathsf{SNR}_2$: In this case, an outer bound on the rate region is given as follows.

$$R_1 \leq \log(1+\mathsf{SNR}_1) \tag{25a}$$
$$R_2 \leq \log(1+\mathsf{SNR}_2) \tag{25b}$$
$$R_3 \leq \log(1+\mathsf{SNR}_3) \tag{25c}$$
$$R_1+R_2 \leq \log(1+\mathsf{SNR}_1)+\log\left(1+\frac{\mathsf{SNR}_2}{1+\mathsf{INR}_2}\right) \tag{25d}$$
$$R_2+R_3 \leq \log(1+\mathsf{SNR}_2)+\log\left(1+\frac{\mathsf{SNR}_3}{1+\mathsf{INR}_3}\right) \tag{25e}$$

Further, if $(\mathsf{INR}_2+1)\mathsf{INR}_3 \leq \mathsf{SNR}_2$

$$R_1+R_2+R_3 \leq \log(1+\mathsf{SNR}_1)+\log\left(1+\frac{\mathsf{SNR}_2}{1+\mathsf{INR}_2}\right)+\log\left(1+\frac{\mathsf{SNR}_3}{1+\mathsf{INR}_3}\right) \tag{26}$$

else if $(\mathsf{INR}_2+1)\mathsf{INR}_3 \geq \mathsf{SNR}_2$

$$R_1+R_2+R_3 \leq \log(1+\mathsf{SNR}_1)+\log(1+\mathsf{INR}_3+\mathsf{SNR}_3). \tag{27}$$

**Theorem 7** (Double-Z Gaussian Channel Outer Bound)**.** *The capacity region of Gaussian double Z-channel is outer bounded by the region formed by* $(R_1, R_2, R_3)$ *satisfying* (20)-(27).

*Proof:* All the single user bounds and the bounds on $R_1+R_2$, $R_2+R_3$ follow from the two user Z-channel. The new bounds for $R_1+R_2+R_3$ in the second and the fourth cases are new, and are shown in Appendix D. ∎

We note that in the case of weak interference for the two Z-channels, $(\mathsf{T}_1, \mathsf{T}_2, \mathsf{R}_1, \mathsf{R}_2)$ and $(\mathsf{T}_2, \mathsf{T}_3, \mathsf{R}_2, \mathsf{R}_3)$, it is not always optimal to treat interference as noise unlike the case in two-user Z-channel. However, there is a region in which the interference is very weak which is when $\mathsf{INR}_3(\mathsf{INR}_2+1) \leq \mathsf{SNR}_2$ where treating interference as noise is optimal.

**Lemma 1.** *Let* $K \geq 3$ *and consider a symmetric K user Z-channel, where (K-1) Z-channels are stacked one over the other, with* $\mathsf{SNR}_i = \mathsf{SNR}$ *and* $\mathsf{INR}_i = \mathsf{INR}$*. When* $\mathsf{INR}(\mathsf{INR}+1) \leq \mathsf{SNR}$*, the sum rate is outer bounded by* $\log(1+\mathsf{SNR})+(K-1)\log\left(1+\frac{\mathsf{SNR}}{1+\mathsf{INR}}\right)$*. Thus in this regime, the sum-capacity can be achieved by treating interference as noise. However as we saw in the case of 3-user system, the above is not true in general for* $\mathsf{INR} < \mathsf{SNR} < \mathsf{INR}(\mathsf{INR}+1)$).

*Proof:* This proof follows using the same techniques as in the special case of very weak interference in Appendix D and is thus omitted. ∎

### *C. Deterministic Model: 1.5 Rounds*

In this section, we will study the achievable sum-rate after 1.5 rounds of message passing. Recall that the local view of each node is given in Figure 4(a) and the nodes have to base their choice of rates and transmission strategy only on their local views. We first describe an achievable rate, building on the Z-channel allocation strategy discussed in Theorem 2. Note that to derive the achievable rate, we will only assume side information about network size, i.e $K = 3$. However, we will show the converse for 1.5 rounds with extra side information about connectivity $\mathsf{SI} = \{\mathsf{E}_{3Z}\}$, which makes our converse stronger than one with only network size as side information.

1) From the point of view of the first transmitter, it knows the upper part of the Z-channel ($n_{11}$, $n_{12}$, $n_{22}$) and only extra cross link possible is a link from the second transmitter to the third receiver. Thus, the first transmitter knows that it is a Z-channel with the third user alone or a double-Z channel. Since $\mathsf{T}_1$ does not know which of the two cases are applicable, it assumes the worst case that second transmitter $\mathsf{T}_2$ will send at full rate and acts as if it is a Z-channel consisting of the first two users only. Note that this strategy is optimal if network connectivity turned out a Z-channel with a decoupled transmitter $\mathsf{T}_3$, instead of double-Z channel. Further this rate can be decoded at the receiver even if it is a double Z-channel.
2) From the point of the second transmitter, it knows $n_{12}$, $n_{22}$, $n_{23}$ and $n_{33}$. Hence, the only cross link possible that it does not know is $n_{31}$. Note that while $n_{31}$ is actually zero, the second transmitter does not know about it. Now suppose that the second transmitter sends as if there are only two users (the second and the third) and hence functions as the upper user of the Z-channel. It assumes that if it is a double-Z, the first transmitter $\mathsf{T}_1$ will send as if the second transmitter $\mathsf{T}_2$ is sending at full rate and thus will take care of interference itself. Hence the data can be decoded at the second receiver $\mathsf{D}_2$. If the network connectivity is cyclic with the presence of $n_{31}$, then every user will use this strategy and back off and thus the data can still be decoded.
3) The third transmitter $\mathsf{T}_3$ knows $n_{33}$ and $n_{23}$. Thus the cross channel gains that the third transmitter does not know about are $n_{12}$ and $n_{21}$. Even though $n_{21} = 0$, the third transmitter does not know about it. The third transmitter transmits at $n_{33}$. In case both the links $n_{12}$ and $n_{21}$ are present, the second transmitter knows the whole state and can shut down or select the rate since it would know the strategy of other users. In case only $n_{21}$ is present, then the second transmitter knows it is one-to-many configuration and there exist an optimal strategy in this case where the first and the third user send at full rate while the second transmitter backs off. The only other case is the double-Z channel where we use this strategy and show that it is still achievable.

To summarize, this achievability scheme reduces for a double Z-channel to using the same strategy at all the users as in the two user Z channel with the relevant knowledge. More specifically, the first transmitter assumes that it is the top transmitter in a two user Z channel (consisting of the first and the second user), the second transmitter assumes that it is the top user of a two user Z channel (consisting of the second and the third user) and the third transmitter assumes that it is the bottom user of a two user Z channel (consisting of the second and the third user).

**Theorem 8** (Achievable Rate with 1.5 Rounds)**.** *The above scheme can achieve the following sum-rate with 1.5 rounds of message passing for a double-Z channel:*

$$\begin{aligned}&\min(\max(n_{22}, n_{23}, n_{33}, n_{22} + n_{33} - n_{23}), n_{22} + n_{33})\\&+\min(n_{11}, \max(n_{11}, n_{12}) - \min(n_{12}, n_{22})).\end{aligned} \tag{28}$$

*Proof:* We show that each transmitter uses network state information obtained from the first round to decide the transmission strategy. The third transmitter sends at full rate, $n_{33}$. The second transmitter does not know $n_{11}$ and uses the strategy as if it was a Z-channel consisting of only the second and the third user. If $n_{23} \leq n_{33}$, the second transmitter will send at a rate of $(n_{22} - n_{23})^+$. If $n_{23} > n_{33}$, the second transmitter sends at a rate of $\min(\max(n_{22}, n_{23}) - n_{33}, n_{22})$. Thus, the second transmitter sends at a rate of $\min(n_{22}, \max(n_{22}, n_{23}) - \min(n_{23}, n_{33})) = \min(\max(n_{22}, n_{23}, n_{33}, n_{22} + n_{33} - n_{23}), n_{22} + n_{33}) - n_{33}$. The first transmitter transmits as if it was a Z-channel consisting of first two users and considering that the second user sends at a rate of $n_{22}$ and hence sends at a rate of $\min(n_{11}, \max(n_{11}, n_{12}) - \min(n_{12}, n_{22}))$. Hence, the above sum rate can be achieved. ■

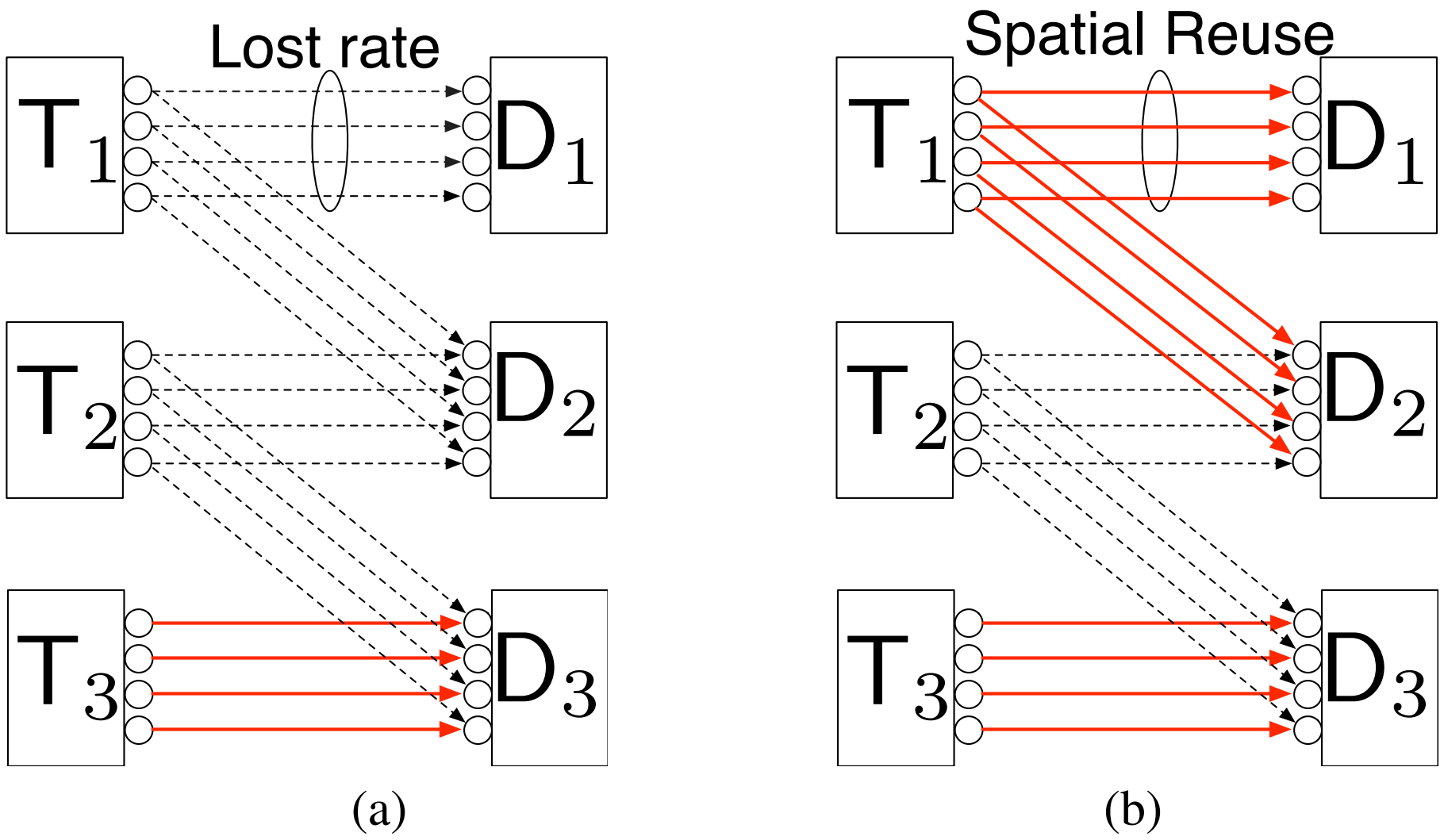

Fig. 5. Loss in spatial reuse due to distributed decisions.

So, the obvious next question is how well does the above scheme perform compared to the sum-capacity in Theorem 6. We show that the gap from the sum-capacity can be anywhere from zero to arbitrarily large. First, we classify all those network states in which the above distributed scheme achieves sum-capacity.

**Theorem 9** (Achieving sum-capacity)**.** *The sum capacity can be achieved with 1.5 rounds of message passing for a double-Z channel if any of the following are true*

- $n_{23} \geq n_{22} + n_{33}$
- $n_{12} \geq n_{11} + n_{22}$
- $n_{23} \leq n_{33}$ *and* $n_{22} \geq n_{23} + n_{12}$

*Proof:*

- $n_{23} \geq n_{22} + n_{33}$: In this case, (28) reduces to the sum rate of $n_{33} + \min(n_{11} + n_{22}, \max(n_{11}, n_{22}, n_{12}, n_{11} + n_{22} - n_{12}))$ which is optimal since the upper bounds on $R_3$ and $R_1 + R_2$ in Theorem 6 are identical to the above expression.
- $n_{12} \geq n_{11} + n_{22}$: In this case, (28) reduces to the sum rate of $n_{11} + \min(\max(n_{22}, n_{23}, n_{33}, n_{22} + n_{33} - n_{23}), n_{22} + n_{33})$ which is optimal since it is identical to upper bounds on $R_1$ and $R_2 + R_3$ in Theorem 6.
- $n_{12} \leq n_{22}$ and $n_{23} \leq n_{33}$ and $n_{22} \geq n_{23} + n_{12}$: In this case, the achievable sum rate and the outer bound both reduce to $n_{22} + n_{33} - n_{23} + (n_{11} - n_{12})^+$. Hence, the sum rate is optimal since this matches the $R_1 + R_2 + R_3$ outer bound in Theorem 6.

■

We next construct an example where the loss can be arbitrarily large. Let $n_{11} = n_{12} = n_{22} = n_{23} = n_{33} = x$. The distributed scheme achieves a sum-rate of $x$ related to rate-tuple of $(0, 0, x)$ as shown in Figure 5(a). However, the rate tuple $(x, 0, x)$ is in the capacity region and relates to the rate allocation shown in Figure 5(b). By taking $x$ large enough, the achievable sum rate can be made arbitrarily far from the outer bound. Thus the above distributed strategy is not universally optimal, i.e, it does not achieve sum-capacity in all network states. The next result shows that no distributed scheme can be universally optimal with 1.5 rounds of information, even with extra side information.

**Theorem 10** (Loss is Inevitable)**.** *There exist no universally optimal strategy with the local information obtained by 1.5 rounds of message passing for double-Z channel even with the side information* $\mathsf{SI} = \mathsf{E}_{3Z}$, *the network connectivity of double-Z channel. Thus, there is no universally optimal strategy with the local information obtained by 1.5 rounds of message passing and the side information of* $K = 3$.

*Proof:* We first assume that the each node is given the information that the network connectivity is a double-Z channel and the information about its relative placement in the network. So, the nodes can choose strategy tailored for double-Z channel. Thus, the set of network states consistent with the local information only have $n_{11}$, $n_{22}$, $n_{33}$, $n_{12}$ and $n_{23}$ as parameters.

We will now prove the Theorem by contradiction. Suppose that there is a universally optimal strategy. Since the third transmitter does not know $n_{22}$ and $n_{11}$ and if $n_{11} = n_{22} = 0$, the only rate allocation that can be used by the third transmitter that would be optimal is $R_3 = n_{33}$. Hence, if there exists a universally optimal strategy, $R_3 = n_{33}$. (Even though the transmitter knows $n_{23}$, it cannot use this extra knowledge to make a decision on the rate allocation.)

The second transmitter does not know $n_{11}$. Hence, if there exist an universally optimal strategy, it should work even if $n_{11} = 0$. For $n_{11} = 0$ and $R_3 = n_{33}$, the only way the second user can use optimal rate allocation is to transmit at a rate as in the proof of Theorem 8.

The first transmitter does not know $n_{33}$. Hence, its optimal strategy should work even if $n_{33} = 0$. If $n_{33} = 0$, $R_2 = n_{22}$ and hence the first user will have to transmit at a rate as in Theorem 8.

Thus, we see that if there exist an universally optimal strategy, the rates of the users have to be the same as in Theorem 8. We note that the sum rate in Theorem 8 is not optimal in general and hence leads to a contradiction. Thus, there is no universally optimal strategy. ■

Theorem 10 is the key result in the paper. It shows that *no* distributed scheme, which only relies on its local view after 1.5 rounds, can be guaranteed to be globally optimal for all channel conditions even with the global knowledge of connectivity. Distributed schemes can be optimal for some values of network matrices $\mathsf{H} \in \mathcal{G}(\mathsf{E}_{3Z})$ but not all of them simultaneously. Thus, there is no *universally optimal* scheme with the local information provided by 1.5 rounds of knowledge and the side information of network connectivity.

The fundamental reason for this unavoidable loss for some channel gains is severely *incomplete* view of the network at different transmitters. As shown in Figure 5, the loss in spatial reuse is due to mismatched knowledge in the deterministic double-Z channel. Here $\mathsf{T}_1$ is backing off for $\mathsf{T}_2$, which in turn is backing off for $\mathsf{T}_3$. In this example, $\mathsf{T}_1$ could have sent at full rate but ended up being too conservative due to its lack of knowledge about state of $\mathsf{T}_3$, as shown in Figure 5(b). As a result, $\mathsf{T}_1$ tailors its action to the worst case scenario, which is $\mathsf{T}_2$ sending at full rate. The reader is reminded that each node is only allowed to adapt its transmission based on its own local knowledge and cannot base its decision on what is *not* known. Thus, once a node defines a rate allocation policy based on its local information (e.g 1.5 rounds), it has to use the same allocation for *all* the possible values of other channel gains that are not known with the local and side information.

We observe that the above proof can also be extended to show that there does not exist any strategy that will perform within a bounded gap (independent of channel gains) from the optimal sum capacity. Thus, an approximately universally optimal strategy also does not exist. To prove that claim, assume $n_{ij} = c_{ij}L$ where $L$ can be taken as large as possible, $c_{ij} \geq 0$. With the local channel knowledge at the nodes, $R_3 \geq c_{ij}L - \Theta(1)$ (where $\Theta(1)$ represents a function that is independent of $L$) since that is the only way an approximately optimal strategy can exist when $c_{22} = c_{11} = 0$. Given this strategy of the third user, $R_2 \geq L\min(c_{22}, \max(c_{22}-c_{33}, c_{23}-c_{33}, c_{22}-c_{23})^+) - \Theta(1)$ since otherwise the strategy will not be approximately universally optimal for $c_{11} = 0$. Similar strategy goes for the first user. We can also show that this strategy will be unbounded rate away from optimal as $L \to \infty$, thus proving non-existence of approximately universally optimal strategies.

### *D. Gaussian Channel: 1.5 Rounds*

The achievability strategy for Gaussian channels follows the same technique as in the deterministic model. As mentioned for the deterministic model, the achievable scheme used by each of the users will work as an achievable strategy for all possible global network connectivities if they see the same local network connectivity. The details are as follows.

The third transmitter makes a codebook of rate

$$R_3(\mathsf{SNR}_3, \mathsf{INR}_3) = \log\left(1 + \mathsf{SNR}_3(1+\mathsf{INR}_3)/(1+2\mathsf{INR}_3)\right), \tag{29}$$

and uses a power level of $P_3 = 1$ to transmit. The second user however uses a common and a private message with rates

$$R_{2,c}(\mathsf{SNR}_2, \mathsf{SNR}_3, \mathsf{INR}_3) = \begin{cases} 0 & \text{if } \mathsf{INR}_3 < \mathsf{SNR}_3 \\ \log\left(1 + \frac{1+\mathsf{INR}_2}{1+2\mathsf{INR}_2} \min\left(\mathsf{SNR}_2, \frac{\mathsf{INR}_3}{1+\frac{\mathsf{SNR}_3\mathsf{INR}_3}{1+\mathsf{SNR}_3+2\mathsf{INR}_3}}\right)\right) & \text{if } \mathsf{INR}_3 \geq \mathsf{SNR}_3, \mathsf{INR}_3 \geq \mathsf{SNR}_2 \\ \log\left(1 + \frac{1+\mathsf{INR}_2}{1+2\mathsf{INR}_2} \frac{\mathsf{INR}_3^2}{1+2\mathsf{INR}_3+\mathsf{SNR}_3(1+\mathsf{INR}_3)}\right) & \text{if } \mathsf{INR}_3 \geq \mathsf{SNR}_3, \mathsf{INR}_3 < \mathsf{SNR}_2 \end{cases} \tag{30}$$

$$R_{2,p}(\mathsf{SNR}_2, \mathsf{SNR}_3, \mathsf{INR}_3) = \begin{cases} 0 & \text{if } \mathsf{INR}_3 \geq \mathsf{SNR}_3, \mathsf{INR}_3 \geq \mathsf{SNR}_2 \\ \log\left(1 + \frac{1+\mathsf{INR}_2}{1+2\mathsf{INR}_2}\mathsf{SNR}_2/(1+\mathsf{INR}_3)\right) & \text{otherwise} \end{cases} \tag{31}$$

Further, the power levels of

$$P_{2,c}(\mathsf{SNR}_2, \mathsf{SNR}_3, \mathsf{INR}_3) = \begin{cases} 0 & \text{if } \mathsf{INR}_3 < \mathsf{SNR}_3 \\ 1 & \text{if } \mathsf{INR}_3 \geq \mathsf{SNR}_3, \mathsf{INR}_3 \geq \mathsf{SNR}_2 \\ \mathsf{INR}_3/(1+\mathsf{INR}_3) & \text{if } \mathsf{INR}_3 \geq \mathsf{SNR}_3, \mathsf{INR}_3 < \mathsf{SNR}_2 \end{cases} \tag{32}$$

$$P_{2,p}(\mathsf{SNR}_2, \mathsf{SNR}_3, \mathsf{INR}_3) = \begin{cases} 0 & \text{if } \mathsf{INR}_3 \geq \mathsf{SNR}_3, \mathsf{INR}_3 \geq \mathsf{SNR}_2 \\ 1/(1+\mathsf{INR}_3) & \text{otherwise} \end{cases} \tag{33}$$

are used for the common and private parts. The common and the private parts are added together and transmitted.

The first user uses a common and a private message with rates

$$R_{1,c}(\mathsf{SNR}_1, \mathsf{SNR}_2, \mathsf{INR}_2) = \begin{cases} 0 & \text{if } \mathsf{INR}_2 < \mathsf{SNR}_2 \\ \log\left(1 + \min(\mathsf{SNR}_1, \frac{\mathsf{INR}_2}{1+\frac{\mathsf{SNR}_2\mathsf{INR}_2}{1+\mathsf{SNR}_2+2\mathsf{INR}_2}})\right) & \text{if } \mathsf{INR}_2 \geq \mathsf{SNR}_2, \mathsf{INR}_2 \geq \mathsf{SNR}_1 \\ \log\left(1 + \frac{\mathsf{INR}_2^2}{1+2\mathsf{INR}_2+\mathsf{SNR}_2(1+\mathsf{INR}_2)}\right) & \text{if } \mathsf{INR}_2 \geq \mathsf{SNR}_2, \mathsf{INR}_2 < \mathsf{SNR}_1 \end{cases} \tag{34}$$

$$R_{1,p}(\mathsf{SNR}_1, \mathsf{SNR}_2, \mathsf{INR}_2) = \begin{cases} 0 & \text{if } \mathsf{INR}_2 \geq \mathsf{SNR}_2, \mathsf{INR}_2 \geq \mathsf{SNR}_1 \\ \log\left(1 + \mathsf{SNR}_1/(1+\mathsf{INR}_2)\right) & \text{otherwise} \end{cases} \tag{35}$$

Further, the power levels of

$$P_{1,c}(\mathsf{SNR}_1, \mathsf{SNR}_2, \mathsf{INR}_2) = \begin{cases} 0 & \text{if } \mathsf{INR}_2 < \mathsf{SNR}_2 \\ 1 & \text{if } \mathsf{INR}_2 \geq \mathsf{SNR}_2, \mathsf{INR}_2 \geq \mathsf{SNR}_1 \\ \mathsf{INR}_2/(1+\mathsf{INR}_2) & \text{if } \mathsf{INR}_2 \geq \mathsf{SNR}_2, \mathsf{INR}_2 < \mathsf{SNR}_1 \end{cases} \tag{36}$$

$$P_{1,p}(\mathsf{SNR}_1, \mathsf{SNR}_2, \mathsf{INR}_2) = \begin{cases} 0 & \text{if } \mathsf{INR}_2 \geq \mathsf{SNR}_2, \mathsf{INR}_2 \geq \mathsf{SNR}_1 \\ 1/(1+\mathsf{INR}_2) & \text{otherwise} \end{cases} \tag{37}$$

are used for the common and private parts. The common and the private parts are added together and transmitted.

We will show in Appendix E that the above rates can be decoded by the receivers. We now see the various cases when the achievable strategy will be bounded distance from the sum-capacity.

**Theorem 11.** *The sum-capacity within 4 bits can be achieved with the first full round and half of the second round of message passing when any of the following conditions hold*

1) $\mathsf{INR}_3 \geq \mathsf{SNR}_3$, $\frac{\mathsf{INR}_3}{1+\frac{\mathsf{SNR}_3\mathsf{INR}_3}{1+\mathsf{SNR}_3+2\mathsf{INR}_3}} \geq \mathsf{SNR}_2$
2) $\mathsf{INR}_2 \geq \mathsf{SNR}_2$, $\frac{\mathsf{INR}_2}{1+\frac{\mathsf{SNR}_2\mathsf{INR}_2}{1+\mathsf{SNR}_2+2\mathsf{INR}_2}} \geq \mathsf{SNR}_1$
3) $\mathsf{INR}_2 < \mathsf{SNR}_2$ *and* $\mathsf{INR}_3 < \mathsf{SNR}_3$, $\mathsf{SNR}_2 \geq \mathsf{INR}_3(\mathsf{INR}_2+1)$

*Proof:*

1) $\mathsf{INR}_3 \geq \mathsf{SNR}_3$, $\frac{\mathsf{INR}_3}{1+\frac{\mathsf{SNR}_3\mathsf{INR}_3}{1+\mathsf{SNR}_3+2\mathsf{INR}_3}} \geq \mathsf{SNR}_2$: In this case, $R_2 = \log\left(1+\frac{1+\mathsf{INR}_2}{1+2\mathsf{INR}_2}\mathsf{SNR}_2\right)$ which yields $R_1+R_2$ within 2 bits of optimal as in the two-user Z-channel and since $R_3$ is within 1 bit of $\log(1+\mathsf{SNR}_3)$, $R_1+R_2+R_3$ is within 3 bits of the outer bound.
2) $\mathsf{INR}_2 \geq \mathsf{SNR}_2$, $\frac{\mathsf{INR}_2}{1+\frac{\mathsf{SNR}_2\mathsf{INR}_2}{1+\mathsf{SNR}_2+2\mathsf{INR}_2}} \geq \mathsf{SNR}_1$: In this case, $R_1 = \log(1+\mathsf{SNR}_1)$. Further, $R_2+R_3$ is within 2 bits of that in the case of two-user Z-channel which is further within 2 bits of optimal; thus leading to the sum rate within 4 bits of optimal.
3) $\mathsf{INR}_2 < \mathsf{SNR}_2$ and $\mathsf{INR}_3 < \mathsf{SNR}_3$, $\mathsf{SNR}_2 \geq \mathsf{INR}_3(\mathsf{INR}_2+1)$: The achievable sum rate in this case reduces to

$$\log\left(1+\frac{\mathsf{SNR}_1}{1+\mathsf{INR}_2}\right)+\log\left(1+\frac{1+\mathsf{INR}_2}{1+2\mathsf{INR}_2}\frac{\mathsf{SNR}_2}{1+\mathsf{INR}_3}\right)+\log\left(1+\mathsf{SNR}_3\frac{1+\mathsf{INR}_3}{1+2\mathsf{INR}_3}\right).$$

If $\mathsf{INR}_2 \geq \mathsf{SNR}_1$ the outer bound is $\log(1+\mathsf{SNR}_2+\mathsf{INR}_2)+\log\left(1+\frac{\mathsf{SNR}_3}{1+\mathsf{INR}_3}\right)$. However, if $\mathsf{INR}_2 \leq \mathsf{SNR}_1$, the outer bound is $\log(1+\mathsf{SNR}_1)+\log\left(1+\frac{\mathsf{SNR}_2}{1+\mathsf{INR}_2}\right)+\log\left(1+\frac{\mathsf{SNR}_3}{1+\mathsf{INR}_3}\right)$. We note that in both these cases, the achievability is within 4 bits in both the cases.

■

As in deterministic model, there are cases when the sum rate achieved with 1.5 rounds can be arbitrarily far from the optimal. As an example, consider $\mathsf{INR}_2 = \mathsf{SNR}_1 = \mathsf{SNR}_2 = \mathsf{INR}_3 = \mathsf{SNR}_3 = x$. The achievable sum rate is $\log[(1+6x+11x^2+5x^3)/(1+3x+x^2)]$. However with full information, rate pair of $(\log(1+x), 0, \log(1+x))$ can be achieved. For $x \geq 2$, $\log[(1+6x+11x^2+5x^3)/(1+3x+x^2)] \leq 2\log(1+x)$ and the difference grows unbounded with $x$ thus proving that the difference between achievability and outer bound can be unbounded in some cases.

We also note that the strong converse mentioned for the deterministic case also holds in the Gaussian case that there is no approximately universally optimal strategy with 1.5 rounds of message passing. Since the proof uses the same ideas, it is omitted.

### *E. Both Channels with 2.5 Rounds*

We first note that with 2.5 rounds of message passing, all the nodes know the state information except the third transmitter which does not know $H_{11}$. Thus, the side information of $\mathsf{SI} = \{K=3\}$ with 2.5 rounds of message passing and the side information of network connectivity, $\mathsf{SI} = \{\mathsf{E}\}$, with 2.5 rounds of message passing are equivalent.

We first consider the capacity region of the deterministic double-Z channel as shown in Figure 6. The region consists of single user constraints, the constraints on $R_1+R_2$, $R_2+R_3$ and a constraint on $R_1 + R_2 + R_3$ depicting the various planes in the figure. Note that there is a segment on the optimal face of $R_1 + R_2 + R_3$ that has $R_3 = n_{33}$ as marked in the figure. Any point on this segment can be achieved with 2.5 rounds of message passing. This is because the third transmitter sends at a rate of $n_{33}$ while the first two transmitters know all the channel gains to select the policy to operate at any point on this line.

Thus, there exist universally optimal strategy with the local information at the nodes obtained by 2.5 rounds of message passing and the side information about network size $\mathsf{SI} = \{K = 3\}$. So with 2.5 rounds, we can prove existence of a universally optimal strategy with less side information compared to the converse for 1.5 rounds in Theorem 10.

**Theorem 12.** *There exists a universally optimal strategy with the local information at the nodes obtained by 2.5 rounds of message passing when each node is provided the side information that there are only three nodes in the network.*

*Proof:* With 2.5 rounds of message passing, all the nodes would know the network connectivity in their connected component. In all the cases except the double-Z channel, all the nodes would know all the channel gains of the connected component in which they are and hence can use an optimal strategy by ordering the nodes based on the node identities and the network states. For double-Z channel connectivity, the strategy described before can be used to get a universally optimal strategy. ■

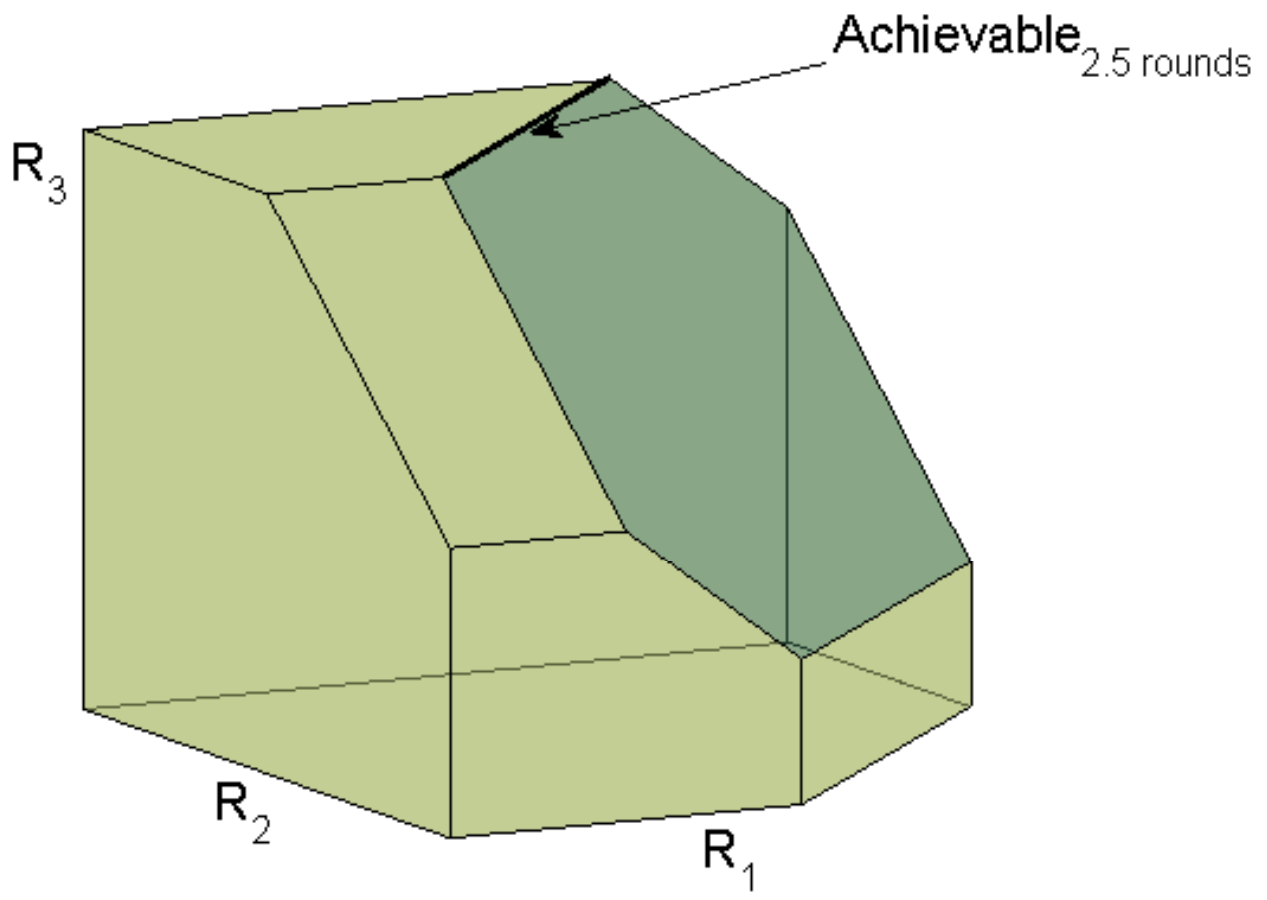

Fig. 6. Capacity region of Deterministic double Z-channel.

We now consider Gaussian channel model. As we saw in the deterministic model that any strategy following the rate allocation on a line is optimal, we consider one of the corner point on the line corresponding to higher $R_2$ and provide an achievability strategy that is within 4 bits of sum capacity for 2.5 rounds of message passing. In this strategy, the second and the third user uses the same policy as with 1.5 rounds and therefore do not change the strategy. However, the first transmitter knows all the channels and hence changes its strategy.

The first user uses a common and a private message with rates

$$R_{1,c} = \begin{cases} \min(\log(1+\mathsf{SNR}_1), \log(1+\mathsf{INR}_2+\mathsf{SNR}_2)-R_2) & \text{if } \gtrless\geq\geq\geq \\ \min(\log(1+\mathsf{SNR}_1), \log(1+\mathsf{INR}_2)-R_2) & \text{if } \geq\geq\geq< \\ \min(\log(1+\frac{\mathsf{INR}_2^2}{1+2\mathsf{INR}_2}), \log(1+\frac{\mathsf{SNR}_2(1+\mathsf{INR}_2)+\mathsf{INR}_2^2}{1+2\mathsf{INR}_2})-R_2) & \text{if } \gtrless\geq<\geq \\ \min(\log(1+\frac{\mathsf{SNR}_1\mathsf{INR}_2}{1+\mathsf{SNR}_1+\mathsf{INR}_2}), \log(1+\frac{\mathsf{INR}_2^2}{1+2\mathsf{INR}_2})-R_2) & \text{if } \geq\geq<< \\ \min(\log(1+\mathsf{SNR}_1), \log(1+\mathsf{INR}_2+\frac{\mathsf{SNR}_2}{1+\mathsf{INR}_3})-R_2) & \text{if } \gtrless<\geq\gtrless \\ \min(\log(1+\frac{\mathsf{SNR}_1\mathsf{INR}_2}{1+\mathsf{SNR}_1+\mathsf{INR}_2}), \log(1+\frac{\mathsf{INR}_2^2}{1+2\mathsf{INR}_2})-R_2) & \text{if } \geq<<\gtrless \\ \min(\log(1+\mathsf{SNR}_1), \log(1+\mathsf{INR}_2+\frac{\mathsf{SNR}_2\mathsf{INR}_3}{1+\mathsf{INR}_3})-R_{2,c}, & \\ \log(1+\mathsf{INR}_2+\frac{\mathsf{SNR}_2}{1+\mathsf{INR}_3})-R_{2,p}, \log(1+\mathsf{INR}_2+\mathsf{SNR}_2)-R_2) & \text{if } <\geq\geq< \\ \min(\log(1+\frac{\mathsf{INR}_2^2}{1+2\mathsf{INR}_2}), \log(1+\frac{\mathsf{INR}_2^2+\mathsf{SNR}_2\mathsf{INR}_3(1+\mathsf{INR}_2)/(1+\mathsf{INR}_3)}{1+2\mathsf{INR}_2})-R_{2,c}, & \\ \log(1+\frac{\mathsf{INR}_2^2+\mathsf{SNR}_2(1+\mathsf{INR}_2)/(1+\mathsf{INR}_3)}{1+2\mathsf{INR}_2})-R_{2,p}, \log(1+\frac{\mathsf{INR}_2^2+\mathsf{SNR}_2(1+\mathsf{INR}_2)}{1+2\mathsf{INR}_2})-R_2) & \text{if } <\geq<< \\ \min(\log(1+\frac{\mathsf{INR}_2^2}{1+2\mathsf{INR}_2}), \log(1+\frac{\mathsf{INR}_2^2+\mathsf{SNR}_2(1+\mathsf{INR}_2)}{1+2\mathsf{INR}_2})-R_2) & \text{if } <<<\gtrless \end{cases} \quad (38)$$

where the four inequalities in the if condition represents the order in $\mathsf{INR}_2$ vs. $\mathsf{SNR}_2$, $\mathsf{INR}_3$ vs. $\mathsf{SNR}_3$, $\mathsf{INR}_2$ vs. $\mathsf{SNR}_1$ and $\mathsf{INR}_3$ vs. $\mathsf{SNR}_2$ respectively. $\gtrless$ represents that it can be either side of inequality. For example, $<\geq<\geq$ represents $\mathsf{INR}_2 < \mathsf{SNR}_2$, $\mathsf{INR}_3 \geq \mathsf{SNR}_3$, $\mathsf{INR}_2 < \mathsf{SNR}_1$ and $\mathsf{INR}_3 \geq \mathsf{SNR}_2$.

$$R_{1,p} = \begin{cases} 0 & \text{if } \mathsf{INR}_2 \geq \mathsf{SNR}_1 \\ \log(1+\mathsf{SNR}_1/(1+\mathsf{INR}_2)) & \text{if } \mathsf{INR}_2 < \mathsf{SNR}_1 \end{cases} \quad (39)$$

Further, the power levels of

$$P_{1,c} = \begin{cases} 1 & \text{if } \mathsf{INR}_2 \geq \mathsf{SNR}_1 \\ \mathsf{INR}_2/(1+\mathsf{INR}_2) & \text{if } \mathsf{INR}_2 < \mathsf{SNR}_1 \end{cases} \quad (40)$$

$$P_{1,p} = \begin{cases} 0 & \text{if } \mathsf{INR}_2 \geq \mathsf{SNR}_1 \\ 1/(1+\mathsf{INR}_2) & \text{if } \mathsf{INR}_2 < \mathsf{SNR}_1 \end{cases} \quad (41)$$

are used for the common and private parts. The common and the private parts are added together and transmitted. It is straightforward to see in all the cases that the above rates can be decoded by all the users. Further as shown in the next theorem, the above strategy achieves the sum capacity within four bits.

**Theorem 13.** *The strategy achieves a sum-rate that is within four bits of the full-information sum-capacity for all* $\mathsf{H} \in \mathcal{G}(\mathsf{E}_{3Z})$.

*Proof:* We will show in Appendix F that the sum-capacity within four bits can be achieved by splitting the channel gain regimes into 14 cases. ■

**Corollary 3.** *There exists an approximately universally optimal strategy with* $\tau = 4$ *with the local information at the nodes obtained by 2.5 rounds of message passing when each node is provided the side information that there are only three nodes in the network, i.e.* $\mathsf{SI} = \{K = 3\}$.

*Proof:* With 2.5 rounds of message passing, all the nodes would know the network connectivity in their connected component. In all the cases except the double-Z channel, all the nodes would know all the channel gains of the connected component in which they are and hence can use an optimal strategy by ordering the nodes based on the node identities and the network states. For double-Z channel connectivity, the strategy described in this section can be used to get approximately universally optimal strategy. ■

# VI. $K$-USER Z-NETWORK

While exact analysis for two and three-user interference channels was tractable, extensions to general interference channel remains out of reach at the current moment. To make progress, we will consider a special case of channel gains which reduces the network parametrization to only three unknowns: the number of nodes in the network, and two channel gain parameters representing the direct link and the cross link. In addition, we will assume that the network connectivity is of the form of $K-1$ Z-channels stacked on top of each other as described below. Our objective is to quantify the achievable sum-rates with limited number of rounds of message passing.

## A. Channel Models and Message Passing Protocol

For the deterministic model characterized by the direct channel gain $n$ and the cross channel gain $m$, the received signal $Y_{ji}$, $j=1,2,\cdots,K$ of $K$-user Z-channel is given by

$$Y_{1i} = \mathbf{S}^{q-n}X_{1i} \quad (42a)$$

$$Y_{ji} = \mathbf{S}^{q-m}X_{(j-1)i} \oplus \mathbf{S}^{q-n}X_{ji} \text{ for all } 2 \le j \le K. \quad (42b)$$

We label the top transmitter as Node 1, the next transmitter as Node 2 and so on. Thus, we would use the phrase "node above" and "node below" with respect to this labeling unless otherwise stated. We assume that all users knows that it is a symmetric Z-channel but do not know $n$, $m$ and their relative position in the network (which also includes the information about $K$). In this case, the message passing protocol can be performed with less message content than sending the whole channel matrix as explained in Section III-C. We will now provide achievability with $d$ rounds of message passing algorithm for $d \le K$. Since the users do not know $K$ and their placement in the network, the decisions have to be made in a distributed fashion.

For the Gaussian model characterized by the direct channel gain $\sqrt{\mathsf{SNR}}$ and the cross channel gain $\sqrt{\mathsf{INR}}$, the received signal at receiver $j$ at time $i$, $Y_{ji}$, $j=1,2,\cdots,K$ of $K$-user Z-channel is given by

$$Y_{1i} = \sqrt{\mathsf{SNR}}X_{1i} + Z_{1i} \quad (43a)$$

$$Y_{ji} = \sqrt{\mathsf{INR}}X_{(j-1)i} + \sqrt{\mathsf{SNR}}X_{ji} + Z_{ji} \text{ for all } 2 \le j \le K. \quad (43b)$$

We also note that in this section, we will also sometimes consider $d$ rounds of message passing protocol for integer $d$ rather than $d.5$ rounds. This is because the additional symmetry in the network (all direct links are identical and all cross links are identical) allows the receiver to decide on their decoding schemes even if they have less knowledge of links than the transmitter. We assume that the nodes have the network connectivity and its reduced parametrization as side information, i.e, $\mathsf{SI} = \{\mathcal{G}', H_{ii} = H_{11}, H_{i,i+1} = H_{12}\}$, where $\mathsf{F} = \{\mathsf{E} : (\mathsf{T}_i, \mathsf{D}_j)\}$ and $\mathcal{G}' = \{H : H_{ij} \equiv 0 \text{ if } \text{ if } j \neq i \text{ or } j \neq i+1, H_{11} = H_{22} = \cdots \in \{0,1,\ldots,q\}$, $H_{12} = H_{23} = \cdots \in \{0,1,\ldots,q\}\}$. However, they do not know the network size $K$ or the two channel gains $\{H_{11}, H_{21}\}$.

## B. Deterministic Z-Network

We first consider the deterministic model and let $\alpha = m/n$. We show that for $\alpha \ge 2$ or $\alpha \le 1/2$, one round of message passing achieves the sum capacity in the following theorem.

**Theorem 14.** *For $\alpha \ge 2$ and $\alpha \le 1/2$, there exists a strategy which achieves optimal sum-rate with one round of message passing protocol for any $K$ if each node knows the side information that the network state is symmetric and is parameterized by three parameters $K$, $n$ and $\alpha = m/n$.*

*Proof:* We consider the two cases in the statement of the theorem separately as follows.

1) $\alpha \ge 2$: Each transmitter knows that it is very strong interference channel in one round and hence the transmitters transmit at full rate. Further, the receivers (except the first) also know that it is very strong interference and are able to decode both the messages. The first receiver decodes the direct message assuming it is being sent at full rate. Thus, the sum capacity is achieved.

2) $\alpha \leq 1/2$: With 1 round of message passing, each node (including the receivers) knows if it is the top-most user or not. The top most transmitter sends at a rate of $n$. However, all the other transmitters know that there is interference at its receiver and hence backs off to send at the top $(n-m)$ levels. Thus, the sum rate of $n+(K-1)(n-m)$ is achieved. Let $V_i$ be the interference produced by user $i$ to user $i+1$. Then, $\sum_{i=1}^{K-1} H(Y_i|V_i) + H(Y_K)$ is an outer bound to the sum capacity and this is upper bounded by $n+(K-1)(n-m)$. Thus, the sum capacity is achieved.

■

In Theorem 14, we proved that for $\alpha \geq 2$ or $\alpha \leq 1/2$, one round was sufficient for the existence of a universally optimal strategy. Further, one round is the minimum needed for the transmitters to know the rate at which they need to transmit. Note that the top receiver $\mathsf{D}_1$ did not need to know $\alpha$ to decode since the transmitter sends at a rate of $n$ which is known to the receiver. These regimes cover the very strong and the very weak interference. We next consider the case when the interference is strong but not very strong, and when the interference is weak but not very weak. In this case, we give a strategy that uses $\mathsf{O}(K)$ rounds to achieve the sum capacity. This achievable strategy uses $\mathsf{O}(K)$ rounds because the nodes base their decisions on their placement in the network. More precisely, the nodes learn if they are the even-numbered nodes or the odd-numbered nodes in the network state to decide on the rate allocation.

**Theorem 15.** 1) *For $1/2 < \alpha \leq 2/3$, there exists a strategy (in terms of sum-capacity) which is optimal with one round of message passing for $K < 3$ and with 2 rounds of message passing protocol for any $K \geq 3$.*

2) *For $2/3 < \alpha < 1$, the sum-rate of $n+(K-1)(n-m)$ can be achieved with 1 round of message passing, which is optimal for $K < 3$. In general, the sum-rate of $n+(K-1)(n-m)+(2m-n)\sum_{i=1}^{\lfloor (d-1)/2 \rfloor}(1_{K\geq 2i+1}) + (n-m)\sum_{i=1}^{\lfloor (d-2)/2 \rfloor}(1_{K\geq 2i+2})$ can be achieved in $d.5$ rounds. Thus, this strategy achieves optimal sum-rate for $K \geq 3$ in $K.5$ rounds.*

3) *Let $1 \leq \alpha < 2$, then the sum-rate of $n+(K-1)(m-n)+(2n-m)\sum_{i=1}^{\lfloor (d-1)/2 \rfloor}(1_{K\geq 2i+1})$ can be achieved in $d \geq 1$ rounds. This strategy is optimal with one round of message passing for $K < 3$. Further, this strategy is optimal for any odd $K \geq 3$ with $K$ rounds, and for any even $K$, $K \geq 4$ with $(K-1)$ rounds.*

*Proof:* We divide the proof into four cases as below.

1) $1/2 < \alpha \leq 1$, $d=1$: After one round, each node knows if it is the User 1 or not. The first node transmits at a rate of $n$ while all other users send at a rate of $n-m$ avoiding interference. This strategy will achieve the sum-capacity with one round of message passing for $K < 3$.

2) For $1/2 < \alpha \leq 2/3$, $d=2$: Each node knows if there are at least 2 nodes above it after 2 rounds of message passing protocol. The top most transmitter ($\mathsf{T}_1$) sends at a rate of $n$. The transmitter that has only one transmitter above it ($\mathsf{T}_2$) sends on the top $n-m$ signal levels so that the signal can be received interference-free at the receiver. Each transmitter that has at least 2 transmitters above transmits at top $n-m$ levels and the bottom $2m-n$ levels. Doing this, there is no interference at any of the users and a sum rate of $n+(n-m)+(K-2)m$ can be achieved. The outer bound of $\sum_{i=1}^{K-1} H(Y_i|V_i)+H(Y_K)$ reduces in this case to $n+(n-m)+(K-2)m$ thereby proving that the optimal sum-rate is achieved.

3) $2/3 < \alpha < 1$, $d > 1$: we first relabel the nodes from bottom to top as $j=1,\cdots,K$. We first consider a strategy with full information. Let the odd transmitters $\mathsf{T}_i$ with odd $i \in \{3,\cdots,K\}$ send at top $2m-n$ levels and bottom $n-m$ levels, while the even transmitters $\mathsf{T}_i$ with even $i \in \{3,\cdots,K\}$ send on all the top $n-m$, bottom $n-m$ levels but repeat the information of top $\min(n-m,3m-2n)$ levels that clashed with the information from transmitter $\mathsf{T}_{i-1}$ on levels $n-m+1$ to $n-m+1+\min(n-m,3m-2n)$. An example for $\alpha = 3/4$ is shown in Figure 7. Transmitter $\mathsf{T}_2$ sends at bottom $n-m$ levels and transmitter $\mathsf{T}_1$ sends at all the $n$ levels. This strategy will achieve the sum capacity which can be shown as follows. If $K$ is even, for all odd $i \in \{1, \cdots, K\}$, $R_i + R_{i+1}$ is upper bounded by $2n - m$ due to the Z-channel constraints and since our scheme achieves all these outer bounds, the sum capacity is achieved. If $K$ is odd, for all even $i \in \{4, \cdots, K\}$, $R_i + R_{i+1}$ is outer bounded by $2n - m$ due to Z-channel constraint and $R_1 + R_2 + R_3$ is outer bounded by $2n$ due to double Z-channel. Since all these inequalities are satisfied with equality with our achievable scheme, the sum capacity is achieved.

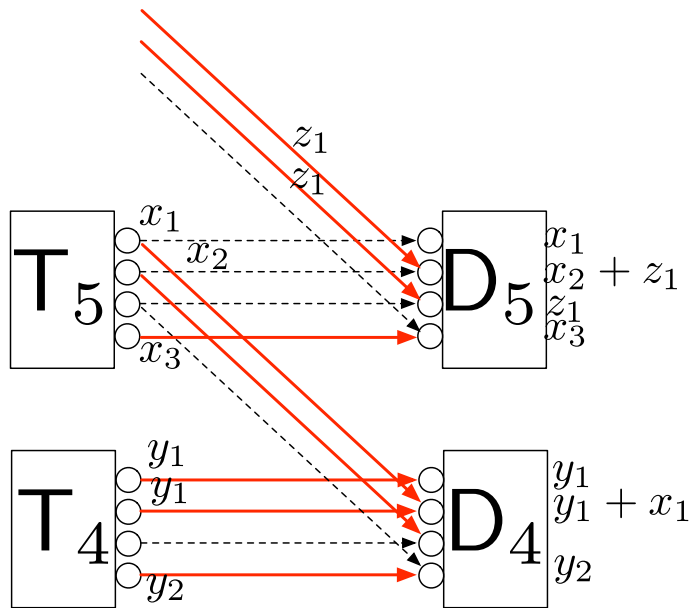

Fig. 7. An example with $m = 3$, $n = 4$, $\alpha = 3/4$. Let there be $\geq 6$ users and we consider users 4,5 and 6.

The above strategy assumed the knowledge of the odd or the even numbering of the users to choose the strategy. With no information of the node being even or odd, each transmitter uses the strategy to avoid interference (send at $n - m$ levels). However, when the node gets to know that it is odd/even numbered, it sends at the strategy of the odd/even node above. This needs larger number of rounds in this case. The third, fourth, fifth, sixth, $\cdots$ transmitters get to know that they are odd/even numbered and $K \geq 3$ after 3.5, 4.5, 5.5, 6.5, $\cdots$ iterations respectively thus proving the result.

4) $1 \leq \alpha < 2$: We again use the above relabeling of the nodes from top to below as $i = 1, \cdots, K$. We first consider a strategy with full information. Let the even transmitters among $\mathsf{T}_i$ with even $i \in \{1, \cdots, K\}$ send at top $m - n$ levels, while the odd-numbered users $\mathsf{T}_i$ with odd $i \in \{1, \cdots, K\}$ send on all the top $n$ levels but repeat the information of $\min(m - n, 2n - m)$ levels that clashed at receiver $\mathsf{D}_{i+1}$ on the lower levels (See Figure 8 for example).
The strategy will achieve the sum-capacity which can be shown as follows. If $K$ is even, then an outer bound can be given by genie-aided Z-channels as $R_i + R_{i+1} \leq m$ for all $1 \leq i < K$ and thus the sum-rate is upper bounded by $Km/2$ and we note that the above strategy achieves this outer bound. If $K$ is odd, then for all all odd-numbered $i \in \{1, \cdots, K - 1\}$, $R_i + R_{i+1}$ is upper bounded by $m$ and $R_K$ is upper bounded by $n$ by the point to point channel. Further, these conditions are satisfied with equality for our rate allocation. Thus, the sum capacity is achieved.

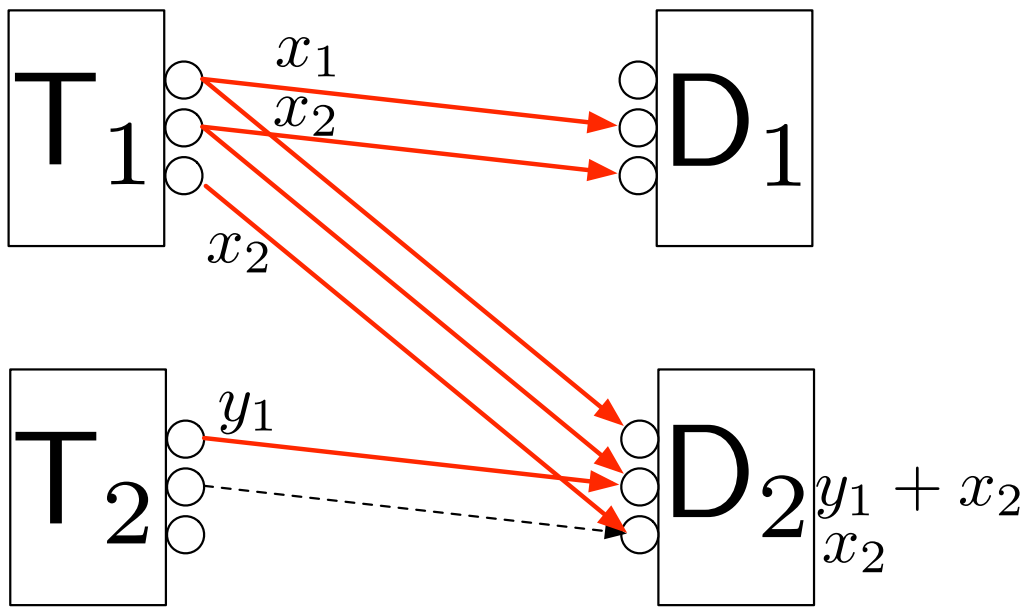

Fig. 8. An example with $m = 3$, $n = 2$, $\alpha = 3/2$. Let there be 5 users and we consider the top two users.

The above strategy assumed the knowledge of the odd or the even numbering of the transmitters to choose the strategy. With no information of the node being even or odd-numbered, each transmitter uses the strategy to avoid interference (send at $m - n$ levels). However, when the node gets to know that it is odd-numbered, it users the strategy of the odd-numbered node above. This needs larger number of rounds in this case. The first, third, fifth, seventh, $\cdots$ transmitters get to know that they are odd-numbered and $K \geq 3$ after 1, 3, 5, 7, $\cdots$ iterations respectively and the nodes know that there are at-least two transmitters below it in 2 rounds, thus proving the result.
Also note that the requirement of the number of rounds for $K = 3, 4$ can be reduced by 1 and for $K \geq 5$ can be further reduced by 2 by changing the strategy so that the bottom most node always send at rate of $n$ and the second node from bottom sends at rate $m - n$. Thus, the sum capacity can be achieved with $K - 2$ and $K - 3$ rounds for odd and even $K$ respectively for $K > 4$. Using similar counting of nodes from below also, the number of rounds needed can be further reduced, but still needing O($K$) rounds so that transmitters get to know even/odd placement information from the bottom or the top.

■

For $1/2 < \alpha \leq 2/3$, 2 rounds of message passing are enough since it is enough for the transmitters to learn if there are more than two nodes above it. However for $2/3 < \alpha < 2$, $O(K)$ rounds were needed to converge to the sum capacity in general so that the transmitters are able to know their placement in the channel connectivity. More precisely, the nodes learnt their relative position in the network being even/odd user. This requires $O(K)$ rounds of message passing. We note that this scheme involved the use of public as well as a private message at the transmitters.

### *C. Gaussian Z-network*

We now turn our attention to Gaussian channel model. The sum capacity for very weak and very strong interference can be achieved with one round of message passing. However, it is easy to show that one round of message passing does not suffice for strong but not very strong interference. This is because after first round of message passing, the users 2 to $K - 1$ have same information and have to send at same rate. If this rate has to be optimal for any $K$, this rate must be $\frac{1}{2}\log(1 + \mathsf{INR} + \mathsf{SNR})$ which limits the rate of the first and the $K^{th}$ user also to $\frac{1}{2}\log(1 + \mathsf{INR} + \mathsf{SNR})$. This sum rate will not be optimal if $K$ is odd. Hence, the optimal strategy needs also to consider if $K$ is even or odd and where each node is placed in the network even with knowledge of SNR, INR.

**Theorem 16.** *Suppose that each node knows the side information that the network is parameterized by three parameters $K$, $\mathsf{SNR}$ and $\mathsf{INR}$ but do not know its relative placement in the network. There exists an optimal strategy (achieving sum rate as with global network state information at all the nodes) with one round of message passing protocol for any $K$ when one of $\mathsf{INR} \geq \mathsf{SNR}(\mathsf{SNR}+1)$ or $\mathsf{INR}(\mathsf{INR}+1) \leq \mathsf{SNR}$ is satisfied. For any $\mathsf{INR}$ and $\mathsf{SNR}$, there exists an optimal strategy with one round of message passing for $K < 3$. For $\mathsf{SNR} \leq \mathsf{INR} < \mathsf{SNR}(\mathsf{SNR}+1)$, the sum rate of $\frac{K}{2}\log(1+\mathsf{INR}+\mathsf{SNR}) + \frac{1}{2}(\log(1+\mathsf{SNR}) - \log(1 + \frac{\mathsf{INR}}{1+\mathsf{SNR}}))1_{K=odd,d\geq K}$ can be achieved for $d > 1$ round of message passing. This strategy is optimal for any even $K$ with 1.5 round of message passing while for odd $K$ with $K$ rounds of message passing.*

*Proof:* We divide the claims of the theorem into four parts as follows.

1) $\mathsf{INR} \geq \mathsf{SNR}(\mathsf{SNR} + 1)$: Each transmitter knows that it is very strong interference in one round and hence the transmitters transmit at a rate of $\log(1 + \mathsf{SNR})$. Further, all the receivers except the first also know that it is very strong interference and are able to decode the messages. We also assume in all our achievability techniques that the first receiver will assume that a rate of $\log(1 + \mathsf{SNR})$ is being used by $\mathsf{T}_1$ and thus $\mathsf{D}_1$ will be able to decode. Thus, the sum-capacity of $K\log(1 + \mathsf{SNR})$ is achieved.
2) $\mathsf{INR}(\mathsf{INR}+1) \leq \mathsf{SNR}$: With one round of message passing, each node knows if it is the top-most user or not. The top-most transmitter sends at a rate of $\log(1+\mathsf{SNR})$. However, all the other transmitters know that there is one transmitter that interferences at the receiver and hence backs off to send at a rate of $\log(1 + \frac{\mathsf{SNR}}{1+\mathsf{INR}})$. Thus, the sum rate of $\log(1 + \mathsf{SNR}) + (K - 1)\log(1 + \frac{\mathsf{SNR}}{1+\mathsf{INR}})$ is achieved.

This can be shown to be an outer bound on the sum capacity using the same technique for the double-Z channel, generalizing the proof of Equation 26 in Appendix D.

3) $\mathsf{INR} \le \mathsf{SNR}$, $d = 1$: Each transmitter knows if it is the top-most transmitter. The top-most transmitter sends at a rate of $\log(1+\mathsf{SNR})$ while all others send at a rate of $\log(1+\frac{\mathsf{SNR}}{1+\mathsf{INR}})$ avoiding interference and thus the sum-capacity is achieved for $K < 3$.
4) $\mathsf{SNR} \le \mathsf{INR} < \mathsf{SNR}(\mathsf{SNR}+1)$, $d > 1$: In this case, the strategy for deterministic can also be extended by changing rates of $m-n$ by $\log(1+\frac{\mathsf{INR}}{1+\mathsf{SNR}})$ and $n$ by $\log(1+\mathsf{SNR})$ to get similar result as in deterministic model. However, we here consider an alternate strategy. Let, after the 1.5 rounds of message passing, each node transmits at a rate of $\frac{1}{2}\log(1+\mathsf{INR}+\mathsf{SNR})$ (0.5 rounds is added because the receiver $\mathsf{D}_1$ needs to know the range of $\mathsf{INR}$ to be able to decode). However, when all the nodes know the whole state which is after $K$ rounds, they change the strategy to that as in the deterministic case if $K$ is odd. This can also be proved to be optimal using a similar proof as in the deterministic case.

■

In the very strong interference, the transmitted message is all public and hence the decoders can decode both the messages since they know the rates. In the strong but not very strong interference, the consideration to even/odd number of users was important due to which the number of rounds of message passing increased to $K$ for odd $K$. Suppose that the nodes do not use node identity in the choice of strategies, but only the relative position known through the local information. The intuition in the use of $\mathsf{O}(K)$ rounds of message passing is that if constant number of neighboring nodes (let us say 10, numbered $2K_1+1, \cdots, 2K_1+10$) know the exact similar structure (which would happen with less than $\mathsf{O}(K)$ rounds), they choose the same rate. If this constant rate is not $\frac{1}{2}\log(1+\mathsf{SNR}+\mathsf{INR})$, the sum rate will not be optimal. Further, pick out even number of users (say $2K_1$) from above and the sum rate of these top nodes has to be $K_1\log(1+\mathsf{SNR}+\mathsf{INR})$, otherwise it won't lead to global optimality. Similarly, choose $2K_2$ users from below such that $K-2K_2 = 2K_1+10$ or $2K_1+11$ depending on $K$ being odd/even. The sum rate for these chosen nodes from below has to be $K_2\log(1+\mathsf{SNR}+\mathsf{INR})$ for optimality. Thus, if remaining nodes $(K-2(K_1+K_2))$ are odd, the sum rate is not optimal. Thus, there cannot exist an optimal rate allocation with less than $\mathsf{O}(K)$ rounds of message passing. In the very weak interference, signal of a transmitter is an interference for the other receivers (to which it is connected). In the remaining region of weak but not very weak interference ($\frac{\mathsf{SNR}}{1+\mathsf{INR}} < \mathsf{INR} < \mathsf{SNR}$), the signals need to be divided into public and private messages and is not considered in this paper for the Gaussian channel model.

## VII. Conclusions

Almost all networks operate with partial network information at different nodes, requiring nodes to make distributed decisions. While a rich literature exists on design of network protocols and their analysis, there is no prior work to understand the impact of distributed decisions on Shannon-theoretic capacity region. In this paper, we laid foundation to characterize partial network information and studied the impact in several network connectivities. Seeking universal optimality, where local decisions with certain side information are always globally optimal, we discovered that there appears to be a critical minimum information required for the network to allow globally optimal decisions. Our current approach is compound capacity based and our next step is to understand impact of partial information on fading interference channels.

## Appendix A
## Proof of Theorem 2

We note that at the end of first round, the first transmitter knows all the channel gains while the second transmitter does not know one of the channel gain $n_{11}$. Let the strategy that the transmitter uses be:

1) Transmitter 2, which is not producing interference, sends at it maximum possible rate of $n_{22}$.

2) Transmitter 1 assumes that transmitter 2 is sending at full rate. Thus, transmitter 1 sends at a rate of $(n_{11}-n_{12})^+$ if $n_{12} \le n_{22}$, thus not sending on any link that produces interference. However, it sends at a rate of $\min(\max(n_{11}, n_{12}) - n_{22}, n_{11})$ if $n_{12} > n_{22}$, transmitting at the non-interfering links to the signal of the second transmitter communicating at the rate $n_{22}$.

We now show that this strategy can achieve the sum rate as follows.

If $n_{12} \le n_{22}$, the sum rate in (7a) simplifies as

$$R_1 + R_2 \quad \le \quad (n_{11} - n_{12})^+ + n_{22}. \tag{44}$$

Hence, the first transmitter will send at a rate of $(n_{11}-n_{12})^+$. Further, since the first transmitter knows $n_{11}$ and $n_{12}$, it can send data on the links at which it is not generating any interference and thus achieve the sum capacity.

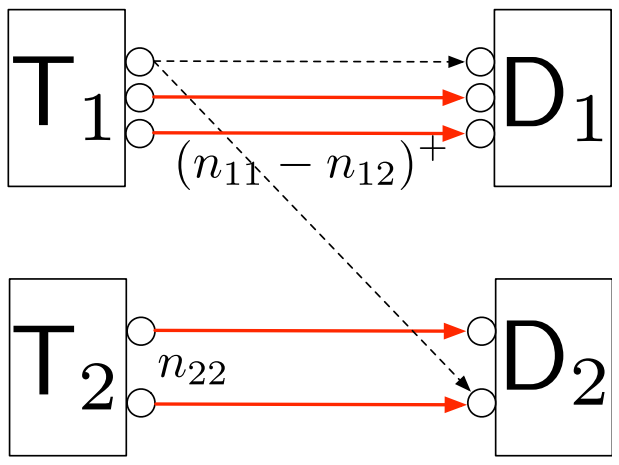

Fig. 9. In this case, $n_{12} \le n_{22}$. The bold lines show the active bits.

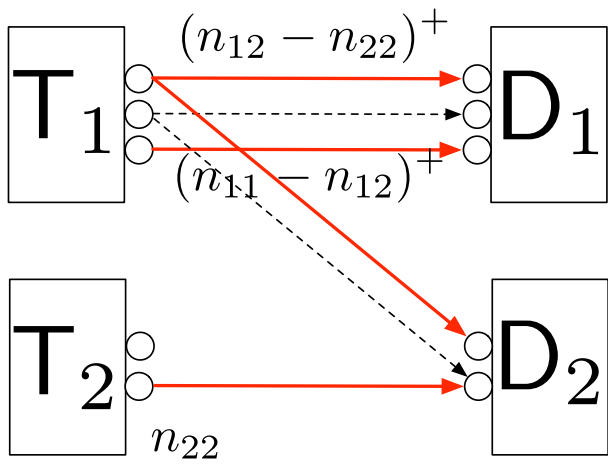

Fig. 10. In this case, $n_{12} > n_{22}$. The bold lines show the active bits.

Now consider the case when $n_{12} > n_{22}$, in which case the sum rate in (7a) simplifies as

$$R_1 + R_2 \quad \le \quad \min(\max(n_{11}, n_{12}), n_{11} + n_{22}). \tag{45}$$

Thus, the first transmitter sends at a rate of $\min(\max(n_{11}, n_{12}) - n_{22}, n_{11}) = (n_{11} - n_{12})^+ + \min(n_{12} - n_{22}, n_{11})$. This is achieved by sending along the dimensions that can not even be heard at the receiver 2 which are $(n_{11}-n_{12})^+$ in number. In addition, among the dimensions that can be heard by the second receiver, the data is sent along the dimensions which do not produce an interference to the direct signal. These are $\min(n_{12} - n_{22}, n_{11})$ in number; see Figure 10. The extra half round is required for the receiver to learn the strategy used by the transmitters. Thus, the maximum sum rate can be achieved with 1.5 rounds.

# Appendix B
# Proof of Theorem 5

## *A. The rates mentioned in* (12)-(16) *can be decoded at the receivers*

In this subsection, we will show that the rates mentioned in (12)-(16) can be decoded at the receivers. In order to see the decoding process at the two users, consider the following scenarios.

1) $\mathsf{INR}_2 < \mathsf{SNR}_2$. The first receiver will be able to decode the data as the rate is supported by the power sent. The second receiver treats the first user's signal as interference. The interference power is $\mathsf{INR}_2/(1+\mathsf{INR}_2)$. Thus, the second receiver is able to decode its data treating this power as interference.

2) $\mathsf{INR}_2 \geq \mathsf{SNR}_2$, $\mathsf{INR}_2 \geq \mathsf{SNR}_1$. Since the received power from the signal of the first transmitter is $\mathsf{SNR}_1$ and rate $\leq \log(1+\mathsf{SNR}_1)$, the first receiver is able to decode. At the second receiver, the two signals are decoded jointly. We have to verify:

$$R_1 \leq \log(1+\mathsf{INR}_2) \tag{46a}$$
$$R_2 \leq \log(1+\mathsf{SNR}_2) \tag{46b}$$
$$R_1+R_2 \leq \log(1+\mathsf{INR}_2+\mathsf{SNR}_2) \tag{46c}$$

The first two hold. For the third, $R_1+R_2 \leq \log\left(1+\frac{\mathsf{INR}_2}{1+\frac{\mathsf{SNR}_2\mathsf{INR}_2}{1+\mathsf{SNR}_2+2\mathsf{INR}_2}}\right)+\log(1+\mathsf{SNR}_2(1+\mathsf{INR}_2)/(1+2\mathsf{INR}_2)) = \log(1+\mathsf{INR}_2+\mathsf{SNR}_2)$. Hence proved.

3) $\mathsf{INR}_2 \geq \mathsf{SNR}_2$, $\mathsf{INR}_2 < \mathsf{SNR}_1$. In this case, there is both a public and private message from the first user. At the first receiver, we need to decode the public data treating the other as noise. Further, the private data can be decoded. To check the first, we need to see

$$R_{1,c} \leq \log\left(1+\frac{\mathsf{SNR}_1\mathsf{INR}_2/(1+\mathsf{INR}_2)}{1+\mathsf{SNR}_1/(1+\mathsf{INR}_2)}\right) \tag{47}$$

Thus,

$$\frac{\mathsf{INR}_2^2}{1+2\mathsf{INR}_2+\mathsf{SNR}_2(1+\mathsf{INR}_2)} \leq \frac{\mathsf{SNR}_1\mathsf{INR}_2}{1+\mathsf{INR}_2+\mathsf{SNR}_1} \tag{48}$$

It is enough to prove $\frac{\mathsf{INR}_2}{1+2\mathsf{INR}_2+\mathsf{SNR}_2(1+\mathsf{INR}_2)} \leq \frac{\mathsf{INR}_2}{1+\mathsf{INR}_2+\mathsf{INR}_2}$ since $\mathsf{INR}_2 \leq \mathsf{SNR}_1$ will prove the rest. The first part trivially holds too.

At the second receiver, we need to show that we can jointly decode the common message of user 1 and the data of user 2 treating the private message from the first user as noise. For this, we need the following:

$$R_{1,c} \leq \log\left(1+\frac{\mathsf{INR}_2^2/(1+\mathsf{INR}_2)}{1+\mathsf{INR}_2/(1+\mathsf{INR}_2)}\right) \tag{49a}$$
$$R_2 \leq \log\left(1+\frac{\mathsf{SNR}_2}{1+\mathsf{INR}_2/(1+\mathsf{INR}_2)}\right) \tag{49b}$$
$$R_{1,c}+R_2 \leq \log\left(1+\frac{\mathsf{SNR}_2+\mathsf{INR}_2^2/(1+\mathsf{INR}_2)}{1+\mathsf{INR}_2/(1+\mathsf{INR}_2)}\right) \tag{49c}$$

We note that all these three conditions are satisfied.

### *B. Difference between achievability and outer bound*

To show that the achievable sum rate is within 2 bits of the outer bound, we consider the following regimes.

1) $\mathsf{INR}_2 < \mathsf{SNR}_2$, $\mathsf{INR}_2 < \mathsf{SNR}_1$.
Achievable sum rate

$$R_{ac} = \log(1+\mathsf{SNR}_1/(1+\mathsf{INR}_2)) + \log(1+\mathsf{SNR}_2(1+\mathsf{INR}_2)/(1+2\mathsf{INR}_2)) \tag{50}$$

Outer bound on sum rate:

$$R_{co} = \log(1+\mathsf{SNR}_1) + \log\left(1+\frac{\mathsf{SNR}_2}{1+\mathsf{INR}_2}\right) \tag{51}$$

$$\begin{aligned} R_{co} - R_{ac} &= \log(1 + \mathsf{SNR}_1) + \log(1 + \mathsf{INR}_2 + \mathsf{SNR}_2) - \log(1 + \mathsf{INR}_2 + \mathsf{SNR}_1) \\ &\quad - \log(1 + \mathsf{SNR}_2(1 + \mathsf{INR}_2)/(1 + 2\mathsf{INR}_2)) && \text{(52a)} \\ &\leq \log(1 + \mathsf{INR}_2 + \mathsf{SNR}_2) - \log(1 + \mathsf{SNR}_2(1 + \mathsf{INR}_2)/(1 + 2\mathsf{INR}_2)) && \text{(52b)} \\ &\leq \log(1 + 2\mathsf{SNR}_2) - \log(1 + \mathsf{SNR}_2/2) && \text{(52c)} \\ &\leq 2 && \text{(52d)} \end{aligned}$$

2) $\mathsf{INR}_2 < \mathsf{SNR}_2$, $\mathsf{INR}_2 \geq \mathsf{SNR}_1$.
   Achievable sum rate

$$R_{ac} = \log(1 + \mathsf{SNR}_1/(1 + \mathsf{INR}_2)) + \log(1 + \mathsf{SNR}_2(1 + \mathsf{INR}_2)/(1 + 2\mathsf{INR}_2)) \tag{53}$$

   Outer bound on sum rate

$$R_{co} = \log(1 + \mathsf{SNR}_2 + \mathsf{INR}_2) \tag{54}$$

$$\begin{aligned} R_{co} - R_{ac} &= \log(1 + \mathsf{INR}_2 + \mathsf{SNR}_2) - \log(1 + \mathsf{INR}_2 + \mathsf{SNR}_1) + \log(1 + \mathsf{INR}_2) \\ &\quad - \log(1 + \mathsf{SNR}_2(1 + \mathsf{INR}_2)/(1 + 2\mathsf{INR}_2)) && \text{(55a)} \\ &\leq \log(1 + \mathsf{INR}_2 + \mathsf{SNR}_2) - \log(1 + \mathsf{SNR}_2(1 + \mathsf{INR}_2)/(1 + 2\mathsf{INR}_2)) && \text{(55b)} \\ &\leq \log(1 + \mathsf{INR}_2 + \mathsf{SNR}_2) - \log(1 + \mathsf{SNR}_2/2) && \text{(55c)} \\ &\leq \log(1 + \mathsf{INR}_2 + \mathsf{SNR}_2) - \log(1 + (\mathsf{SNR}_2 + \mathsf{INR}_2)/4) && \text{(55d)} \\ &\leq 2 && \text{(55e)} \end{aligned}$$

3) $\mathsf{INR}_2 \geq \mathsf{SNR}_2$, $\mathsf{INR}_2 \geq \mathsf{SNR}_1$.
   Achievable sum rate

$$R_{ac} = \log\left(1 + \min\left(\mathsf{SNR}_1, \frac{\mathsf{INR}_2}{1 + \frac{\mathsf{SNR}_2\mathsf{INR}_2}{1+\mathsf{SNR}_2+2\mathsf{INR}_2}}\right)\right) + \log(1 + \mathsf{SNR}_2(1 + \mathsf{INR}_2)/(1 + 2\mathsf{INR}_2)) \tag{56}$$

   Outer bound on sum rate

$$R_{co} = \log(1 + \mathsf{SNR}_2 + \mathsf{INR}_2) \tag{57}$$

   If $\frac{\mathsf{INR}_2}{1+\frac{\mathsf{SNR}_2\mathsf{INR}_2}{1+\mathsf{SNR}_2+2\mathsf{INR}_2}} \leq \mathsf{SNR}_1$, achievability matches the outer bound.
   However, if $\frac{\mathsf{INR}_2}{1+\frac{\mathsf{SNR}_2\mathsf{INR}_2}{1+\mathsf{SNR}_2+2\mathsf{INR}_2}} \geq \mathsf{SNR}_1$, consider $R_{co} = \log(1 + \mathsf{SNR}_1) + \log(1 + \mathsf{SNR}_2)$ to show that $R_{co} - R_{ac} \leq 1$.

4) $\mathsf{INR}_2 \geq \mathsf{SNR}_2$, $\mathsf{INR}_2 < \mathsf{SNR}_1$.
   Achievable sum rate

$$\begin{aligned} R_{ac} &= \log\left(1 + \frac{\mathsf{INR}_2^2}{1 + 2\mathsf{INR}_2 + \mathsf{SNR}_2(1 + \mathsf{INR}_2)}\right) + \log(1 + \mathsf{SNR}_1/(1 + \mathsf{INR}_2)) \\ &\quad + \log(1 + \mathsf{SNR}_2(1 + \mathsf{INR}_2)/(1 + 2\mathsf{INR}_2)) && \text{(58)} \end{aligned}$$

   Outer bound on sum rate:

$$R_{co} = \log(1 + \mathsf{SNR}_1) + \log\left(1 + \frac{\mathsf{SNR}_2}{1 + \mathsf{INR}_2}\right) \tag{59}$$

Note that

$$
\begin{aligned}
R_{ac} &= \log\left(1+\frac{\mathsf{INR}_2^2}{1+2\mathsf{INR}_2+\mathsf{SNR}_2(1+\mathsf{INR}_2)}\right)+\log(1+\mathsf{SNR}_1/(1+\mathsf{INR}_2)) \\
&\quad +\log(1+\mathsf{SNR}_2(1+\mathsf{INR}_2)/(1+2\mathsf{INR}_2)) && \text{(60a)}\\
&\geq \log\left(1+\frac{\mathsf{SNR}_2+\mathsf{INR}_2^2/(1+\mathsf{INR}_2)}{1+\mathsf{INR}_2/(1+\mathsf{INR}_2)}\right)+\log(1+\mathsf{SNR}_1/(1+\mathsf{INR}_2)) && \text{(60b)}\\
&\geq \log\left(1+\frac{\mathsf{INR}_2^2/(1+\mathsf{INR}_2)}{1+\mathsf{INR}_2/(1+\mathsf{INR}_2)}\right)+\log(1+\mathsf{SNR}_1/(1+\mathsf{INR}_2)) && \text{(60c)}\\
&= \log\left(1+\frac{\mathsf{INR}_2^2}{1+2\mathsf{INR}_2}\right)+\log(1+\mathsf{SNR}_1/(1+\mathsf{INR}_2)) && \text{(60d)}\\
&\geq \log(1+2\mathsf{INR}_2+\mathsf{INR}_2^2)-\log(1+2\mathsf{INR}_2)+\log(1+\mathsf{SNR}_1)\\
&\quad -\log(1+\mathsf{INR}_2) && \text{(60e)}\\
&= \log(1+\mathsf{INR}_2)-\log(1+2\mathsf{INR}_2)+\log(1+\mathsf{SNR}_1) && \text{(60f)}\\
&\geq \log(1+\mathsf{SNR}_1)-1 && \text{(60g)}
\end{aligned}
$$

Thus,

$$
\begin{aligned}
R_{co}-R_{ac} &\leq \log(1+\mathsf{SNR}_1)+\log\left(1+\frac{\mathsf{SNR}_2}{1+\mathsf{INR}_2}\right)-\log(1+\mathsf{SNR}_1)+1 && \text{(61a)}\\
&\leq \log\left(1+\frac{\mathsf{SNR}_2}{1+\mathsf{INR}_2}\right)+1 && \text{(61b)}\\
&\leq 2 && \text{(61c)}
\end{aligned}
$$

## APPENDIX C
## A CLASS OF THREE-USER DETERMINISTIC CHANNELS

In this Appendix, we will introduce a class of deterministic double Z channels and find a capacity region for this class of channels.

We consider a class of deterministic discrete memoryless IFC's in which the outputs $Y_1$, $Y_2$, and $Y_3$, and the interferences $V_1$, and $V_2$ are (deterministic) functions of the inputs $X_1$, $X_2$ and $X_3$ as follows:

$$
\begin{aligned}
Y_1 &= f_1(X_1) && \text{(62a)}\\
Y_2 &= f_2(X_2,V_1) && \text{(62b)}\\
Y_3 &= f_3(X_3,V_2) && \text{(62c)}\\
V_1 &= g_1(X_1) && \text{(62d)}\\
V_2 &= g_2(X_2), && \text{(62e)}
\end{aligned}
$$

where $f_1(.)$, $f_2(.)$, $f_3(.)$, $g_1(.)$ and $g_2(.)$ are deterministic functions. Further, let $f_1(.)$, $f_2(.)$ and $f_3(.)$ satisfy

$$
\begin{aligned}
H(Y_1|X_1) &= 0 && \text{(63a)}\\
H(Y_2|X_2) &= H(V_1) && \text{(63b)}\\
H(Y_3|X_3) &= H(V_2). && \text{(63c)}
\end{aligned}
$$

**Theorem 17.** *The capacity region for the class of deterministic double Z channels is given by*

$$\begin{aligned}
R_1 &\leq H(Y_1) && \text{(64a)}\\
R_2 &\leq H(Y_2|V_1) && \text{(64b)}\\
R_3 &\leq H(Y_3|V_2) && \text{(64c)}\\
R_1+R_2 &\leq H(Y_1|V_1)+H(Y_2) && \text{(64d)}\\
R_2+R_3 &\leq H(Y_2|V_1V_2)+H(Y_3) && \text{(64e)}\\
R_1+R_2+R_3 &\leq H(Y_1|V_1)+H(Y_2|V_2)+H(Y_3) && \text{(64f)}
\end{aligned}$$

The remaining part of the section is devoted to the proof of this Theorem.

## *A. Han-Kobayashi Region for a general DMC*

The achievability of the above theorem will follow by specializing the Han-Kobayashi region to the class of deterministic channels. In this subsection, we will describe the Han-Kobayashi region for a general discrete memoryless channel. Let $U_{11}$, $U_{12}$, $U_{22}$, $U_{23}$ and $U_3$ be the auxiliary variables and $Q$ be the time sharing auxiliary variable satisfying the following conditions

1) $U_{11}$, $U_{12}$, $U_{22}$, $U_{23}$ and $U_3$ are conditionally independent given $Q$.
2) $X_1 = f_1(U_{11}, U_{12}|Q)$, $X_2 = f_2(U_{22}, U_{23}|Q)$ and $X_3 = f_3(U_3|Q)$.
3) $\Pr\{Y_1 = y_1, Y_2 = y_2, Y_3 = y_3|X_1 = x_1, X_2 = x_2, X_3 = x_3\} = w(y_1, y_2, y_3|x_1, x_2, x_3)$

**Theorem 18.** *Under the above constraints, the Han-Kobayashi rate region is given by* $(R_{11}+R_{12}, R_{22}+R_{23}, R_3)$*, where* $R_{11}, R_{12}, R_{22}, R_{23}, R_3$ *satisfy following set of inequalities:*

$$\begin{aligned}
R_{11} &\leq a_1 && \text{(65a)}\\
R_{12} &\leq b_1 && \text{(65b)}\\
R_{11}+R_{12} &\leq c_1 && \text{(65c)}\\
R_{12} &\leq d_1 && \text{(65d)}\\
R_{22} &\leq e_1 && \text{(65e)}\\
R_{23} &\leq f_1 && \text{(65f)}\\
R_{12}+R_{22} &\leq g_1 && \text{(65g)}\\
R_{12}+R_{23} &\leq h_1 && \text{(65h)}\\
R_{22}+R_{23} &\leq i_1 && \text{(65i)}\\
R_{12}+R_{22}+R_{23} &\leq j_1 && \text{(65j)}\\
R_{23}+R_3 &\leq k_1 && \text{(65k)}\\
R_{23} &\leq l_1 && \text{(65l)}\\
R_3 &\leq m_1 && \text{(65m)}\\
-R_{11} &\leq 0 && \text{(65n)}\\
-R_{12} &\leq 0 && \text{(65o)}\\
-R_{22} &\leq 0 && \text{(65p)}\\
-R_{23} &\leq 0 && \text{(65q)}\\
-R_3 &\leq 0, && \text{(65r)}
\end{aligned}$$

*where*

$$
\begin{aligned}
a_1 &= I(Y_1; U_{11}|U_{12}, Q) && \text{(66a)}\\
b_1 &= I(Y_1; U_{12}|U_{11}, Q) && \text{(66b)}\\
c_1 &= I(Y_1; U_{12}, U_{11}|Q) && \text{(66c)}\\
d_1 &= I(Y_2; U_{12}|U_{22}, U_{23}, Q) && \text{(66d)}\\
e_1 &= I(Y_2; U_{22}|U_{12}, U_{23}, Q) && \text{(66e)}\\
f_1 &= I(Y_2; U_{23}|U_{12}, U_{22}, , Q) && \text{(66f)}\\
g_1 &= I(Y_2; U_{12}, U_{22}|U_{23}, Q) && \text{(66g)}\\
h_1 &= I(Y_2; U_{12}, U_{23}|U_{22}, Q) && \text{(66h)}\\
i_1 &= I(Y_2; U_{22}, U_{23}|U_{12}, Q) && \text{(66i)}\\
j_1 &= I(Y_2; U_{12}, U_{22}, U_{23}|Q) && \text{(66j)}\\
k_1 &= I(Y_3; U_{23}, U_3|Q) && \text{(66k)}\\
l_1 &= I(Y_3; U_{23}|U_3, Q) && \text{(66l)}\\
m_1 &= I(Y_3; U_3|U_{23}, Q) && \text{(66m)}
\end{aligned}
$$

*Proof:* The proof is an easy extension of [14] and is therefore omitted. ■

**Theorem 19.** *The above rate region can be simplified to*

$$
\begin{aligned}
R_1 &\leq \min(c_1, a_1 + d_1) && \text{(67a)}\\
R_2 &\leq \min(i_1, e_1 + l_1) && \text{(67b)}\\
R_3 &\leq m_1 && \text{(67c)}\\
R_1 + R_2 &\leq a_1 + \min(j_1, g_1 + l_1) && \text{(67d)}\\
R_2 + R_3 &\leq k_1 + e_1 && \text{(67e)}\\
2R_1 + R_2 &\leq 2a_1 + g_1 + h_1 && \text{(67f)}\\
R_1 + R_2 + R_3 &\leq a_1 + g_1 + k_1 && \text{(67g)}\\
-R_1 &\leq 0 && \text{(67h)}\\
-R_2 &\leq 0 && \text{(67i)}\\
-R_3 &\leq 0 && \text{(67j)}
\end{aligned}
$$

*Proof:* The proof follows by repeated use of Fourier Motzkin Elimination Algorithm as described below. In Equations (65a)-(65r), replace $R_{11}$ by $R_1 - R_{12}$. Further, we separate the terms not containing $R_{12}$, containing $R_{12}$ with positive sign on the left of $\leq$ and containing negative sign on the left of $\leq$ as follows.

Terms not containing $R_{12}$:

$$R_1 \leq c_1 \tag{68a}$$
$$R_{22} \leq e_1 \tag{68b}$$
$$R_{23} \leq f_1 \tag{68c}$$
$$R_{22} + R_{23} \leq i_1 \tag{68d}$$
$$R_{22} + R_3 \leq k_1 \tag{68e}$$
$$R_{23} \leq l_1 \tag{68f}$$
$$R_3 \leq m_1 \tag{68g}$$
$$-R_{22} \leq 0 \tag{68h}$$
$$-R_{23} \leq 0 \tag{68i}$$
$$-R_3 \leq 0 \tag{68j}$$

Terms containing $R_{12}$ with positive sign:

$$R_{12} \leq \min(b_1, d_1) \tag{69a}$$
$$R_{12} + R_{22} \leq g_1 \tag{69b}$$
$$R_{12} + R_{23} \leq h_1 \tag{69c}$$
$$R_{12} + R_{22} + R_{23} \leq j_1 \tag{69d}$$
$$R_{12} - R_1 \leq 0 \tag{69e}$$

Terms containing $R_{12}$ with negative sign:

$$R_1 - R_{12} \leq a_1 \tag{70a}$$
$$-R_{12} \leq 0 \tag{70b}$$

Hence, the Han-Kobayashi rate region is equivalently given by Equations (68a)-(70b). $R_{12}$ can be eliminated from Equations (69a)-(70b) using Fourier Motzkin Elimination technique to get the following

$$R_1 \leq a_1 + \min(b_1, d_1) \tag{71a}$$
$$R_1 + R_{22} \leq a_1 + g_1 \tag{71b}$$
$$R_1 + R_{23} \leq a_1 + h_1 \tag{71c}$$
$$R_{22} \leq g_1 \tag{71d}$$
$$R_{23} \leq h_1 \tag{71e}$$
$$R_1 + R_{22} + R_{23} \leq a_1 + j_1 \tag{71f}$$
$$R_{22} + R_{23} \leq j_1 \tag{71g}$$
$$-R_1 \leq 0 \tag{71h}$$

The rate region is now equivalently given by Equations (68a)-(68j), (71a)-(71h). In these equations, substitute $R_{22} = R_2 - R_{23}$ and separate the terms not containing $R_{23}$, containing $R_{23}$ with positive sign on the left hand side of $\leq$, and the terms containing negative sign on the left of $\leq$ as shown below.

Terms not containing $R_{23}$:

$$R_1 \leq \min(c_1, a_1 + b_1, a_1 + d_1) \tag{72a}$$
$$R_2 \leq \min(i_1, j_1) = i_1 \tag{72b}$$
$$R_1 + R_2 \leq a_1 + j_1 \tag{72c}$$
$$-R_1 \leq 0 \tag{72d}$$
$$R_3 \leq m_1 \tag{72e}$$
$$-R_3 \leq 0 \tag{72f}$$

Terms containing $R_{23}$ with a positive sign:

$$R_{23} \leq \min(f_1, h_1, l_1) \tag{73a}$$
$$R_3 + R_{23} \leq k_1 \tag{73b}$$
$$-R_2 + R_{23} \leq 0 \tag{73c}$$
$$R_1 + R_{23} \leq a_1 + h_1 \tag{73d}$$

Terms containing $R_{23}$ with a negative sign:

$$R_2 - R_{23} \leq \min(e_1, g_1) = e_1 \tag{74a}$$
$$R_1 + R_2 - R_{23} \leq a_1 + g_1 \tag{74b}$$
$$-R_{23} \leq 0 \tag{74c}$$

The rate region is now equivalently given by Equations (72a)-(74c). Eliminating $R_{23}$ from Equations (73a)-(74c) using Fourier Motzkin Elimination, we get

$$R_2 \leq \min(e_1 + f_1, e_1 + h_1, e_1 + l_1) \tag{75a}$$
$$R_1 + R_2 \leq a_1 + g_1 + \min(f_1, h_1, l_1) \tag{75b}$$
$$R_1 + R_2 \leq e_1 + a_1 + h_1 \tag{75c}$$
$$2R_1 + R_2 \leq a_1 + g_1 + a_1 + h_1 \tag{75d}$$
$$R_1 \leq a_1 + h_1 \tag{75e}$$
$$R_2 + R_3 \leq k_1 + e_1 \tag{75f}$$
$$R_1 + R_2 + R_3 \leq a_1 + g_1 + k_1 \tag{75g}$$
$$R_3 \leq k_1 \tag{75h}$$
$$R_1 \leq a_1 + g_1 \tag{75i}$$
$$-R_2 \leq 0 \tag{75j}$$

The rate region is now given by Equations (72a)-(72f), (75a)-(75j). Combining these terms, the rate region can be described as follows.

$$R_1 \leq \min(c_1, a_1+b_1, a_1+d_1, a_1+h_1, a_1+g_1) \quad (76a)$$
$$R_2 \leq \min(i_1, e_1+f_1, e_1+h_1, e_1+l_1) \quad (76b)$$
$$R_3 \leq \min(k_1, m_1) \quad (76c)$$
$$R_1+R_2 \leq \min(a_1+j_1, a_1+f_1+g_1, a_1+g_1+h_1, a_1+e_1+h_1, a_1+g_1+l_1) \quad (76d)$$
$$R_2+R_3 \leq k_1+e_1 \quad (76e)$$
$$2R_1+R_2 \leq 2a_1+g_1+h_1 \quad (76f)$$
$$R_1+R_2+R_2 \leq a_1+g_1+k_1 \quad (76g)$$
$$-R_1 \leq 0 \quad (76h)$$
$$-R_2 \leq 0 \quad (76i)$$
$$-R_3 \leq 0 \quad (76j)$$

Since $c_1 \leq a_1+b_1$, $d_1 \leq h_1$ and $d_1 \leq g_1$, the $R_1$ equation is same as in the statement of the Theorem. Since $i_1 \leq e_1+f_1 \leq e_1+h_1$, the $R_2$ equation is same as in the statement of the Theorem. Further, $m_1 \leq k_1$ gives $R_3$ same as in the statement of the Theorem. As $j_1 \leq f_1+g_1$, $j_1 \leq e_1+h_1$ and $f_1 \leq h_1$, $R_1+R_2$ bound is also same as in the statement of the Theorem. The rest of the statements directly follow. ■

### *B. Specializing Han-Kobayashi Achievability for model in Theorem 17*

To show the achievability, we specialize the Han-Kobayashi Rate region by taking the time-sharing variable as trivial, and using $U_{12}=V_1$, $U_{23}=V_2$, $X_2=h_1(U_{22},U_{23})$, $X_1=h_2(U_{11},U_{12})$, $U_{22}=g_3(X_2)$, $U_{11}=g_4(X_1)$ and $U_3=X_3$ for some deterministic functions $h_1$, $h_2$, $g_3$ and $g_4$. With these substitutions, we now show that the Han-Kobayashi rate region reduces to that in the statement of the Theorem 17.

We consider all the equations in Han-Kobayashi region one by one.

From (67), $R_1 \leq \min(c_1, a_1+d_1)$. Here, $c_1 = I(Y_1;X_1) = H(Y_1)$ and $a_1+d_1 = H(Y_1|V_1)+H(Y_2|X_2) = H(Y_1,V_1)$. Thus, the above reduces to $R_1 \leq H(Y_1)$ which is same as in the statement of the Theorem 17.

From (67), $R_2 \leq \min(i_1, e_1+l_1)$. Here, $i_1 = H(Y_2|V_1)$ and $e_1+l_1 = H(Y_2|V_1V_2)+H(Y_3|X_3) = H(Y_2|V_1V_2)+H(V_2) \geq H(Y_2,V_2|V_1) \geq H(Y_2|V_1)$. Thus, this bound also reduces to the same as in the statement of the Theorem 17.

From (67), $R_3 \leq m_1 = H(Y_3|V_2)$.

From (67), $R_1+R_2 \leq a_1+\min(j_1, g_1+l_1)$. Here, $j_1 = H(Y_2) \leq H(Y_2,V_2) = H(Y_2|V_2)+H(Y_3|X_3) = g_1+l_1$. Thus, this reduces to $R_1+R_2 \leq H(Y_1|V_1)+H(Y_2)$ as in the statement of the Theorem 17.

From (67), $R_2+R_3 \leq k_1+e_1 = H(Y_3)+H(Y_2|V_1V_2)$ is the same as in the statement of the Theorem 17.

From (67), $2R_1+R_2 \leq 2a_1+g_1+h_1 = 2H(Y_1|V_1)+H(Y_2|V_2)+H(Y_2|U_{22})$. We will now show that this constraint is looser than the sum of the constraints on $R_1$ and $R_1+R_2$, which proves that this is not a limiting condition on the rate region.

$$2H(Y_1|V_1)+H(Y_2|V_2)+H(Y_2|U_{22}) \quad (77a)$$
$$= 2H(Y_1|V_1)+H(Y_2|V_2)+H(Y_2|U_{22})-H(Y_2|U_{22}V_2)+H(Y_2|U_{22}V_2) \quad (77b)$$
$$= 2H(Y_1|V_1)+I(Y_2;U_{22}|V_2)+H(Y_2|U_{22})+H(Y_2|X_2) \quad (77c)$$
$$\overset{(a)}{=} 2H(Y_1|V_1)+I(Y_2;U_{22})+H(Y_2|U_{22})+H(V_1) \quad (77d)$$
$$= H(Y_1,V_1)+H(Y_2)+H(Y_1|V_1) \quad (77e)$$
$$\geq H(Y_1)+H(Y_2)+H(Y_1|V_1), \quad (77f)$$

where (a) follows since $V_2$ and $U_{22}$ are independent. Thus, the bound of $2R_1 + R_2$ is redundant.

From (67), $R_1 + R_2 + R_3 \leq a_1 + g_1 + k_1 = H(Y_1|V_1) + H(Y_2|V_2) + H(Y_3)$ which proves the last condition in Theorem 17. This completes the proof of the achievability of the rate region.

### C. Converse for Theorem 17

The individual bounds on $R_1$, $R_2$, $R_3$ and $R_1 + R_2$ follow the same steps as in [12] and are omitted. For $R_2 + R_3$,

$$\begin{aligned} n(R_2+R_3) &\leq I(X_2^n;Y_2^n)+I(X_3^n;Y_3^n)+n\epsilon && \text{(78a)}\\ &\overset{(b)}{\leq} I(X_2^n;Y_2^n|V_1^n)+I(X_3^n;Y_3^n)+n\epsilon && \text{(78b)}\\ &\leq I(X_2^n;V_2^nY_2^n|V_1^n)+H(Y_3^n)-H(Y_3^n|X_3^n)+n\epsilon && \text{(78c)}\\ &= I(X_2^n;V_2^n|V_1^n)+I(X_2^n;Y_2^n|V_2^n,V_1^n)+H(Y_3^n)-H(V_2^n)+n\epsilon && \text{(78d)}\\ &\overset{(c)}{=} H(V_2^n)+H(Y_2^n|V_2^n,V_1^n)+H(Y_3^n)-H(V_2^n)+n\epsilon && \text{(78e)}\\ &= H(Y_2^n|V_2^n,V_1^n)+H(Y_3^n)+n\epsilon, && \text{(78f)} \end{aligned}$$

where (b) follows since $H(Y_2) \leq H(Y_2, V_1)$ implies $H(Y_2) \leq H(V_1) + H(Y_2|V_1) = H(Y_2|X_2) + H(Y_2|V_1) = H(Y_2|X_2) + H(Y_2|V_1) - H(Y_2|X_2, V_1)$ , and (c) follows since $V_1$ and $V_2$ are independent. Taking $n \to \infty$ and by the convexity properties of the region, the bound in the statement of Theorem 17 is obtained.

For the bound on $R_1 + R_2 + R_3$,

$$\begin{aligned} & n(R_1+R_2+R_3)\\ \leq\ & I(X_1^n;Y_1^n)+I(X_2^n;Y_2^n)+I(X_3^n;Y_3^n)+n\epsilon && \text{(79a)}\\ =\ & H(Y_1^n)+H(Y_2^n)-H(V_1^n)+H(Y_3^n)-H(V_2^n)+n\epsilon && \text{(79b)}\\ \leq\ & H(Y_1^n)-H(V_1^n)+H(V_1^n|Y_1^n)+H(Y_2^n)-H(V_2^n)+H(V_2^n|Y_2^n)+H(Y_3^n)+n\epsilon && \text{(79c)}\\ =\ & H(Y_1^n|V_1^n)+H(Y_2^n|V_2^n)+H(Y_3^n)+n\epsilon && \text{(79d)} \end{aligned}$$

Taking $n \to \infty$ and by the convexity properties of the region, the bound in the statement of Theorem 17 is obtained.

## APPENDIX D
## OUTER BOUNDS FOR GAUSSIAN 3-USER DOUBLE Z-CHANNEL

In this Section, we prove the sum rate bounds in the second and the fourth cases, more precisely, Equations (22)-(23), (26)-(27).

### A. $\mathsf{INR}_2 \geq \mathsf{SNR}_1$ *and* $\mathsf{INR}_3 \leq \mathsf{SNR}_2$

Note that there is a strong interference between first two users. Hence both $X_1$ and $X_2$ can be decoded from $Y_2$. Thus,

$$n(R_1+R_2+R_3) \leq I(X_1^n,X_2^n;Y_2^n)+I(X_3^n;Y_3^n)+n\epsilon \tag{80}$$

Note that $Y_2 = \sqrt{\mathsf{INR}_2}X_1 + \sqrt{\mathsf{SNR}_2}X_2 + Z_2$ and $Y_3 = \sqrt{\mathsf{INR}_3}X_2 + \sqrt{\mathsf{SNR}_3}X_3 + Z_3$. Let $S = \sqrt{\mathsf{INR}_3}X_2 + V$ where $V$ is zero mean unit variance complex Gaussian random variable independent of $X_1$, $X_2$ and $X_3$

(can depend on Z's though). By providing genie $S^n$ to the second receiver,

$$\begin{aligned} n(R_1+R_2+R_3) &\leq I(X_1^n, X_2^n; Y_2^n, S^n) + I(X_3^n; Y_3^n) + n\epsilon && \text{(81a)}\\ &= I(X_1^n, X_2^n; Y_2^n, S^n) + h(Y_3^n) - h(S^n) + n\epsilon && \text{(81b)}\\ &= I(X_1^n, X_2^n; S^n) + I(X_1^n, X_2^n; Y_2^n|S^n) + h(Y_3^n) - h(S^n) + n\epsilon && \text{(81c)}\\ &= h(S^n) - h(V^n) + h(Y_2^n|S^n) - h(Z_2^n|V^n) + h(Y_3^n) \\ &\quad -h(S^n) + n\epsilon && \text{(81d)}\\ &= -h(V^n) + h(Y_2^n|S^n) - h(Z_2^n|V^n) + h(Y_3^n) + n\epsilon && \text{(81e)}\\ &\leq -h(V^n) + nh(Y_{2,G}|S_G) - h(Z_2^n|V^n) + nh(Y_{3,G}) + n\epsilon && \text{(81f)} \end{aligned}$$

The last step follows since Gaussian maximizes conditional entropy as well as marginal entropy [15]. Further, choosing $\epsilon$ arbitrarily small,

$$\begin{aligned} (R_1+R_2+R_3) &\leq -h(V) + h(Y_{2,G}|S_G) - h(Z_2|V) + h(Y_{3,G}) && \text{(82a)}\\ &= h(Y_{3,G}) - h(S_G) + h(Y_{2,G}|S_G) + h(S_G) - h(Z_2, V) && \text{(82b)}\\ &= I(Y_{3,G}; X_{3,G}) + h(Y_{2,G}, S_G) - h(Z_2, V) && \text{(82c)} \end{aligned}$$

The last step followed by choosing $V$ independent of $X_{3,G}$.

$$\begin{aligned} (R_1+R_2+R_3) &\leq I(Y_{3,G}; X_{3,G}) + h(Y_{2,G}, S_G) - h(Z_2, V) && \text{(83a)}\\ &\leq I(Y_{3,G}; X_{3,G}) + h(Y_{2,G}, S_G) - h(Z_2, V|X_{1,G}X_{2,G}) && \text{(83b)}\\ &= I(Y_{3,G}; X_{3,G}) + I(X_{1,G}X_{2,G}; Y_{2,G}, S_G) && \text{(83c)} \end{aligned}$$

For $(\mathsf{INR}_2+1)\mathsf{INR}_3 \leq \mathsf{SNR}_2$, we can choose $V = \sqrt{\frac{\mathsf{INR}_2\mathsf{INR}_3}{\mathsf{SNR}_2}}X_{1,G} + \sqrt{\frac{\mathsf{INR}_3}{\mathsf{SNR}_2}}Z_2 + Z_a$ where $Z_a$ is additional random variable independent of all other random variables, complex normal to make variance of $V$ unity. With this choice, $I(X_{1,G}X_{2,G}; Y_{2,G}, S_G) = I(X_{1,G}X_{2,G}; Y_{2,G})$. Thus, the above formula can be obtained.

For $(\mathsf{INR}_2+1)\mathsf{INR}_3 \geq \mathsf{SNR}_2$, we can choose $V = c(\sqrt{\frac{\mathsf{INR}_2\mathsf{INR}_3}{\mathsf{SNR}_2}}X_{1,G} + \sqrt{\frac{\mathsf{INR}_3}{\mathsf{SNR}_2}}Z_2)$ where $c \leq 1$ is chosen to make the above unit variance. With this choice,

$$\begin{aligned} & I(X_{1,G}X_{2,G}; Y_{2,G}, S_G) \\ = \; & I(X_{1,G}; Y_{2,G}, S_G|X_{2,G}) + I(X_{2,G}; Y_{2,G}, S_G) && \text{(84a)}\\ = \; & I(X_{1,G}; Y_{2,G}|X_{2,G}) + I(X_{1,G}; S_G|X_{2,G}Y_{2,G}) + I(X_{2,G}; Y_{2,G}, S_G) && \text{(84b)} \end{aligned}$$

For the second term $I(X_{1,G}; S_G|X_{2,G}Y_{2,G})$, $S_G$ is deterministic function of $X_{2,G}$ and $Y_{2,G}$ and this term is therefore 0. The third term $I(X_{2,G}; Y_{2,G}, S_G) = I(X_{2,G}; S_G) + I(X_{2,G}; Y_{2,G}|S_G)$. But, $I(X_{2,G}; Y_{2,G}|S_G) = 0$ as $X_{2,G} - S_G - Y_{2,G}$ is a Markov Chain. Thus,

$$I(X_{1,G}X_{2,G}; Y_{2,G}, S_G) = I(X_{1,G}; Y_{2,G}|X_{2,G}) + I(X_{2,G}; S_G) \tag{85}$$

This gives the above sum rate.

## *B.* $\mathsf{INR}_2 \leq \mathsf{SNR}_1$ *and* $\mathsf{INR}_3 \leq \mathsf{SNR}_2$

By Fano's inequality,

$$n(R_1+R_2+R_3) \leq I(X_1^n; Y_1^n) + I(X_2^n; Y_2^n) + I(X_3^n; Y_3^n) + n\epsilon \tag{86}$$

Note that $Y_3 = \sqrt{\mathsf{INR}_3}X_2 + \sqrt{\mathsf{SNR}_3}X_3 + Z_3$, $Y_2 = \sqrt{\mathsf{INR}_2}X_1 + \sqrt{\mathsf{SNR}_2}X_2 + Z_2$ and $Y_1 = \sqrt{\mathsf{SNR}_1}X_1 + Z_1$. Let $S_1 = \sqrt{\mathsf{INR}_2}X_1 + V_1$ and $S_2 = \sqrt{\mathsf{INR}_3}X_2 + V_2$ where $V_1$ and $V_2$ are mutually independent complex Gaussians of unit variance and independent of all $X_i$'s. Note that $I(X_2^n; Y_2^n) = I(X_2^n; S_1^n + \sqrt{\mathsf{SNR}_2}X_2^n)$ since the distributions remain the same.

$$\begin{aligned}
n(R_1+R_2+R_3) &\le I(X_1^n;Y_1^n)+I(X_2^n;Y_2^n)+I(X_3^n;Y_3^n)+n\epsilon && \text{(87a)}\\
&\le I(X_1^n;Y_1^n,S_1^n)+I(X_2^n;S_1^n+\sqrt{\mathsf{SNR}_2}X_2^n)+I(X_3^n;Y_3^n)+n\epsilon && \text{(87b)}\\
&\le I(X_1^n;Y_1^n,S_1^n)+I(X_2^n;S_1^n+\sqrt{\mathsf{SNR}_2}X_2^n,S_2^n)\\
&\quad +I(X_3^n;Y_3^n)+n\epsilon && \text{(87c)}\\
&= I(X_1^n;S_1^n)+I(X_1^n;Y_1^n|S_1^n)+I(X_2^n;S_2^n)\\
&\quad +I(X_2^n;S_1^n+\sqrt{\mathsf{SNR}_2}X_2^n|S_2^n)+h(Y_3^n)-h(S_2^n)+n\epsilon && \text{(87d)}\\
&= h(S_1^n)-h(V_1^n)+I(X_1^n;Y_1^n|S_1^n)+h(S_2^n)-h(V_2^n)\\
&\quad +I(X_2^n;S_1^n+\sqrt{\mathsf{SNR}_2}X_2^n|S_2^n)+h(Y_3^n)-h(S_2^n)+n\epsilon && \text{(87e)}\\
&= h(S_1^n)-h(V_1^n)+h(Y_1^n|S_1^n)-h(Z_1^n|V_1^n)-h(V_2^n)\\
&\quad +h(S_1^n+\sqrt{\mathsf{SNR}_2}X_2^n|S_2^n)-h(S_1^n|V_2^n)+h(Y_3^n)+n\epsilon && \text{(87f)}\\
&= h(S_1^n)+h(Y_1^n|S_1^n)-h(Z_1^n,V_1^n)-h(V_2^n)\\
&\quad +h(S_1^n+\sqrt{\mathsf{SNR}_2}X_2^n|S_2^n)-h(S_1^n)+h(Y_3^n)+n\epsilon && \text{(87g)}
\end{aligned}$$

The last step follows by choosing $V_1^n$ independent of $Z_2^n$. Further since conditional and marginal entropies are maximized by Gaussians,

$$\begin{aligned}
n(R_1+R_2+R_3) &\le h(Y_1^n|S_1^n)-h(Z_1^n,V_1^n)-h(V_2^n)\\
&\quad +h(S_1^n+\sqrt{\mathsf{SNR}_2}X_2^n|S_2^n)+h(Y_3^n)+n\epsilon && \text{(88a)}\\
&\le nh(Y_{1G}|S_{1G})-nh(Z_1,V_1)-nh(V_2)\\
&\quad +nh(S_{1,G}+\sqrt{\mathsf{SNR}_2}X_{2,G}|S_{2,G})+nh(Y_{3,G})+n\epsilon && \text{(88b)}
\end{aligned}$$

Letting $\epsilon\to 0$, we get

$$\begin{aligned}
R_1+R_2+R_3 &\le h(Y_{1G}|S_{1G})-h(Z_1,V_1)-h(V_2)\\
&\quad +h(S_{1,G}+\sqrt{\mathsf{SNR}_2}X_{2,G}|S_{2,G})+h(Y_{3,G}) && \text{(89a)}\\
&= h(Y_{1G}|S_{1G})-h(Z_1|V_1)-h(V_1)-h(V_2)\\
&\quad +h(S_{1,G}+\sqrt{\mathsf{SNR}_2}X_{2,G}|S_{2,G})+h(Y_{3,G}) && \text{(89b)}\\
&= h(Y_{1G}|S_{1G})-h(Y_{1G}|S_{1G},X_{1,G})-h(V_1)-h(V_2)\\
&\quad +h(S_{1,G}+\sqrt{\mathsf{SNR}_2}X_{2,G}|S_{2,G})+h(Y_{3,G}) && \text{(89c)}\\
&= I(Y_{1G};X_{1,G}|S_{1G})-h(V_1)-h(V_2)\\
&\quad +h(S_{1,G}+\sqrt{\mathsf{SNR}_2}X_{2,G}|S_{2,G})+h(Y_{3,G}) && \text{(89d)}\\
&= I(X_{1G};Y_{1,G},S_{1G})-h(S_{1G})-h(V_2)\\
&\quad +h(S_{1,G}+\sqrt{\mathsf{SNR}_2}X_{2,G}|S_{2,G})+h(Y_{3,G}) && \text{(89e)}
\end{aligned}$$

The last step follows by choosing $V_1$ independent of $X_{1G}$. We choose $V_1=\sqrt{\frac{\mathsf{INR}_2}{\mathsf{SNR}_1}}Z_1+Z_{1a}$ where $Z_{1a}$ is complex normal independent of all other variables in order to make $V_1$ unit norm. With this choice, $X_{1,G} - S_{1,G} - Y_{1,G}$ becomes a Markov Chain and thus $I(X_{1G}; Y_{1,G}, S_{1G}) = I(X_{1G}; Y_{1,G}) = \log(1+\mathsf{SNR}_1)$.

$$\begin{aligned} R_1 + R_2 + R_3 &\leq \log(1+\mathsf{SNR}_1) + h(Y_{3,G}) - h(S_{1G}) - h(V_2) \\ &\quad + h(S_{1,G} + \sqrt{\mathsf{SNR}_2} X_{2,G} | S_{2,G}) && \text{(90a)} \\ &= \log(1+\mathsf{SNR}_1) + h(Y_{3,G}) - h(S_{1G}) - h(V_2) - h(S_{2G}) \\ &\quad + h(S_{1,G} + \sqrt{\mathsf{SNR}_2} X_{2,G}, S_{2,G}) && \text{(90b)} \\ &= \log(1+\mathsf{SNR}_1) + h(Y_{3,G}) - h(S_{1G}) - h(S_{2G}) + h(V_1) - h(V_1) - h(V_2) \\ &\quad + h(S_{1,G} + \sqrt{\mathsf{SNR}_2} X_{2,G}, S_{2,G}) && \text{(90c)} \\ &\leq \log(1+\mathsf{SNR}_1) + h(Y_{3,G}) - h(S_{1G}) - h(S_{2G}) + h(V_1) - h(V_1, V_2 | X_{1,G} X_{2,G}) \\ &\quad + h(S_{1,G} + \sqrt{\mathsf{SNR}_2} X_{2,G}, S_{2,G}) && \text{(90d)} \\ &= \log(1+\mathsf{SNR}_1) + h(Y_{3,G}) - h(S_{1G}) - h(S_{2G}) + h(V_1) \\ &\quad + I(X_{1,G}, X_{2,G}; S_{1,G} + \sqrt{\mathsf{SNR}_2} X_{2,G}, S_{2,G}) && \text{(90e)} \end{aligned}$$

Note that the last expression $I(X_{1,G}, X_{2,G}; S_{1,G} + \sqrt{\mathsf{SNR}_2} X_{2,G}, S_{2,G})$ is similar to that in subsection A and thus $V_2$ can be chosen as in subsection A. ($S_{1,G} + \sqrt{\mathsf{SNR}_2} X_{2,G}$ plays the role of $Y_{2,G}$, only difference that $Z_2$ is replaced by $V_1$.) Thus, the same steps give the required bounds on $R_1 + R_2 + R_3$.

## Appendix E
## Proof that rates can be decoded with 1.5 rounds of message passing

To show this, we divide the range of $\mathsf{INR}_2$, $\mathsf{SNR}_2$, $\mathsf{INR}_3$ and $\mathsf{SNR}_3$ into the following nine cases.

*1)* $\mathsf{INR}_2 \geq \mathsf{SNR}_2$, $\mathsf{INR}_3 \geq \mathsf{SNR}_3$, $\mathsf{INR}_2 \geq \mathsf{SNR}_1$, $\mathsf{INR}_3 \geq \mathsf{SNR}_2$*:* The first transmitter makes a codebook of rate

$$R_1 = \log\left(1 + \min\left(\mathsf{SNR}_1, \frac{\mathsf{INR}_2}{1 + \frac{\mathsf{SNR}_2 \mathsf{INR}_2}{1+\mathsf{SNR}_2+2\mathsf{INR}_2}}\right)\right), \tag{91}$$

and uses a power level of 1 to transmit. The second transmitter makes a codebook of rate

$$R_2 = \log\left(1 + \frac{1+\mathsf{INR}_2}{1+2\mathsf{INR}_2} \min\left(\mathsf{SNR}_2, \frac{\mathsf{INR}_3}{1 + \frac{\mathsf{SNR}_3 \mathsf{INR}_3}{1+\mathsf{SNR}_3+2\mathsf{INR}_3}}\right)\right), \tag{92}$$

and uses a power level of 1 to transmit. The third transmitter makes a codebook of rate

$$R_3 = \log\left(1 + \mathsf{SNR}_3(1+\mathsf{INR}_3)/(1+2\mathsf{INR}_3)\right) \tag{93}$$

and uses a power level of 1 to transmit.

At the receiver 1, $R_1$ can be decoded since $R_1 \leq \log(1+\mathsf{SNR}_1)$. At the receiver 2, $R_1$ and $R_2$ are decoded jointly. We see that

$$\begin{aligned} R_1 &\leq \log(1+\mathsf{INR}_2) && \text{(94a)} \\ R_2 &\leq \log(1+\mathsf{SNR}_2) && \text{(94b)} \\ R_1 + R_2 &\leq \log(1+\mathsf{INR}_2+\mathsf{SNR}_2) && \text{(94c)} \end{aligned}$$

$R_1 + R_2$ equation holds as

$$R_1 + R_2 \leq \log\left(1 + \frac{\mathsf{INR}_2}{1+\frac{\mathsf{SNR}_2\mathsf{INR}_2}{1+\mathsf{SNR}_2+2\mathsf{INR}_2}}\right) + \log\left(1 + \frac{1+\mathsf{INR}_2}{1+2\mathsf{INR}_2}\mathsf{SNR}_2\right) = \log(1+\mathsf{INR}_2+\mathsf{SNR}_2).$$

At the receiver 3, $R_2$ and $R_3$ are decoded jointly. We see that

$$\begin{aligned} R_2 &\leq \log(1+\mathsf{INR}_3) & \text{(95a)} \\ R_3 &\leq \log(1+\mathsf{SNR}_3) & \text{(95b)} \\ R_2+R_3 &\leq \log(1+\mathsf{INR}_3+\mathsf{SNR}_3) & \text{(95c)} \end{aligned}$$

$R_2+R_3$ equation holds as $R_2+R_3 \leq \log\left(1+\frac{\mathsf{INR}_3}{1+\frac{\mathsf{SNR}_3\mathsf{INR}_3}{1+\mathsf{SNR}_3+2\mathsf{INR}_3}}\right)+\log\left(1+\frac{1+\mathsf{INR}_3}{1+2\mathsf{INR}_3}\mathsf{SNR}_3\right) = \log(1+\mathsf{INR}_3+\mathsf{SNR}_3)$.

*2)* $\mathsf{INR}_2 \geq \mathsf{SNR}_2$*,* $\mathsf{INR}_3 \geq \mathsf{SNR}_3$*,* $\mathsf{INR}_2 \geq \mathsf{SNR}_1$*,* $\mathsf{INR}_3 < \mathsf{SNR}_2$*:* The first transmitter makes a codebook of rate

$$R_1 = \log\left(1+\min\left(\mathsf{SNR}_1, \frac{\mathsf{INR}_2}{1+\frac{\mathsf{SNR}_2\mathsf{INR}_2}{1+\mathsf{SNR}_2+2\mathsf{INR}_2}}\right)\right), \tag{96}$$

and uses a power level of $1$ to transmit.

The second transmitter makes two codebooks, the first one of rate

$$R_{2,c} = \log\left(1+\frac{1+\mathsf{INR}_2}{1+2\mathsf{INR}_2}\frac{\mathsf{INR}_3^2}{1+2\mathsf{INR}_3+\mathsf{SNR}_3(1+\mathsf{INR}_3)}\right), \tag{97}$$

and the second of rate

$$R_{2,p} = \log\left(1+\frac{1+\mathsf{INR}_2}{1+2\mathsf{INR}_2}\mathsf{SNR}_2/(1+\mathsf{INR}_3)\right). \tag{98}$$

The transmitter send the first one at a power of $\mathsf{INR}_3/(1+\mathsf{INR}_3)$ and the second one at a power of $1/(1+\mathsf{INR}_3)$, adds them up and transmit.

The third transmitter makes a codebook of rate

$$R_3 = \log(1+\mathsf{SNR}_3(1+\mathsf{INR}_3)/(1+2\mathsf{INR}_3)) \tag{99}$$

and uses a power level of $1$ to transmit.

At the receiver 1, $R_1$ can be decoded since $R_1 \leq \log(1+\mathsf{SNR}_1)$. At the receiver 2, $R_1$, $R_{2,c}$ and $R_{2,p}$ are decoded jointly. We see that

$$\begin{aligned} R_1 &\leq \log(1+\mathsf{INR}_2) & \text{(100a)} \\ R_{2,c} &\leq \log(1+\mathsf{SNR}_2\mathsf{INR}_3/(1+\mathsf{INR}_3)) & \text{(100b)} \\ R_{2,p} &\leq \log(1+\mathsf{SNR}_2/(1+\mathsf{INR}_3)) & \text{(100c)} \\ R_1+R_{2,c} &\leq \log(1+\mathsf{INR}_2+\mathsf{SNR}_2\mathsf{INR}_3/(1+\mathsf{INR}_3)) & \text{(100d)} \\ R_1+R_{2,p} &\leq \log(1+\mathsf{INR}_2+\mathsf{SNR}_2/(1+\mathsf{INR}_3)) & \text{(100e)} \\ R_{2,c}+R_{2,p} &\leq \log(1+\mathsf{SNR}_2) & \text{(100f)} \\ R_1+R_{2,c}+R_{2,p} &\leq \log(1+\mathsf{INR}_2+\mathsf{SNR}_2) & \text{(100g)} \end{aligned}$$

The first three sub-equations hold trivially. For the fourth, we see that

$$\begin{aligned} & R_1 + R_{2,c} \\ \leq\ & \log\left(1 + \frac{\mathsf{INR}_2}{1 + \frac{\mathsf{SNR}_2\mathsf{INR}_2}{1+\mathsf{SNR}_2+2\mathsf{INR}_2}}\right) + \log\left(1 + \frac{\mathsf{INR}_3^2}{1 + 2\mathsf{INR}_3 + \mathsf{SNR}_3(1+\mathsf{INR}_3)}\right) && (101) \\ =\ & \log\left(1 + \frac{\mathsf{INR}_2}{1 + \frac{\mathsf{SNR}_2\mathsf{INR}_2}{1+\mathsf{SNR}_2+2\mathsf{INR}_2}} + \frac{\mathsf{INR}_3^2}{1 + 2\mathsf{INR}_3 + \mathsf{SNR}_3(1+\mathsf{INR}_3)}\right. \\ & \left. + \frac{\mathsf{INR}_2}{1 + \frac{\mathsf{SNR}_2\mathsf{INR}_2}{1+\mathsf{SNR}_2+2\mathsf{INR}_2}} \frac{\mathsf{INR}_3^2}{1 + 2\mathsf{INR}_3 + \mathsf{SNR}_3(1+\mathsf{INR}_3)}\right) && (102) \\ & && (103) \end{aligned}$$

Thus, we need to show that

$$\frac{\mathsf{INR}_3^2}{1 + 2\mathsf{INR}_3 + \mathsf{SNR}_3(1+\mathsf{INR}_3)} + \frac{\mathsf{INR}_2}{1 + \frac{\mathsf{SNR}_2\mathsf{INR}_2}{1+\mathsf{SNR}_2+2\mathsf{INR}_2}} \frac{\mathsf{INR}_3^2}{1 + 2\mathsf{INR}_3 + \mathsf{SNR}_3(1+\mathsf{INR}_3)}$$
$$-\frac{\mathsf{INR}_2^2\mathsf{SNR}_2}{1 + \mathsf{SNR}_2 + 2\mathsf{INR}_2 + \mathsf{SNR}_2\mathsf{INR}_2} - \frac{\mathsf{SNR}_2\mathsf{INR}_3}{1+\mathsf{INR}_3} \leq 0. \tag{104}$$

Note that LHS decreases with $\mathsf{SNR}_3$ and hence it is enough to show that above is negative for $\mathsf{SNR}_3 = 0$. With $\mathsf{SNR}_3 = 0$, we find that the above decreases with $\mathsf{SNR}_2$ and thus we only need to show for low value of $\mathsf{SNR}_2$ and as $\mathsf{SNR}_2 \geq \mathsf{INR}_3$, it is enough to show for $\mathsf{INR}_3 = \mathsf{SNR}_2$. The above reduces to

$$A = -\frac{\mathsf{INR}_3^2(1 + \mathsf{INR}_3 + 2\mathsf{INR}_2 + \mathsf{INR}_3\mathsf{INR}_2)}{(1+2\mathsf{INR}_3)(1+\mathsf{INR}_3)} + \frac{\mathsf{INR}_3\mathsf{INR}_2(1 + \mathsf{INR}_3 + 2\mathsf{INR}_2)}{1+2\mathsf{INR}_3} - \mathsf{INR}_2^2 \leq 0. \tag{105}$$

We now show that this decreases with $\mathsf{INR}_2$. To see that, differentiate w.r.t. $\mathsf{INR}_2$, we get

$$\begin{aligned} dA/d\mathsf{INR}_2 &= -\frac{\mathsf{INR}_3^2(2+\mathsf{INR}_3)}{(1+2\mathsf{INR}_3)(1+\mathsf{INR}_3)} + \frac{\mathsf{INR}_3(1+\mathsf{INR}_3+4\mathsf{INR}_2)}{1+2\mathsf{INR}_3} - 2\mathsf{INR}_2 && (106a) \\ &= -\frac{\mathsf{INR}_3}{(1+2\mathsf{INR}_3)(1+\mathsf{INR}_3)}\left(\mathsf{INR}_3(2+\mathsf{INR}_3) - (1+\mathsf{INR}_3)^2\right) \\ &\quad + \frac{\mathsf{INR}_3(4\mathsf{INR}_2) - 2\mathsf{INR}_2(1+2\mathsf{INR}_3)}{1+2\mathsf{INR}_3} && (106b) \\ &= \frac{\mathsf{INR}_3}{(1+2\mathsf{INR}_3)(1+\mathsf{INR}_3)} - \frac{2\mathsf{INR}_2}{1+2\mathsf{INR}_3} && (106c) \\ &= \frac{1}{(1+2\mathsf{INR}_3)}(\mathsf{INR}_3/(1+\mathsf{INR}_3) - 2\mathsf{INR}_2) && (106d) \end{aligned}$$

As $\mathsf{INR}_3 \leq \mathsf{INR}_2$, the above is negative; thus it is enough to prove the above for $\mathsf{INR}_2 = \mathsf{INR}_3$. Thus, it is enough to show that

$$-\frac{(1+3\mathsf{INR}_3+\mathsf{INR}_3^2)}{(1+2\mathsf{INR}_3)(1+\mathsf{INR}_3)} + \frac{(1+3\mathsf{INR}_3)}{1+2\mathsf{INR}_3} - 1 \leq 0. \tag{107}$$

which is true.

To show $R_1 + R_{2,p} \leq \log(1 + \mathsf{INR}_2 + \mathsf{SNR}_2/(1+\mathsf{INR}_3))$ , we will show that $R_1 + R_{2,p} \leq \log(1 + \mathsf{INR}_2 + \mathsf{SNR}_2/(1+\mathsf{INR}_3) - \frac{\mathsf{INR}_2^2\mathsf{SNR}_2\mathsf{INR}_3}{(1+\mathsf{SNR}_2+2\mathsf{INR}_2+\mathsf{SNR}_2\mathsf{INR}_2)(1+\mathsf{INR}_3)})$ which will prove the claim. To see this, it is enough to prove that

$$\frac{\mathsf{INR}_2}{1+\frac{\mathsf{SNR}_2\mathsf{INR}_2}{1+\mathsf{SNR}_2+2\mathsf{INR}_2}}+\frac{1+\mathsf{INR}_2}{1+2\mathsf{INR}_2}\mathsf{SNR}_2/(1+\mathsf{INR}_3)$$
$$+\frac{\mathsf{INR}_2}{1+\frac{\mathsf{SNR}_2\mathsf{INR}_2}{1+\mathsf{SNR}_2+2\mathsf{INR}_2}}\frac{1+\mathsf{INR}_2}{1+2\mathsf{INR}_2}\mathsf{SNR}_2/(1+\mathsf{INR}_3)$$
$$= \mathsf{INR}_2+\mathsf{SNR}_2/(1+\mathsf{INR}_3)-\frac{\mathsf{INR}_2^2\mathsf{SNR}_2\mathsf{INR}_3}{(1+\mathsf{SNR}_2+2\mathsf{INR}_2+\mathsf{SNR}_2\mathsf{INR}_2)(1+\mathsf{INR}_3)} \tag{108}$$

This is equivalent to proving

$$(1+\mathsf{SNR}_2+2\mathsf{INR}_2)\frac{1+\mathsf{INR}_2}{1+2\mathsf{INR}_2}\frac{\mathsf{SNR}_2}{1+\mathsf{INR}_3}$$
$$= \mathsf{INR}_2\mathsf{SNR}_2+\frac{1+\mathsf{SNR}_2+2\mathsf{INR}_2+\mathsf{SNR}_2\mathsf{INR}_2}{1+2\mathsf{INR}_2}\frac{\mathsf{SNR}_2}{1+\mathsf{INR}_3}-\frac{\mathsf{INR}_2\mathsf{SNR}_2\mathsf{INR}_3}{(1+\mathsf{INR}_3)} \tag{109}$$

which is equivalent to

$$\mathsf{INR}_2\mathsf{SNR}_2/(1+\mathsf{INR}_3)=\mathsf{INR}_2\mathsf{SNR}_2-\frac{\mathsf{INR}_2\mathsf{SNR}_2\mathsf{INR}_3}{(1+\mathsf{INR}_3)} \tag{110}$$

which is true. Hence proved.

To show $R_{2,c}+R_{2,p}\le\log(1+\mathsf{SNR}_2)$, it is enough to prove that

$$\frac{\mathsf{INR}_3^2}{1+2\mathsf{INR}_3+\mathsf{SNR}_3(1+\mathsf{INR}_3)}+\mathsf{SNR}_2/(1+\mathsf{INR}_3)+\frac{\mathsf{INR}_3^2}{1+2\mathsf{INR}_3+\mathsf{SNR}_3(1+\mathsf{INR}_3)}\mathsf{SNR}_2/(1+\mathsf{INR}_3)\le\mathsf{SNR}_2 \tag{111}$$

Or,

$$\frac{\mathsf{INR}_3}{1+2\mathsf{INR}_3+\mathsf{SNR}_3(1+\mathsf{INR}_3)}+\frac{\mathsf{INR}_3}{1+2\mathsf{INR}_3+\mathsf{SNR}_3(1+\mathsf{INR}_3)}\mathsf{SNR}_2/(1+\mathsf{INR}_3)\le\mathsf{SNR}_2/(1+\mathsf{INR}_3) \tag{112}$$

Or,

$$\mathsf{INR}_3\le\mathsf{SNR}_2(1+\mathsf{SNR}_3) \tag{113}$$

which holds since $\mathsf{INR}_3\le\mathsf{SNR}_2$.

To show $R_1+R_{2,c}+R_{2,p}\le\log(1+\mathsf{INR}_2+\mathsf{SNR}_2)$, it is sufficient to show that

$$\mathsf{INR}_2+\mathsf{SNR}_2/(1+\mathsf{INR}_3)-d+\frac{1+\mathsf{INR}_2}{1+2\mathsf{INR}_2}\frac{\mathsf{INR}_3^2}{1+2\mathsf{INR}_3+\mathsf{SNR}_3(1+\mathsf{INR}_3)}$$
$$+(\mathsf{INR}_2+\mathsf{SNR}_2/(1+\mathsf{INR}_3)-d)\frac{1+\mathsf{INR}_2}{1+2\mathsf{INR}_2}\frac{\mathsf{INR}_3^2}{1+2\mathsf{INR}_3+\mathsf{SNR}_3(1+\mathsf{INR}_3)}$$
$$\le \mathsf{INR}_2+\mathsf{SNR}_2, \tag{114a}$$

where $d=\frac{\mathsf{INR}_2^2\mathsf{SNR}_2\mathsf{INR}_3}{(1+\mathsf{SNR}_2+2\mathsf{INR}_2+\mathsf{SNR}_2\mathsf{INR}_2)(1+\mathsf{INR}_3)}$. Note that the above is equivalent to show

$$-\mathsf{SNR}_2\mathsf{INR}_3/(1+\mathsf{INR}_3)-d+\frac{1+\mathsf{INR}_2}{1+2\mathsf{INR}_2}\frac{\mathsf{INR}_3^2}{1+2\mathsf{INR}_3+\mathsf{SNR}_3(1+\mathsf{INR}_3)}$$
$$+(\mathsf{INR}_2+\mathsf{SNR}_2/(1+\mathsf{INR}_3)-d)\frac{1+\mathsf{INR}_2}{1+2\mathsf{INR}_2}\frac{\mathsf{INR}_3^2}{1+2\mathsf{INR}_3+\mathsf{SNR}_3(1+\mathsf{INR}_3)}\le 0 \tag{115}$$

Since this decreases with $\mathsf{SNR}_3$, it is enough to show for $\mathsf{SNR}_3=0$. Note that the above is equivalent to

show

$$-\mathsf{SNR}_2\mathsf{INR}_3/(1+\mathsf{INR}_3)(1-\frac{1+\mathsf{INR}_2}{1+2\mathsf{INR}_2}\frac{\mathsf{INR}_3}{1+2\mathsf{INR}_3})+\frac{(1+\mathsf{INR}_2)^2}{1+2\mathsf{INR}_2}\frac{\mathsf{INR}_3^2}{1+2\mathsf{INR}_3}$$
$$-d(1+\frac{1+\mathsf{INR}_2}{1+2\mathsf{INR}_2}\frac{\mathsf{INR}_3^2}{1+2\mathsf{INR}_3})\leq 0 \tag{116}$$

Since this decreases with $\mathsf{SNR}_2$, it is enough to show the above for $\mathsf{SNR}_2=\mathsf{INR}_3$ as $\mathsf{SNR}_2\geq\mathsf{INR}_3$.

$$-\left(1-\frac{1+\mathsf{INR}_2}{1+2\mathsf{INR}_2}\frac{\mathsf{INR}_3}{1+2\mathsf{INR}_3}\right)+\frac{(1+\mathsf{INR}_2)^2(1+\mathsf{INR}_3)}{1+2\mathsf{INR}_2}\frac{1}{1+2\mathsf{INR}_3}$$
$$-\frac{\mathsf{INR}_2^2}{(1+\mathsf{INR}_3+2\mathsf{INR}_2+\mathsf{INR}_3\mathsf{INR}_2)}\left(1+\frac{1+\mathsf{INR}_2}{1+2\mathsf{INR}_2}\frac{\mathsf{INR}_3^2}{1+2\mathsf{INR}_3}\right)\leq 0 \tag{117}$$

Thus, it suffices to show

$$\begin{aligned}A &= -(1+\mathsf{INR}_3+2\mathsf{INR}_2+\mathsf{INR}_3\mathsf{INR}_2)(1+2\mathsf{INR}_2)(1+2\mathsf{INR}_3)+(1+\mathsf{INR}_3+2\mathsf{INR}_2+\mathsf{INR}_3\mathsf{INR}_2)\times\\ &\quad(1+\mathsf{INR}_2)(\mathsf{INR}_3+(1+\mathsf{INR}_2)(1+\mathsf{INR}_3))-\mathsf{INR}_2^2((1+2\mathsf{INR}_2)(1+2\mathsf{INR}_3)+(1+\mathsf{INR}_2)\mathsf{INR}_3^2)\leq 0\end{aligned}$$

Note that

$$\begin{aligned}dA/d(\mathsf{INR}_2) &= -(1+\mathsf{INR}_3+2\mathsf{INR}_2+\mathsf{INR}_2\mathsf{INR}_3)2(1+2\mathsf{INR}_3)-(2+\mathsf{INR}_3)(1+2\mathsf{INR}_2)(1+2\mathsf{INR}_3)\\ &\quad+(2+\mathsf{INR}_3)(1+\mathsf{INR}_2)(\mathsf{INR}_3+(1+\mathsf{INR}_2)(1+\mathsf{INR}_3))\\ &\quad+(1+\mathsf{INR}_3+2\mathsf{INR}_2+\mathsf{INR}_2\mathsf{INR}_3)(\mathsf{INR}_3+(1+\mathsf{INR}_2)(1+\mathsf{INR}_3))\\ &\quad+(1+\mathsf{INR}_3+2\mathsf{INR}_2+\mathsf{INR}_2\mathsf{INR}_3)(1+\mathsf{INR}_2)(1+\mathsf{INR}_3)\\ &\quad-2\mathsf{INR}_2((1+2\mathsf{INR}_2)(1+2\mathsf{INR}_3)+\mathsf{INR}_3^2(1+\mathsf{INR}_2))-\mathsf{INR}_2^2(2+4\mathsf{INR}_3+\mathsf{INR}_3^2)\\ &= -\mathsf{INR}_3(1+3\mathsf{INR}_2^2+\mathsf{INR}_3+2\mathsf{INR}_2(2+\mathsf{INR}_3))<0 \qquad (118a)\end{aligned}$$

Thus, it is enough to prove $A\leq 0$ for $\mathsf{INR}_2=\mathsf{INR}_3$, and hence we need to show that

$$-(1+3\mathsf{INR}_3+\mathsf{INR}_3^2)(1+2\mathsf{INR}_3)^2+(1+3\mathsf{INR}_3+\mathsf{INR}_3^2)^2(1+\mathsf{INR}_3)$$
$$-\mathsf{INR}_3^2((1+2\mathsf{INR}_3)^2+(1+\mathsf{INR}_3)\mathsf{INR}_3^2)\leq 0 \tag{119}$$

which holds since the above reduces to $-\mathsf{INR}_3^2(1+3\mathsf{INR}_3+2\mathsf{INR}_3^2)\leq 0$.

At the receiver 3, $R_{2,c}$ and $R_3$ are decoded jointly treating $R_{2,p}$ as noise. We see that

$$\begin{aligned}R_{2,c} &\leq \log(1+\mathsf{INR}_3^2/(1+2\mathsf{INR}_3)) &(120a)\\ R_3 &\leq \log(1+\mathsf{SNR}_3(1+\mathsf{INR}_3)/(1+2\mathsf{INR}_3)) &(120b)\\ R_{2,c}+R_3 &\leq \log(1+\mathsf{INR}_3^2/(1+2\mathsf{INR}_3)+\mathsf{SNR}_3(1+\mathsf{INR}_3)/(1+2\mathsf{INR}_3)) &(120c)\end{aligned}$$

To show $R_{2,c}+R_3\leq\log(1+\mathsf{INR}_3^2/(1+2\mathsf{INR}_3)+\mathsf{SNR}_3(1+\mathsf{INR}_3)/(1+2\mathsf{INR}_3))$, we see that $R_{2,c}+R_3\leq\log(1+\frac{\mathsf{INR}_3^2}{1+2\mathsf{INR}_3+\mathsf{SNR}_3(1+\mathsf{INR}_3)})+\log(1+\mathsf{SNR}_3(1+\mathsf{INR}_3)/(1+2\mathsf{INR}_3))\leq\log(1+\mathsf{INR}_3^2/(1+2\mathsf{INR}_3)+\mathsf{SNR}_3(1+\mathsf{INR}_3)/(1+2\mathsf{INR}_3))$ as was in the case of 2-user channel.

*3)* $\mathsf{INR}_2\geq\mathsf{SNR}_2$*,* $\mathsf{INR}_3\geq\mathsf{SNR}_3$*,* $\mathsf{INR}_2<\mathsf{SNR}_1$*,* $\mathsf{INR}_3\geq\mathsf{SNR}_2$*:* The first transmitter makes two codebooks, the first one of rate

$$R_{1,c}=\log\left(1+\frac{\mathsf{INR}_2^2}{1+2\mathsf{INR}_2+\mathsf{SNR}_2(1+\mathsf{INR}_2)}\right), \tag{121}$$

and the second of rate

$$R_{1,p}=\log(1+\mathsf{SNR}_1/(1+\mathsf{INR}_2)). \tag{122}$$

The transmitter send the first one at a power of $\mathsf{INR}_2/(1+\mathsf{INR}_2)$ and the second one at a power of $1/(1+\mathsf{INR}_2)$, adds them up and transmit. The second transmitter makes a codebook of rate

$$R_2 = \log\left(1+\frac{1+\mathsf{INR}_2}{1+2\mathsf{INR}_2}\min\left(\mathsf{SNR}_2, \frac{\mathsf{INR}_3}{1+\frac{\mathsf{SNR}_3\mathsf{INR}_3}{1+\mathsf{SNR}_3+2\mathsf{INR}_3}}\right)\right), \tag{123}$$

and uses a power level of 1 to transmit. The third transmitter makes a codebook of rate

$$R_3 = \log(1+\mathsf{SNR}_3(1+\mathsf{INR}_3)/(1+2\mathsf{INR}_3)) \tag{124}$$

and uses a power level of 1 to transmit.

The first receiver is able to decode following the same steps as in two user interference channel given above. The second receiver decodes $R_{1,c}$ and $R_2$ treating $R_{2,p}$ as noise. The decoding happens when

$$R_{1,c} \leq \log\left(1+\frac{\mathsf{INR}_2^2/(1+\mathsf{INR}_2)}{1+\mathsf{INR}_2/(1+\mathsf{INR}_2)}\right) \tag{125a}$$

$$R_2 \leq \log\left(1+\frac{\mathsf{SNR}_2}{1+\mathsf{INR}_2/(1+\mathsf{INR}_2)}\right) \tag{125b}$$

$$R_{1,c}+R_2 \leq \log\left(1+\frac{\mathsf{SNR}_2+\mathsf{INR}_2^2/(1+\mathsf{INR}_2)}{1+\mathsf{INR}_2/(1+\mathsf{INR}_2)}\right) \tag{125c}$$

These conditions can be easily shown to be satisfied.

The third receiver does a joint decoding and can decode $R_3$, and the rate constraints are satisfied as in 2-user since powers remain same as there while rates now are even lesser.

*4)* $\mathsf{INR}_2 \geq \mathsf{SNR}_2$*,* $\mathsf{INR}_3 \geq \mathsf{SNR}_3$*,* $\mathsf{INR}_2 < \mathsf{SNR}_1$*,* $\mathsf{INR}_3 < \mathsf{SNR}_2$*:* The first transmitter makes two codebooks, the first one of rate

$$R_{1,c} = \log\left(1+\frac{\mathsf{INR}_2^2}{1+2\mathsf{INR}_2+\mathsf{SNR}_2(1+\mathsf{INR}_2)}\right), \tag{126}$$

and the second of rate

$$R_{1,p} = \log(1+\mathsf{SNR}_1/(1+\mathsf{INR}_2)). \tag{127}$$

The transmitter send the first one at a power of $\mathsf{INR}_2/(1+\mathsf{INR}_2)$ and the second one at a power of $1/(1+\mathsf{INR}_2)$, adds them up and transmit.

The second transmitter makes two codebooks, the first one of rate

$$R_{2,c} = \log\left(1+\frac{1+\mathsf{INR}_2}{1+2\mathsf{INR}_2}\frac{\mathsf{INR}_3^2}{1+2\mathsf{INR}_3+\mathsf{SNR}_3(1+\mathsf{INR}_3)}\right), \tag{128}$$

and the second of rate

$$R_{2,p} = \log\left(1+\frac{1+\mathsf{INR}_2}{1+2\mathsf{INR}_2}\mathsf{SNR}_2/(1+\mathsf{INR}_3)\right). \tag{129}$$

The transmitter send the first one at a power of $\mathsf{INR}_3/(1+\mathsf{INR}_3)$ and the second one at a power of $1/(1+\mathsf{INR}_3)$, adds them up and transmit.

The third transmitter makes a codebook of rate

$$R_3 = \log(1+\mathsf{SNR}_3(1+\mathsf{INR}_3)/(1+2\mathsf{INR}_3)) \tag{130}$$

and uses a power level of 1 to transmit.

As in 2-user case, the first receiver is able to decode the public message treating private as noise and then decode the private message. The third receiver can decode the $R_{2,c}$ and $R_3$ jointly treating $R_{2,p}$ as noise which can be shown following the same steps as in 2-user. However, the calculations at the second receiver are different. The second receiver decodes $R_{1,c}$, $R_{2,c}$ and $R_{2,p}$ jointly treating $R_{1,p}$ as noise. The decoding happens when the following are satisfied.

$$
\begin{aligned}
R_{1,c} &\le \log\left(1+\frac{\mathsf{INR}_2^2/(1+\mathsf{INR}_2)}{1+\mathsf{INR}_2/(1+\mathsf{INR}_2)}\right) && \text{(131a)}\\
R_{2,c} &\le \log\left(1+\frac{\mathsf{SNR}_2}{1+\mathsf{INR}_2/(1+\mathsf{INR}_2)}\mathsf{INR}_3/(1+\mathsf{INR}_3)\right) && \text{(131b)}\\
R_{2,p} &\le \log\left(1+\frac{\mathsf{SNR}_2/(1+\mathsf{INR}_3)}{1+\mathsf{INR}_2/(1+\mathsf{INR}_2)}\right) && \text{(131c)}\\
R_{1,c}+R_{2,c} &\le \log\left(1+\frac{\mathsf{INR}_2^2/(1+\mathsf{INR}_2)+\mathsf{SNR}_2\mathsf{INR}_3/(1+\mathsf{INR}_3)}{1+\mathsf{INR}_2/(1+\mathsf{INR}_2)}\right) && \text{(131d)}\\
R_{1,c}+R_{2,p} &\le \log\left(1+\frac{\mathsf{INR}_2^2/(1+\mathsf{INR}_2)+\mathsf{SNR}_2/(1+\mathsf{INR}_3)}{1+\mathsf{INR}_2/(1+\mathsf{INR}_2)}\right) && \text{(131e)}\\
R_{2,c}+R_{2,p} &\le \log\left(1+\frac{\mathsf{SNR}_2}{1+\mathsf{INR}_2/(1+\mathsf{INR}_2)}\right) && \text{(131f)}\\
R_{1,c}+R_{2,c}+R_{2,p} &\le \log\left(1+\frac{\mathsf{INR}_2^2/(1+\mathsf{INR}_2)+\mathsf{SNR}_2}{1+\mathsf{INR}_2/(1+\mathsf{INR}_2)}\right) && \text{(131g)}
\end{aligned}
$$

It is straightforward to see that the individual constraints are satisfied. To see the constraint on $R_{1,c}+R_{2,c}$, we write the satisfying equation and note as in second case that it is enough to use $\mathsf{SNR}_3 = 0$ and $\mathsf{SNR}_2 = \mathsf{INR}_3$. Thus, it is sufficient to prove that

$$
\begin{aligned}
&\frac{\mathsf{INR}_2^2}{1+2\mathsf{INR}_2+\mathsf{INR}_3(1+\mathsf{INR}_2)}+\frac{1+\mathsf{INR}_2}{1+2\mathsf{INR}_2}\frac{\mathsf{INR}_3^2}{1+2\mathsf{INR}_3}\\
&+\frac{\mathsf{INR}_2^2}{1+2\mathsf{INR}_2+\mathsf{INR}_3(1+\mathsf{INR}_2)}\frac{1+\mathsf{INR}_2}{1+2\mathsf{INR}_2}\frac{\mathsf{INR}_3^2}{1+2\mathsf{INR}_3}\\
\le\ &\frac{1+\mathsf{INR}_2}{1+2\mathsf{INR}_2}\left(\frac{\mathsf{INR}_2^2}{1+\mathsf{INR}_2}+\frac{\mathsf{INR}_3^2}{1+\mathsf{INR}_3}\right) && \text{(132a)}
\end{aligned}
$$

This is equivalent to

$$
\begin{aligned}
&-1+\frac{\mathsf{INR}_3}{1+2\mathsf{INR}_3}\\
&-(1+2\mathsf{INR}_2+\mathsf{INR}_3(1+\mathsf{INR}_2))\frac{\mathsf{INR}_3^2}{\mathsf{INR}_2^2(1+\mathsf{INR}_3)(1+2\mathsf{INR}_3)}\le 0 && \text{(133)}
\end{aligned}
$$

The above expression increases with $\mathsf{INR}_3$ and hence it is sufficient to show that the above holds as $\mathsf{INR}_2 \to \infty$, at which also, this is $\le 0$ and hence holds.

To show $R_{1,c}+R_{2,p} \le \log\left(1+\frac{\mathsf{INR}_2^2/(1+\mathsf{INR}_2)+\mathsf{SNR}_2/(1+\mathsf{INR}_3)}{1+\mathsf{INR}_2/(1+\mathsf{INR}_2)}\right)$, it is sufficient to show that

$$
\frac{\mathsf{INR}_2^2}{1+2\mathsf{INR}_2+\mathsf{SNR}_2(1+\mathsf{INR}_2)}\left(1+\frac{1+\mathsf{INR}_2}{1+2\mathsf{INR}_2}\frac{\mathsf{SNR}_2}{1+\mathsf{INR}_3}\right)\le\frac{1+\mathsf{INR}_2}{1+2\mathsf{INR}_2}\frac{\mathsf{INR}_2^2}{1+\mathsf{INR}_2} \tag{134}
$$

This reduces to

$$
\frac{1+\mathsf{INR}_2}{1+2\mathsf{INR}_2}\frac{\mathsf{SNR}_2}{1+\mathsf{INR}_3}\le\frac{1}{1+2\mathsf{INR}_2}(\mathsf{SNR}_2(1+\mathsf{INR}_2)), \tag{135}
$$

which trivially holds. Note that more precisely,

$$R_{1,c}+R_{2,p}\leq\log\left(1+\frac{\mathsf{INR}_2^2/(1+\mathsf{INR}_2)+\mathsf{SNR}_2/(1+\mathsf{INR}_3)}{1+\mathsf{INR}_2/(1+\mathsf{INR}_2)}-d\right)$$

where $d=\frac{\mathsf{INR}_2^2\mathsf{SNR}_2\mathsf{INR}_3(1+\mathsf{INR}_2)}{(1+\mathsf{INR}_3)(1+2\mathsf{INR}_2)(1+2\mathsf{INR}_2+\mathsf{SNR}_2(1+\mathsf{INR}_2))}$.

The result $R_{2,c}+R_{2,p}\leq\log(1+\frac{\mathsf{SNR}_2}{1+\mathsf{INR}_2/(1+\mathsf{INR}_2)})$ holds by using the same steps as in the 2-user Z since cancelling $\frac{1+\mathsf{INR}_2}{1+2\mathsf{INR}_2}$ on both sides, the LHS is now even smaller than was before.

Lastly, we need to show $R_{1,c}+R_{2,c}+R_{2,p}\leq\log\left(1+\frac{\mathsf{INR}_2^2/(1+\mathsf{INR}_2)+\mathsf{SNR}_2}{1+\mathsf{INR}_2/(1+\mathsf{INR}_2)}\right)$. To show this, it is sufficient to prove that

$$\begin{aligned}&\log\left(1+\frac{\mathsf{INR}_2^2/(1+\mathsf{INR}_2)+\mathsf{SNR}_2/(1+\mathsf{INR}_3)}{1+\mathsf{INR}_2/(1+\mathsf{INR}_2)}-d\right)\\&+\log\left(1+\frac{1+\mathsf{INR}_2}{1+2\mathsf{INR}_2}\frac{\mathsf{INR}_3^2}{1+2\mathsf{INR}_3+\mathsf{SNR}_3(1+\mathsf{INR}_3)}\right)\\\leq\;&\log\left(1+\frac{\mathsf{INR}_2^2/(1+\mathsf{INR}_2)+\mathsf{SNR}_2}{1+\mathsf{INR}_2/(1+\mathsf{INR}_2)}\right).\end{aligned}\tag{136}$$

As before, it is sufficient to prove this for $\mathsf{SNR}_3=0$.

Hence, it is sufficient to prove that

$$\begin{aligned}&-d\left(1+\frac{1+\mathsf{INR}_2}{1+2\mathsf{INR}_2}\frac{\mathsf{INR}_3^2}{1+2\mathsf{INR}_3}\right)+\frac{1+\mathsf{INR}_2}{1+2\mathsf{INR}_2}\frac{\mathsf{INR}_3^2}{1+2\mathsf{INR}_3}\\&+\frac{\mathsf{INR}_2^2(1+\mathsf{INR}_2)}{(1+2\mathsf{INR}_2)^2}\frac{\mathsf{INR}_3^2}{1+2\mathsf{INR}_3}\\&-\frac{\mathsf{SNR}_2(1+\mathsf{INR}_2)\mathsf{INR}_3}{(1+\mathsf{INR}_3)(1+2\mathsf{INR}_2)}\left(1-\frac{1+\mathsf{INR}_2}{1+2\mathsf{INR}_2}\frac{\mathsf{INR}_3}{1+2\mathsf{INR}_3}\right)\leq 0\end{aligned}\tag{137}$$

As $d$ increases with $\mathsf{SNR}_2$, the above expression decreases with $\mathsf{SNR}_2$ and thus it is enough to show that above holds for $\mathsf{SNR}_2=\mathsf{INR}_3$. Thus,

$$\begin{aligned}&-\frac{\mathsf{INR}_2^2(1+\mathsf{INR}_2)}{(1+\mathsf{INR}_3)(1+2\mathsf{INR}_2)(1+2\mathsf{INR}_2+\mathsf{INR}_3(1+\mathsf{INR}_2))}\left(1+\frac{1+\mathsf{INR}_2}{1+2\mathsf{INR}_2}\frac{\mathsf{INR}_3^2}{1+2\mathsf{INR}_3}\right)\\&+\frac{1+\mathsf{INR}_2}{1+2\mathsf{INR}_2}\frac{1}{1+2\mathsf{INR}_3}+\frac{\mathsf{INR}_2^2(1+\mathsf{INR}_2)}{(1+2\mathsf{INR}_2)^2}\frac{1}{1+2\mathsf{INR}_3}\\&-\frac{(1+\mathsf{INR}_2)}{(1+\mathsf{INR}_3)(1+2\mathsf{INR}_2)}\left(1-\frac{1+\mathsf{INR}_2}{1+2\mathsf{INR}_2}\frac{\mathsf{INR}_3}{1+2\mathsf{INR}_3}\right)\leq 0\end{aligned}\tag{138}$$

Or,

$$\begin{aligned}&-\mathsf{INR}_2^2(1+\mathsf{INR}_2)((1+2\mathsf{INR}_2)(1+2\mathsf{INR}_3)+(1+\mathsf{INR}_2)\mathsf{INR}_3^2)+(1+\mathsf{INR}_2)(1+2\mathsf{INR}_2)(1+\mathsf{INR}_3)\\&\times(1+2\mathsf{INR}_2+\mathsf{INR}_3(1+\mathsf{INR}_2))+\mathsf{INR}_2^2(1+\mathsf{INR}_2)(1+\mathsf{INR}_3)(1+2\mathsf{INR}_2+\mathsf{INR}_3(1+\mathsf{INR}_2))\\&-(1+\mathsf{INR}_2)(1+2\mathsf{INR}_2+\mathsf{INR}_3(1+\mathsf{INR}_2))(1+2\mathsf{INR}_2+\mathsf{INR}_3+3\mathsf{INR}_2\mathsf{INR}_3)\leq 0\end{aligned}\tag{139}$$

which is true since the above reduces to $-\mathsf{INR}_2(1+\mathsf{INR}_2)^2\mathsf{INR}_3(1+\mathsf{INR}_2+\mathsf{INR}_3)\leq 0$

*5) $\mathsf{INR}_2\geq\mathsf{SNR}_2$, $\mathsf{INR}_3<\mathsf{SNR}_3$, $\mathsf{INR}_2\geq\mathsf{SNR}_1$:* The first transmitter makes a codebook of rate

$$R_1=\log\left(1+\min\left(\mathsf{SNR}_1,\frac{\mathsf{INR}_2}{1+\frac{\mathsf{SNR}_2\mathsf{INR}_2}{1+\mathsf{SNR}_2+2\mathsf{INR}_2}}\right)\right),\tag{140}$$

and uses a power level of $1$ to transmit.

The second transmitter makes a codebook of rate

$$R_2 = \log\left(1 + \frac{1+\mathsf{INR}_2}{1+2\mathsf{INR}_2}\mathsf{SNR}_2/(1+\mathsf{INR}_3)\right), \tag{141}$$

and uses a power level of $1/(1+\mathsf{INR}_3)$ to transmit.

The third transmitter makes a codebook of rate

$$R_3 = \log(1+\mathsf{SNR}_3(1+\mathsf{INR}_3)/(1+2\mathsf{INR}_3)) \tag{142}$$

and uses a power level of $1$ to transmit.

The first receiver is able to decode as rate $\le \log(1+\mathsf{SNR}_1)$. The third user is able to decode treating $R_2$ as noise. Further, the second user decodes $R_1$ and $R_2$ jointly. The decoding happens since

$$R_1 \le \log(1+\mathsf{INR}_2) \tag{143a}$$
$$R_2 \le \log(1+\mathsf{SNR}_2/(1+\mathsf{INR}_3)) \tag{143b}$$
$$R_1 + R_2 \le \log(1+\mathsf{INR}_2+\mathsf{SNR}_2/(1+\mathsf{INR}_3)) \tag{143c}$$

To show the sum rate, it is enough to prove

$$\begin{aligned}&\frac{\mathsf{INR}_2}{1+\frac{\mathsf{SNR}_2\mathsf{INR}_2}{1+\mathsf{SNR}_2+2\mathsf{INR}_2}} + \frac{1+\mathsf{INR}_2}{1+2\mathsf{INR}_2}\mathsf{SNR}_2/(1+\mathsf{INR}_3)\\ &+\frac{\mathsf{INR}_2}{1+\frac{\mathsf{SNR}_2\mathsf{INR}_2}{1+\mathsf{SNR}_2+2\mathsf{INR}_2}}\frac{1+\mathsf{INR}_2}{1+2\mathsf{INR}_2}\mathsf{SNR}_2/(1+\mathsf{INR}_3) \le \mathsf{INR}_2+\mathsf{SNR}_2/(1+\mathsf{INR}_3)\end{aligned} \tag{144a}$$

This is equivalent to proving

$$\begin{aligned}&(1+\mathsf{INR}_2)(1+\mathsf{SNR}_2+2\mathsf{INR}_2)\\ \le\ &\mathsf{INR}_2(1+2\mathsf{INR}_2)(1+\mathsf{INR}_3)+1+\mathsf{SNR}_2+2\mathsf{INR}_2+\mathsf{SNR}_2\mathsf{INR}_2\end{aligned} \tag{145}$$

which trivially holds.

*6)* $\mathsf{INR}_2 \ge \mathsf{SNR}_2$*,* $\mathsf{INR}_3 < \mathsf{SNR}_3$*,* $\mathsf{INR}_2 \le \mathsf{SNR}_1$*:* The first transmitter makes two codebooks, the first one of rate

$$R_{1,c} = \log\left(1 + \frac{\mathsf{INR}_2^2}{1+2\mathsf{INR}_2+\mathsf{SNR}_2(1+\mathsf{INR}_2)}\right), \tag{146}$$

and the second of rate

$$R_{1,p} = \log(1+\mathsf{SNR}_1/(1+\mathsf{INR}_2)). \tag{147}$$

The transmitter send the first one at a power of $\mathsf{INR}_2/(1+\mathsf{INR}_2)$ and the second one at a power of $1/(1+\mathsf{INR}_2)$, adds them up and transmit.

The second transmitter makes a codebook of rate

$$R_2 = \log\left(1 + \frac{1+\mathsf{INR}_2}{1+2\mathsf{INR}_2}\mathsf{SNR}_2/(1+\mathsf{INR}_3)\right), \tag{148}$$

and uses a power level of $1/(1+\mathsf{INR}_3)$ to transmit.

The third transmitter makes a codebook of rate

$$R_3 = \log\left(1+\mathsf{SNR}_3(1+\mathsf{INR}_3)/(1+2\mathsf{INR}_3)\right) \tag{149}$$

and uses a power level of $1$ to transmit.

The first decoder can decode in the same way as in 2-user Z, and the third user is able to decode treating $R_2$ as noise. The second receiver can decode $R_{1,c}$ and $R_2$ jointly treating $R_{1,p}$ as noise since

$$R_{1,c} \leq \log(1+\mathsf{INR}_2^2/(1+2\mathsf{INR}_2)) \tag{150a}$$

$$R_2 \leq \log\left(1+\mathsf{SNR}_2/(1+\mathsf{INR}_3)\frac{1+\mathsf{INR}_2}{1+2\mathsf{INR}_2}\right) \tag{150b}$$

$$R_{1,c}+R_2 \leq \log\left(1+\mathsf{INR}_2^2/(1+2\mathsf{INR}_2)+\mathsf{SNR}_2/(1+\mathsf{INR}_3)\frac{1+\mathsf{INR}_2}{1+2\mathsf{INR}_2}\right) \tag{150c}$$

These conditions are straightforward to verify and are therefore omitted.

*7)* $\mathsf{INR}_2 < \mathsf{SNR}_2$*,* $\mathsf{INR}_3 \geq \mathsf{SNR}_3$*,* $\mathsf{INR}_3 \geq \mathsf{SNR}_2$*:* The first transmitter makes a codebook of rate

$$R_1 = \log(1+\mathsf{SNR}_1/(1+\mathsf{INR}_2)), \tag{151}$$

and uses a power level of $1/(1+\mathsf{INR}_2)$ to transmit.

The second transmitter makes a codebook of rate

$$R_2 = \log\left(1+\frac{1+\mathsf{INR}_2}{1+2\mathsf{INR}_2}\min\left(\mathsf{SNR}_2, \frac{\mathsf{INR}_3}{1+\frac{\mathsf{SNR}_3\mathsf{INR}_3}{1+\mathsf{SNR}_3+2\mathsf{INR}_3}}\right)\right), \tag{152}$$

and uses a power level of 1 to transmit.

The third transmitter makes a codebook of rate

$$R_3 = \log\left(1+\mathsf{SNR}_3(1+\mathsf{INR}_3)/(1+2\mathsf{INR}_3)\right) \tag{153}$$

and uses a power level of 1 to transmit.

The 1st receiver and 3rd are able to decode by the same scheme of 2-user. The second receiver can decode treating $R_1$ as noise.

*8)* $\mathsf{INR}_2 < \mathsf{SNR}_2$*,* $\mathsf{INR}_3 \geq \mathsf{SNR}_3$*,* $\mathsf{INR}_3 < \mathsf{SNR}_2$*:* The first transmitter makes a codebook of rate

$$R_1 = \log(1+\mathsf{SNR}_1/(1+\mathsf{INR}_2)), \tag{154}$$

and uses a power level of $1/(1+\mathsf{INR}_2)$ to transmit.

The second transmitter makes two codebooks, the first one of rate

$$R_{2,c} = \log\left(1+\frac{1+\mathsf{INR}_2}{1+2\mathsf{INR}_2}\frac{\mathsf{INR}_3^2}{1+2\mathsf{INR}_3+\mathsf{SNR}_3(1+\mathsf{INR}_3)}\right), \tag{155}$$

and the second of rate

$$R_{2,p} = \log\left(1+\frac{1+\mathsf{INR}_2}{1+2\mathsf{INR}_2}\mathsf{SNR}_2/(1+\mathsf{INR}_3)\right). \tag{156}$$

The transmitter send the first one at a power of $\mathsf{INR}_3/(1+\mathsf{INR}_3)$ and the second one at a power of $1/(1+\mathsf{INR}_3)$, adds them up and transmit.

The third transmitter makes a codebook of rate

$$R_3 = \log\left(1+\mathsf{SNR}_3(1+\mathsf{INR}_3)/(1+2\mathsf{INR}_3)\right) \tag{157}$$

and uses a power level of 1 to transmit.

The second receiver treats $R_1$ as noise and the equation for decoding at second receiver follow the 2-user equations.

*9)* $\mathsf{INR}_2 < \mathsf{SNR}_2$ *and* $\mathsf{INR}_3 < \mathsf{SNR}_3$*:* The first transmitter makes a codebook of rate

$$R_1 = \log(1+\mathsf{SNR}_1/(1+\mathsf{INR}_2)), \tag{158}$$

and uses a power level of $1/(1+\mathsf{INR}_2)$ to transmit.

The second transmitter makes a codebook of rate

$$R_2 = \log\left(1+\frac{1+\mathsf{INR}_2}{1+2\mathsf{INR}_2}\mathsf{SNR}_2/(1+\mathsf{INR}_3)\right), \tag{159}$$

and uses a power level of $1/(1+\mathsf{INR}_3)$ to transmit. The third transmitter makes a codebook of rate

$$R_3 = \log\left(1+\mathsf{SNR}_3(1+\mathsf{INR}_3)/(1+2\mathsf{INR}_3)\right) \tag{160}$$

and uses a power level of $1$ to transmit.

Here, second user can treat first user's data as noise. Thus, all are able to decode.

## Appendix F
## The sum rate is within 4 bits with 2.5 rounds of message passing

### A. $\mathsf{INR}_2 \geq \mathsf{SNR}_2$, $\mathsf{INR}_3 \geq \mathsf{SNR}_3$, $\mathsf{INR}_2 \geq \mathsf{SNR}_1$, $\mathsf{INR}_3 \geq \mathsf{SNR}_2$

Note that we only need to consider the case when $\frac{\mathsf{INR}_2}{1+\frac{\mathsf{SNR}_2\mathsf{INR}_2}{1+\mathsf{SNR}_2+2\mathsf{INR}_2}} \leq \mathsf{SNR}_1$ and $\frac{\mathsf{INR}_3}{1+\frac{\mathsf{SNR}_3\mathsf{INR}_3}{1+\mathsf{SNR}_3+2\mathsf{INR}_3}} \leq \mathsf{SNR}_2$, since otherwise the 1.5 round scheme was within 4 bits of optimal.

The rates allocated to the users in this case are:

$$R_1 = \min(\log(1+\mathsf{SNR}_1), \log(1+\mathsf{INR}_2+\mathsf{SNR}_2)-R_2) \tag{161a}$$

$$R_2 = \log\left(1+\frac{1+\mathsf{INR}_2}{1+2\mathsf{INR}_2}\min\left(\mathsf{SNR}_2, \frac{\mathsf{INR}_3}{1+\frac{\mathsf{SNR}_3\mathsf{INR}_3}{1+\mathsf{SNR}_3+2\mathsf{INR}_3}}\right)\right) \tag{161b}$$

$$R_3 = \log(1+\mathsf{SNR}_3(1+\mathsf{INR}_3)/(1+2\mathsf{INR}_3)) \tag{161c}$$

For $\log(1+\mathsf{INR}_2+\mathsf{SNR}_2)-R_2 \geq \log(1+\mathsf{SNR}_1)$, the sum rate is within 3 bits since $R_1$ is optimal and $R_2+R_3$ is within 3 bits (2 due to Z, 1 additional due to user 2 backing off). For $\log(1+\mathsf{INR}_2+\mathsf{SNR}_2)-R_2 \leq \log(1+\mathsf{SNR}_1)$, $R_1+R_2$ is optimal and $R_3$ is within 1 bit.

### B. $\mathsf{INR}_2 \geq \mathsf{SNR}_2$, $\mathsf{INR}_3 \geq \mathsf{SNR}_3$, $\mathsf{INR}_2 \geq \mathsf{SNR}_1$, $\mathsf{INR}_3 < \mathsf{SNR}_2$

The rates allocated to the users in this case are:

$$R_1 = \min(\log(1+\mathsf{SNR}_1), \log(1+\mathsf{INR}_2)-R_2) \tag{162a}$$

$$\begin{aligned} R_2 = {} & \log\left(1+\frac{1+\mathsf{INR}_2}{1+2\mathsf{INR}_2}\frac{\mathsf{INR}_3^2}{1+2\mathsf{INR}_3+\mathsf{SNR}_3(1+\mathsf{INR}_3)}\right) \\ & +\log\left(1+\frac{1+\mathsf{INR}_2}{1+2\mathsf{INR}_2}\mathsf{SNR}_2/(1+\mathsf{INR}_3)\right) \end{aligned} \tag{162b}$$

$$R_3 = \log(1+\mathsf{SNR}_3(1+\mathsf{INR}_3)/(1+2\mathsf{INR}_3)) \tag{162c}$$

For $\log(1+\mathsf{INR}_2)-R_2 \geq \log(1+\mathsf{SNR}_1)$, the sum rate is within 3 bits since $R_1$ is optimal and $R_2+R_3$ is within 4 bits (2 due to Z, 2 additional due to user 2 backing off). For $\log(1+\mathsf{INR}_2)-R_2 \leq \log(1+\mathsf{SNR}_1)$, $R_1+R_2$ is within 1 bit and $R_3$ is within 1 bit making overall sum rate within 2 bits.

*C.* $\mathsf{INR}_2 \geq \mathsf{SNR}_2$, $\mathsf{INR}_3 \geq \mathsf{SNR}_3$, $\mathsf{INR}_2 < \mathsf{SNR}_1$, $\mathsf{INR}_3 \geq \mathsf{SNR}_2$

The rates allocated to the users in this case are:

$$\begin{aligned} R_1 &= \log\left(1+\frac{\mathsf{SNR}_1}{1+\mathsf{INR}_2}\right)+\min\left(\log\left(1+\frac{\mathsf{INR}_2^2}{1+2\mathsf{INR}_2}\right),\right.\\ &\left.\log\left(1+\frac{\mathsf{SNR}_2(1+\mathsf{INR}_2)+\mathsf{INR}_2^2}{1+2\mathsf{INR}_2}\right)-R_2\right) \quad (163a)\\ R_2 &= \log\left(1+\frac{1+\mathsf{INR}_2}{1+2\mathsf{INR}_2}\min\left(\mathsf{SNR}_2, \frac{\mathsf{INR}_3}{1+\frac{\mathsf{SNR}_3\mathsf{INR}_3}{1+\mathsf{SNR}_3+2\mathsf{INR}_3}}\right)\right) \quad (163b)\\ R_3 &= \log(1+\mathsf{SNR}_3(1+\mathsf{INR}_3)/(1+2\mathsf{INR}_3)) \quad (163c)\end{aligned}$$

For $\log\left(1+\frac{\mathsf{INR}_2^2}{1+2\mathsf{INR}_2}\right) \leq \log\left(1+\frac{\mathsf{SNR}_2(1+\mathsf{INR}_2)+\mathsf{INR}_2^2}{1+2\mathsf{INR}_2}\right)-R_2$:

$$R_1 \geq \log\left(1+\mathsf{SNR}_1/(1+\mathsf{INR}_2)\left(1+\frac{\mathsf{INR}_2^2}{1+2\mathsf{INR}_2}\right)\right) \geq \log(1+\mathsf{SNR}_1/2).$$

Thus, $R_1$ is within 1 bit of optimal and $R_2+R_3$ is within 3 bits thus making overall sum-rate within 4 bits.

For $\log\left(1+\frac{\mathsf{INR}_2^2}{1+2\mathsf{INR}_2}\right) \geq \log\left(1+\frac{\mathsf{SNR}_2(1+\mathsf{INR}_2)+\mathsf{INR}_2^2}{1+2\mathsf{INR}_2}\right)-R_2$,

$$\begin{aligned} R_1+R_2 &\geq \log\left(\left(1+\frac{\mathsf{SNR}_1}{1+\mathsf{INR}_2}\right)\left(1+\frac{\mathsf{SNR}_2(1+\mathsf{INR}_2)+\mathsf{INR}_2^2}{1+2\mathsf{INR}_2}\right)\right)\\ &\geq \log\left((1+\mathsf{SNR}_1)/(1+\mathsf{INR}_2)\left(\frac{(1+\mathsf{INR}_2)(1+\mathsf{INR}_2+\mathsf{SNR}_2)}{1+2\mathsf{INR}_2}\right)\right)\\ &\geq \log((1+\mathsf{SNR}_1)/(1+\mathsf{INR}_2)((1+\mathsf{INR}_2+\mathsf{SNR}_2)/2)) \quad (164)\end{aligned}$$

and thus $R_1+R_2$ is within 1 bit of optimal and $R_3$ is within 1 bit of optimal. Thus, the sum rate is within 2 bits.

*D.* $\mathsf{INR}_2 \geq \mathsf{SNR}_2$, $\mathsf{INR}_3 \geq \mathsf{SNR}_3$, $\mathsf{INR}_2 < \mathsf{SNR}_1$, $\mathsf{INR}_3 < \mathsf{SNR}_2$

The rates allocated to the users in this case are:

$$\begin{aligned} R_1 &= \log(1+\mathsf{SNR}_1/(1+\mathsf{INR}_2))+\min\left(\log\left(1+\frac{\mathsf{SNR}_1\mathsf{INR}_2}{1+\mathsf{SNR}_1+\mathsf{INR}_2}\right),\right.\\ &\left.\log\left(1+\frac{\mathsf{INR}_2^2}{1+2\mathsf{INR}_2}\right)-R_2\right) \quad (165a)\\ R_2 &= \log\left(1+\frac{1+\mathsf{INR}_2}{1+2\mathsf{INR}_2}\frac{\mathsf{INR}_3^2}{1+2\mathsf{INR}_3+\mathsf{SNR}_3(1+\mathsf{INR}_3)}\right)\\ &+\log\left(1+\frac{1+\mathsf{INR}_2}{1+2\mathsf{INR}_2}\mathsf{SNR}_2/(1+\mathsf{INR}_3)\right) \quad (165b)\\ R_3 &= \log(1+\mathsf{SNR}_3(1+\mathsf{INR}_3)/(1+2\mathsf{INR}_3)) \quad (165c)\end{aligned}$$

If $\log\left(1+\frac{\mathsf{SNR}_1\mathsf{INR}_2}{1+\mathsf{SNR}_1+\mathsf{INR}_2}\right) \leq \log\left(1+\frac{\mathsf{INR}_2^2}{1+2\mathsf{INR}_2}\right)-R_2$, $R_1 = \log\left(1+\frac{\mathsf{SNR}_1\mathsf{INR}_2}{1+\mathsf{SNR}_1+\mathsf{INR}_2}\right)+\log(1+\mathsf{SNR}_1/(1+\mathsf{INR}_2)) = \log(1+\mathsf{SNR}_1)$. Thus, the first transmitter transmits at optimal rate and $R_2+R_3$ is within 4 bits.

If $\log\left(1+\frac{\mathsf{SNR}_1\mathsf{INR}_2}{1+\mathsf{SNR}_1+\mathsf{INR}_2}\right) \geq \log\left(1+\frac{\mathsf{INR}_2^2}{1+2\mathsf{INR}_2}\right)-R_2$, $R_1+R_2 \geq \log(1+\mathsf{SNR}_1)+\log\left(1+\frac{\mathsf{SNR}_2}{1+\mathsf{INR}_2}\right)-2$. Thus, the overall sum rate is within 3 bits of optimal.

*E.* $\mathsf{INR}_2 \geq \mathsf{SNR}_2$, $\mathsf{INR}_3 < \mathsf{SNR}_3$, $\mathsf{INR}_2 \geq \mathsf{SNR}_1$

The rates allocated to the users in this case are:

$$R_1 = \min\left(\log(1+\mathsf{SNR}_1), \log\left(1+\mathsf{INR}_2+\frac{\mathsf{SNR}_2}{1+\mathsf{INR}_3}\right)-R_2\right) \tag{166a}$$

$$R_2 = \log\left(1+\frac{1+\mathsf{INR}_2}{1+2\mathsf{INR}_2}\mathsf{SNR}_2/(1+\mathsf{INR}_3)\right) \tag{166b}$$

$$R_3 = \log(1+\mathsf{SNR}_3(1+\mathsf{INR}_3)/(1+2\mathsf{INR}_3)) \tag{166c}$$

If $\log(1+\mathsf{SNR}_1) \leq \log\left(1+\mathsf{INR}_2+\frac{\mathsf{SNR}_2}{1+\mathsf{INR}_3}\right)-R_2$, then user 1 transmits at the optimal rate and $R_2+R_3$ is within 3 bits of optimal sum-rate, thus, we achieve within 3 bits of the sum capacity.

If $\log(1+\mathsf{SNR}_1) \geq \log\left(1+\mathsf{INR}_2+\frac{\mathsf{SNR}_2}{1+\mathsf{INR}_3}\right)-R_2$, $R_1+R_2 \geq \log\left(1+\mathsf{INR}_2+\frac{\mathsf{SNR}_2}{1+\mathsf{INR}_3}\right) \geq \log(1+\mathsf{INR}_2) \geq \log(1+\mathsf{INR}_2+\mathsf{SNR}_2)-1$. Thus, $R_1+R_2$ is within 1 bit of optimal and $R_3$ is within 1 bit and thus the sum rate is within 2 bits of optimal.

*F.* $\mathsf{INR}_2 \geq \mathsf{SNR}_2$, $\mathsf{INR}_3 < \mathsf{SNR}_3$, $\mathsf{INR}_2 < \mathsf{SNR}_1$

The rates allocated to the users in this case are:

$$\begin{aligned} R_1 = {} & \log(1+\mathsf{SNR}_1/(1+\mathsf{INR}_2)) + \min\left(\log\left(1+\frac{\mathsf{SNR}_1\mathsf{INR}_2}{1+\mathsf{SNR}_1+\mathsf{INR}_2}\right),\right. \\ & \left.\log\left(1+\frac{\mathsf{INR}_2^2}{1+2\mathsf{INR}_2}\right)-R_2\right) \end{aligned} \tag{167a}$$

$$R_2 = \log\left(1+\frac{1+\mathsf{INR}_2}{1+2\mathsf{INR}_2}\mathsf{SNR}_2/(1+\mathsf{INR}_3)\right) \tag{167b}$$

$$R_3 = \log(1+\mathsf{SNR}_3(1+\mathsf{INR}_3)/(1+2\mathsf{INR}_3)) \tag{167c}$$

If $\log\left(1+\frac{\mathsf{SNR}_1\mathsf{INR}_2}{1+\mathsf{SNR}_1+\mathsf{INR}_2}\right) \leq \log\left(1+\frac{\mathsf{INR}_2^2}{1+2\mathsf{INR}_2}\right)-R_2$, $R_1 = \log(1+\mathsf{SNR}_1)$ and $R_2+R_3$ is within 3 bits of optimal and thus the sum rate is within 3 bits of optimal

If $\log\left(1+\frac{\mathsf{SNR}_1\mathsf{INR}_2}{1+\mathsf{SNR}_1+\mathsf{INR}_2}\right) \geq \log\left(1+\frac{\mathsf{INR}_2^2}{1+2\mathsf{INR}_2}\right)-R_2$, $R_1+R_2 = \log\left(1+\frac{\mathsf{INR}_2^2}{1+2\mathsf{INR}_2}\right)+\log(1+\mathsf{SNR}_1/(1+\mathsf{INR}_2))$ which is within 2 bits of optimal as in case D. Thus, the overall sum rate is within 3 bits of optimal.

*G.* $\mathsf{INR}_2 < \mathsf{SNR}_2$, $\mathsf{INR}_3 \geq \mathsf{SNR}_3$, $\mathsf{INR}_2 \geq \mathsf{SNR}_1$, $\mathsf{INR}_3 \geq \mathsf{SNR}_2$

The rates allocated to the users in this case are:

$$R_1 = \min(\log(1+\mathsf{SNR}_1), \log(1+\mathsf{INR}_2+\mathsf{SNR}_2)-R_2) \tag{168a}$$

$$R_2 = \log\left(1+\frac{1+\mathsf{INR}_2}{1+2\mathsf{INR}_2}\min\left(\mathsf{SNR}_2, \frac{\mathsf{INR}_3}{1+\frac{\mathsf{SNR}_3\mathsf{INR}_3}{1+\mathsf{SNR}_3+2\mathsf{INR}_3}}\right)\right) \tag{168b}$$

$$R_3 = \log(1+\mathsf{SNR}_3(1+\mathsf{INR}_3)/(1+2\mathsf{INR}_3)) \tag{168c}$$

If $\log(1+\mathsf{SNR}_1) \leq \log(1+\mathsf{INR}_2+\mathsf{SNR}_2)-R_2$, the first user sends at optimal rate and $R_2+R_3$ is within 3 bits of optimal and thus the sum capacity is within 3 bits of optimal.

If $\log(1+\mathsf{SNR}_1) \geq \log(1+\mathsf{INR}_2+\mathsf{SNR}_2)-R_2$, $R_1+R_2$ is optimal and thus the sum capacity is within 1 bit of optimal.

*H.* $\mathsf{INR}_2 < \mathsf{SNR}_2$, $\mathsf{INR}_3 \geq \mathsf{SNR}_3$, $\mathsf{INR}_2 < \mathsf{SNR}_1$, $\mathsf{INR}_3 \geq \mathsf{SNR}_2$

The rates allocated to the users in this case are:

$$\begin{aligned} R_1 &= \log\left(1+\frac{\mathsf{SNR}_1}{1+\mathsf{INR}_2}\right)+\min\left(\log\left(1+\frac{\mathsf{INR}_2^2}{1+2\mathsf{INR}_2}\right),\right. \\ &\quad \left.\log\left(1+\frac{\mathsf{INR}_2^2+\mathsf{SNR}_2(1+\mathsf{INR}_2)}{1+2\mathsf{INR}_2}\right)-R_2\right) && \text{(169a)} \\ R_2 &= \log\left(1+\frac{1+\mathsf{INR}_2}{1+2\mathsf{INR}_2}\min\left(\mathsf{SNR}_2, \frac{\mathsf{INR}_3}{1+\frac{\mathsf{SNR}_3\mathsf{INR}_3}{1+\mathsf{SNR}_3+2\mathsf{INR}_3}}\right)\right) && \text{(169b)} \\ R_3 &= \log(1+\mathsf{SNR}_3(1+\mathsf{INR}_3)/(1+2\mathsf{INR}_3)) && \text{(169c)} \end{aligned}$$

If $\log\left(1+\frac{\mathsf{INR}_2^2}{1+2\mathsf{INR}_2}\right) \leq \log\left(1+\frac{\mathsf{INR}_2^2+\mathsf{SNR}_2(1+\mathsf{INR}_2)}{1+2\mathsf{INR}_2}\right) - R_2$, then

$$R_1 \geq \log\left(1+\frac{\mathsf{SNR}_1}{1+\mathsf{INR}_2}\left(1+\frac{\mathsf{INR}_2^2}{1+2\mathsf{INR}_2}\right)\right) \geq \log(1+\mathsf{SNR}_1)-1.$$

Thus, the sum rate is within 4 bits of the optimal.

If $\log\left(1+\frac{\mathsf{INR}_2^2}{1+2\mathsf{INR}_2}\right) \geq \log\left(1+\frac{\mathsf{INR}_2^2+\mathsf{SNR}_2(1+\mathsf{INR}_2)}{1+2\mathsf{INR}_2}\right) - R_2$,

$$\begin{aligned} R_1+R_2 &= \log\left(1+\frac{\mathsf{INR}_2^2+\mathsf{SNR}_2(1+\mathsf{INR}_2)}{1+2\mathsf{INR}_2}\right)+\log\left(1+\frac{\mathsf{SNR}_1}{1+\mathsf{INR}_2}\right) \\ &\geq \log(1+\mathsf{SNR}_2/2)+\log(1+\mathsf{SNR}_1)-\log(1+\mathsf{INR}_2) \\ &\geq \log(1+\mathsf{SNR}_2+\mathsf{INR}_2)-2+\log(1+\mathsf{SNR}_1)-\log(1+\mathsf{INR}_2). && \text{(170)} \end{aligned}$$

Thus, the sum rate is within 3 bits of the optimal.

*I.* $\mathsf{INR}_2 < \mathsf{SNR}_2$, $\mathsf{INR}_3 \geq \mathsf{SNR}_3$, $\mathsf{INR}_2 \geq \mathsf{SNR}_1$, $\mathsf{INR}_3 < \mathsf{SNR}_2$

The rates allocated to the users in this case are:

$$\begin{aligned} R_1 &= \min\left(\log(1+\mathsf{SNR}_1), \log\left(1+\mathsf{INR}_2+\frac{\mathsf{SNR}_2\mathsf{INR}_3}{1+\mathsf{INR}_3}\right)-R_{2,c},\right. \\ &\quad \left.\log\left(1+\mathsf{INR}_2+\frac{\mathsf{SNR}_2}{1+\mathsf{INR}_3}\right)-R_{2,p}, \log(1+\mathsf{INR}_2+\mathsf{SNR}_2)-R_2\right) && \text{(171a)} \\ R_2 &= \log\left(1+\frac{1+\mathsf{INR}_2}{1+2\mathsf{INR}_2}\frac{\mathsf{INR}_3^2}{1+2\mathsf{INR}_3+\mathsf{SNR}_3(1+\mathsf{INR}_3)}\right) \\ &\quad +\log\left(1+\frac{1+\mathsf{INR}_2}{1+2\mathsf{INR}_2}\mathsf{SNR}_2/(1+\mathsf{INR}_3)\right) && \text{(171b)} \\ R_3 &= \log(1+\mathsf{SNR}_3(1+\mathsf{INR}_3)/(1+2\mathsf{INR}_3)) && \text{(171c)} \end{aligned}$$

We consider the four cases when the corresponding term in $R_1$ is the minimum:

1) Minimum is $\log(1+\mathsf{SNR}_1)$: $R_1$ is optimal and $R_2+R_3$ is within 4 bits.
2) Minimum is $\log\left(1+\mathsf{INR}_2+\frac{\mathsf{SNR}_2\mathsf{INR}_3}{1+\mathsf{INR}_3}\right)-R_{2,c}$:

$$\begin{aligned} R_1+R_2 &\geq \log\left(1+\mathsf{INR}_2+\frac{\mathsf{SNR}_2\mathsf{INR}_3}{1+\mathsf{INR}_3}\right)+\log\left(1+\frac{1+\mathsf{INR}_2}{1+2\mathsf{INR}_2}\mathsf{SNR}_2/(1+\mathsf{INR}_3)\right) \\ &\geq \log(1+\mathsf{INR}_2+\mathsf{SNR}_2)-1. \end{aligned}$$

Thus, $R_1+R_2$ and $R_3$ are both within 1 bit resulting in sum capacity within 2 bits.

3) Minimum is $\log\left(1+\mathsf{INR}_2+\frac{\mathsf{SNR}_2}{1+\mathsf{INR}_3}\right)-R_{2,p}$:

$$\begin{aligned} & R_1+R_2+R_3 \\ = \; & \log\left(1+\mathsf{INR}_2+\frac{\mathsf{SNR}_2}{1+\mathsf{INR}_3}\right)+\log\left(1+\frac{1+\mathsf{INR}_2}{1+2\mathsf{INR}_2}\frac{\mathsf{INR}_3^2}{1+2\mathsf{INR}_3+\mathsf{SNR}_3(1+\mathsf{INR}_3)}\right) \\ & +\log(1+\mathsf{SNR}_3(1+\mathsf{INR}_3)/(1+2\mathsf{INR}_3)) \qquad (172) \\ \geq \; & \log\left(1+\mathsf{INR}_2+\frac{\mathsf{SNR}_2}{1+\mathsf{INR}_3}\right)+\log(1+\mathsf{INR}_3+\mathsf{SNR}_3)-2. \qquad (173) \end{aligned}$$

The first term $\log\left(1+\mathsf{INR}_2+\frac{\mathsf{SNR}_2}{1+\mathsf{INR}_3}\right)\geq\log(1+\mathsf{INR}_2)$ and also $\geq\log(1+\mathsf{INR}_2+\mathsf{SNR}_2)-\log(1+\mathsf{INR}_3)$. Thus, the above is within 2 bits of optimal sum capacity.

4) Minimum is $\log(1+\mathsf{INR}_2+\mathsf{SNR}_2)-R_2$: $R_1+R_2$ is optimal and $R_3$ is within 1 bit of optimal; thus the sum capacity is achieved within 1 bit.

*J.* $\mathsf{INR}_2<\mathsf{SNR}_2$, $\mathsf{INR}_3\geq\mathsf{SNR}_3$, $\mathsf{INR}_2<\mathsf{SNR}_1$, $\mathsf{INR}_3<\mathsf{SNR}_2$

The rates allocated to the users in this case are:

$$\begin{aligned} R_1 &= \log\left(1+\frac{\mathsf{SNR}_1}{1+\mathsf{INR}_2}\right)+\min\left(\log\left(1+\frac{\mathsf{INR}_2^2}{1+2\mathsf{INR}_2}\right),\right. \\ & \log\left(1+\frac{\mathsf{INR}_2^2+\mathsf{SNR}_2\mathsf{INR}_3(1+\mathsf{INR}_2)/(1+\mathsf{INR}_3)}{1+2\mathsf{INR}_2}\right)-R_{2,c}, \\ & \log\left(1+\frac{\mathsf{INR}_2^2+\mathsf{SNR}_2(1+\mathsf{INR}_2)/(1+\mathsf{INR}_3)}{1+2\mathsf{INR}_2}\right)-R_{2,p}, \\ & \left.\log\left(1+\frac{\mathsf{INR}_2^2+\mathsf{SNR}_2(1+\mathsf{INR}_2)}{1+2\mathsf{INR}_2}\right)-R_2\right) \qquad (174a) \\ R_2 &= \log\left(1+\frac{1+\mathsf{INR}_2}{1+2\mathsf{INR}_2}\frac{\mathsf{INR}_3^2}{1+2\mathsf{INR}_3+\mathsf{SNR}_3(1+\mathsf{INR}_3)}\right) \\ & +\log\left(1+\frac{1+\mathsf{INR}_2}{1+2\mathsf{INR}_2}\mathsf{SNR}_2/(1+\mathsf{INR}_3)\right) \qquad (174b) \\ R_3 &= \log(1+\mathsf{SNR}_3(1+\mathsf{INR}_3)/(1+2\mathsf{INR}_3)) \qquad (174c) \end{aligned}$$

We consider the four cases when the corresponding term in $R_{1,c}$ is the minimum:

1) Minimum is $\log\left(1+\frac{\mathsf{INR}_2^2}{1+2\mathsf{INR}_2}\right)$: $R_1=\log\left(1+\frac{\mathsf{INR}_2^2}{1+2\mathsf{INR}_2}\right)+\log\left(1+\frac{\mathsf{SNR}_1}{1+\mathsf{INR}_2}\right)$ is within 1 bit of $\log(1+\mathsf{SNR}_1)$ (will be shown in Appendix F.N). Moreover, $R_{2,c}+R_3$ is within 2 bits of $\log(1+\mathsf{INR}_3+\mathsf{SNR}_3)$ (shown in Appendix F.I). $R_{2,p}$ is within 1 bit of $\log(1+\mathsf{SNR}_2)-\log(1+\mathsf{INR}_3)$. Thus, $R_2+R_3$ is within 3 bits of optimal and $R_1$ within 1 bit of optimal; thus the overall sum rate is within 4 bits of optimal.

2) Minimum is $\log\left(1+\frac{\mathsf{INR}_2^2+\mathsf{SNR}_2\mathsf{INR}_3(1+\mathsf{INR}_2)/(1+\mathsf{INR}_3)}{1+2\mathsf{INR}_2}\right)-R_{2,c}$: In this case,

$$R_{1,c}+R_2\geq\log\left(1+\frac{\mathsf{INR}_2^2+\mathsf{SNR}_2(1+\mathsf{INR}_2)}{1+2\mathsf{INR}_2}\right).$$

This sum rate will be larger than in the fourth case, and is therefore omitted here.

3) Minimum is $\log\left(1+\frac{\mathsf{INR}_2^2+\mathsf{SNR}_2(1+\mathsf{INR}_2)/(1+\mathsf{INR}_3)}{1+2\mathsf{INR}_2}\right)-R_{2,p}$: In this case, $R_1+R_2+R_3=\log\left(1+\frac{\mathsf{SNR}_1}{1+\mathsf{INR}_2}\right)+\log\left(1+\frac{\mathsf{INR}_2^2+\mathsf{SNR}_2(1+\mathsf{INR}_2)/(1+\mathsf{INR}_3)}{1+2\mathsf{INR}_2}\right)+\log\left(1+\frac{1+\mathsf{INR}_2}{1+2\mathsf{INR}_2}\frac{\mathsf{INR}_3^2}{1+2\mathsf{INR}_3+\mathsf{SNR}_3(1+\mathsf{INR}_3)}\right)+\log(1+\mathsf{SNR}_3(1+\mathsf{INR}_3)/(1+2\mathsf{INR}_3))\geq\log\left(1+\frac{\mathsf{SNR}_1}{1+\mathsf{INR}_2}\right)+\log\left(1+\frac{\mathsf{INR}_2^2+\mathsf{SNR}_2(1+\mathsf{INR}_2)/(1+\mathsf{INR}_3)}{1+2\mathsf{INR}_2}\right)+\log(1+\mathsf{INR}_3+$

$\mathsf{SNR}_3) - 2$.

- If $\mathsf{INR}_3(\mathsf{INR}_2 + 1) \geq \mathsf{SNR}_2$:

$$\begin{aligned} & \log\left(1 + \frac{\mathsf{SNR}_1}{1 + \mathsf{INR}_2}\right) + \log\left(1 + \frac{\mathsf{INR}_2^2 + \mathsf{SNR}_2(1 + \mathsf{INR}_2)/(1 + \mathsf{INR}_3)}{1 + 2\mathsf{INR}_2}\right) \\ \geq & \log\left(1 + \frac{\mathsf{SNR}_1}{1 + \mathsf{INR}_2}\right) + \log\left(1 + \frac{\mathsf{INR}_2^2}{1 + 2\mathsf{INR}_2}\right) \\ \geq & \log(1 + \mathsf{SNR}_1) - 1. \end{aligned}$$

Thus, the rate within 3 bits of sum capacity can be achieved.

- If $\mathsf{INR}_3(\mathsf{INR}_2 + 1) \leq \mathsf{SNR}_2$: $\log\left(1 + \frac{\mathsf{SNR}_1}{1+\mathsf{INR}_2}\right) + \log\left(1 + \frac{\mathsf{INR}_2^2 + \mathsf{SNR}_2(1+\mathsf{INR}_2)/(1+\mathsf{INR}_3)}{1+2\mathsf{INR}_2}\right) \geq \log(1 + \mathsf{SNR}_1) - \log(1 + \mathsf{INR}_2) + \log(1 + \mathsf{INR}_2 + \mathsf{SNR}_2/(1 + \mathsf{INR}_3)) - 1 \geq \log(1 + \mathsf{SNR}_1) - \log(1 + \mathsf{INR}_2) + \log(1 + \mathsf{INR}_2 + \mathsf{SNR}_2) - \log(1 + \mathsf{INR}_3) - 1$. Thus, the rate within 3 bits of sum capacity can be achieved.

4) Minimum is $\log\left(1 + \frac{\mathsf{INR}_2^2 + \mathsf{SNR}_2(1+\mathsf{INR}_2)}{1+2\mathsf{INR}_2}\right) - R_2$: In this case, $R_1 + R_2$ is within 1 bit of optimal ($\log(1+\mathsf{SNR}_1) + \log\left(1 + \frac{\mathsf{SNR}_2}{1+\mathsf{INR}_2}\right)$) and $R_3$ is also within 1 bit of optimal; thus sum capacity within 2 bits can be achieved.

*K.* $\mathsf{INR}_2 < \mathsf{SNR}_2$, $\mathsf{INR}_3 < \mathsf{SNR}_3$, $\mathsf{INR}_2 \geq \mathsf{SNR}_1$, $\mathsf{INR}_3 \geq \mathsf{SNR}_2$

The rates allocated to the users in this case are:

$$R_1 = \min\left(\log(1 + \mathsf{SNR}_1), \log\left(1 + \mathsf{INR}_2 + \frac{\mathsf{SNR}_2}{1 + \mathsf{INR}_3}\right) - R_2\right) \quad (175a)$$

$$R_2 = \log\left(1 + \frac{1 + \mathsf{INR}_2}{1 + 2\mathsf{INR}_2}\mathsf{SNR}_2/(1 + \mathsf{INR}_3)\right) \quad (175b)$$

$$R_3 = \log(1 + \mathsf{SNR}_3(1 + \mathsf{INR}_3)/(1 + 2\mathsf{INR}_3)) \quad (175c)$$

If $\log(1 + \mathsf{SNR}_1) \leq \log\left(1 + \mathsf{INR}_2 + \frac{\mathsf{SNR}_2}{1+\mathsf{INR}_3}\right) - R_2$, $R_2 + R_3$ is within 3 bits of optimal and thus the sum rate within 3 bits can be achieved.

If $\log(1 + \mathsf{SNR}_1) \geq \log\left(1 + \mathsf{INR}_2 + \frac{\mathsf{SNR}_2}{1+\mathsf{INR}_3}\right) - R_2$, $R_1 + R_2 + R_3 \geq \log\left(1 + \mathsf{INR}_2 + \frac{\mathsf{SNR}_2}{1+\mathsf{INR}_3}\right) + R_3 \geq \log(1 + \mathsf{SNR}_1) + R_3 \geq \log(1 + \mathsf{SNR}_1) + \log(1 + \mathsf{SNR}_3) - 1 \geq \log(1 + \mathsf{SNR}_1) + \log(1 + \mathsf{SNR}_3 + \mathsf{INR}_3) - 2$. Thus, sum capacity within 2 bits can be achieved.

*L.* $\mathsf{INR}_2 < \mathsf{SNR}_2$, $\mathsf{INR}_3 < \mathsf{SNR}_3$, $\mathsf{INR}_2 \geq \mathsf{SNR}_1$, $\mathsf{INR}_3 < \mathsf{SNR}_2$

Note that we will focus on $\mathsf{SNR}_2 \leq \mathsf{INR}_3(\mathsf{INR}_2 + 1)$, since for $\mathsf{SNR}_2 \geq \mathsf{INR}_3(\mathsf{INR}_2 + 1)$, one and a half round was already within 4 bits.

The rates allocated to the users in this case are:

$$R_1 = \min\left(\log(1 + \mathsf{SNR}_1), \log\left(1 + \mathsf{INR}_2 + \frac{\mathsf{SNR}_2}{1 + \mathsf{INR}_3}\right) - R_2\right) \quad (176a)$$

$$R_2 = \log\left(1 + \frac{1 + \mathsf{INR}_2}{1 + 2\mathsf{INR}_2}\mathsf{SNR}_2/(1 + \mathsf{INR}_3)\right) \quad (176b)$$

$$R_3 = \log(1 + \mathsf{SNR}_3(1 + \mathsf{INR}_3)/(1 + 2\mathsf{INR}_3)) \quad (176c)$$

If $\log(1 + \mathsf{SNR}_1) \leq \log\left(1 + \mathsf{INR}_2 + \frac{\mathsf{SNR}_2}{1+\mathsf{INR}_3}\right) - R_2$, $R_2 + R_3$ is within 3 bits of optimal and thus the sum rate within 3 bits can be achieved.

If $\log(1 + \mathsf{SNR}_1) \geq \log\left(1 + \mathsf{INR}_2 + \frac{\mathsf{SNR}_2}{1+\mathsf{INR}_3}\right) - R_2$, $R_1 + R_2 + R_3 \geq \log\left(1 + \mathsf{INR}_2 + \frac{\mathsf{SNR}_2}{1+\mathsf{INR}_3}\right) + R_3 \geq \log(1 + \mathsf{INR}_2) + R_3 \geq \log(1 + \mathsf{INR}_2) + \log(1 + \mathsf{SNR}_3) - 1 \geq \log(1 + \mathsf{INR}_2) + \log(1 + \mathsf{SNR}_3 + \mathsf{INR}_3) - 2$. Thus, sum capacity within 2 bits can be achieved.

*M.* $\mathsf{INR}_2 < \mathsf{SNR}_2$, $\mathsf{INR}_3 < \mathsf{SNR}_3$, $\mathsf{INR}_2 < \mathsf{SNR}_1$, $\mathsf{INR}_3 \geq \mathsf{SNR}_2$

The rates allocated to the users in this case are:

$$\begin{aligned} R_1 &= \log\left(1+\frac{\mathsf{SNR}_1}{1+\mathsf{INR}_2}\right)+\min\left(\log\left(1+\frac{\mathsf{INR}_2^2}{1+2\mathsf{INR}_2}\right),\right. \\ &\quad \left.\log\left(1+\frac{\mathsf{INR}_2^2+\mathsf{SNR}_2(1+\mathsf{INR}_2)}{1+2\mathsf{INR}_2}\right)-R_2\right) && \text{(177a)} \\ R_2 &= \log\left(1+\frac{1+\mathsf{INR}_2}{1+2\mathsf{INR}_2}\mathsf{SNR}_2/(1+\mathsf{INR}_3)\right) && \text{(177b)} \\ R_3 &= \log(1+\mathsf{SNR}_3(1+\mathsf{INR}_3)/(1+2\mathsf{INR}_3)) && \text{(177c)} \end{aligned}$$

If $\log\left(1+\frac{\mathsf{INR}_2^2}{1+2\mathsf{INR}_2}\right) \leq \log\left(1+\frac{\mathsf{INR}_2^2+\mathsf{SNR}_2(1+\mathsf{INR}_2)}{1+2\mathsf{INR}_2}\right)-R_2$, then $R_1 \geq \log\left(1+\frac{\mathsf{SNR}_1}{1+\mathsf{INR}_2}(1+\frac{\mathsf{INR}_2^2}{1+2\mathsf{INR}_2})\right) \geq \log(1+\mathsf{SNR}_1)-1$. Thus, the sum rate is within 4 bits of the optimal.

If $\log\left(1+\frac{\mathsf{INR}_2^2}{1+2\mathsf{INR}_2}\right) \geq \log\left(1+\frac{\mathsf{INR}_2^2+\mathsf{SNR}_2(1+\mathsf{INR}_2)}{1+2\mathsf{INR}_2}\right)-R_2$, $R_1+R_2 = \log\left(1+\frac{\mathsf{INR}_2^2+\mathsf{SNR}_2(1+\mathsf{INR}_2)}{1+2\mathsf{INR}_2}\right)+\log\left(1+\frac{\mathsf{SNR}_1}{1+\mathsf{INR}_2}\right) \geq \log(1+\mathsf{SNR}_1)+\log\left(1+\frac{\mathsf{INR}_2^2+\mathsf{SNR}_2(1+\mathsf{INR}_2)}{1+2\mathsf{INR}_2}\right)-\log(1+\mathsf{INR}_2) \geq \log(1+\mathsf{SNR}_1)$. Thus, $R_1+R_2+R_3 \geq \log(1+\mathsf{SNR}_1)+\log(1+\mathsf{SNR}_3) \geq \log(1+\mathsf{SNR}_1)+\log(1+\mathsf{SNR}_3+\mathsf{INR}_3)-1$. Thus, the sum rate within 1 bit can be achieved.

*N.* $\mathsf{INR}_2 < \mathsf{SNR}_2$, $\mathsf{INR}_3 < \mathsf{SNR}_3$, $\mathsf{INR}_2 < \mathsf{SNR}_1$, $\mathsf{INR}_3 < \mathsf{SNR}_2$

Note that we will focus on $\mathsf{SNR}_2 \leq \mathsf{INR}_3(\mathsf{INR}_2+1)$, since for $\mathsf{SNR}_2 \geq \mathsf{INR}_3(\mathsf{INR}_2+1)$, one and a half round was already within 4 bits.

The rates allocated to the users in this case are:

$$\begin{aligned} R_1 &= \log\left(1+\frac{\mathsf{SNR}_1}{1+\mathsf{INR}_2}\right)+\min\left(\log\left(1+\frac{\mathsf{INR}_2^2}{1+2\mathsf{INR}_2}\right),\right. \\ &\quad \left.\log\left(1+\frac{\mathsf{INR}_2^2+\mathsf{SNR}_2(1+\mathsf{INR}_2)}{1+2\mathsf{INR}_2}\right)-R_2\right) && \text{(178a)} \\ R_2 &= \log\left(1+\frac{1+\mathsf{INR}_2}{1+2\mathsf{INR}_2}\mathsf{SNR}_2/(1+\mathsf{INR}_3)\right) && \text{(178b)} \\ R_3 &= \log(1+\mathsf{SNR}_3(1+\mathsf{INR}_3)/(1+2\mathsf{INR}_3)) && \text{(178c)} \end{aligned}$$

If $\log\left(1+\frac{\mathsf{INR}_2^2}{1+2\mathsf{INR}_2}\right) \leq \log\left(1+\frac{\mathsf{INR}_2^2+\mathsf{SNR}_2(1+\mathsf{INR}_2)}{1+2\mathsf{INR}_2}\right)-R_2$, then

$$R_1 \geq \log\left(1+\frac{\mathsf{SNR}_1}{1+\mathsf{INR}_2}\left(1+\frac{\mathsf{INR}_2^2}{1+2\mathsf{INR}_2}\right)\right) \geq \log(1+\mathsf{SNR}_1)-1.$$

Thus, the sum rate is within 4 bits of the optimal.

If $\log\left(1+\frac{\mathsf{INR}_2^2}{1+2\mathsf{INR}_2}\right) \geq \log\left(1+\frac{\mathsf{INR}_2^2+\mathsf{SNR}_2(1+\mathsf{INR}_2)}{1+2\mathsf{INR}_2}\right)-R_2$,

$$\begin{aligned} R_1+R_2 &= \log\left(1+\frac{\mathsf{INR}_2^2+\mathsf{SNR}_2(1+\mathsf{INR}_2)}{1+2\mathsf{INR}_2}\right)+\log\left(1+\frac{\mathsf{SNR}_1}{1+\mathsf{INR}_2}\right) \\ &\geq \log(1+\mathsf{SNR}_1)+\log\left(1+\frac{\mathsf{INR}_2^2+\mathsf{SNR}_2(1+\mathsf{INR}_2)}{1+2\mathsf{INR}_2}\right)-\log(1+\mathsf{INR}_2) \\ &\geq \log(1+\mathsf{SNR}_1). \end{aligned}$$

Thus, $R_1+R_2+R_3 \geq \log(1+\mathsf{SNR}_1)+\log(1+\mathsf{SNR}_3) \geq \log(1+\mathsf{SNR}_1)+\log(1+\mathsf{SNR}_3+\mathsf{INR}_3)-1$. Thus, the sum rate within 1 bit can be achieved.